\documentclass[botnum, fleqn, final]{unmeethesis}
\pdfoutput=1

\usepackage{amsmath,amsfonts,amssymb,amstext}
\usepackage{graphicx}
\usepackage[bookmarks = true, pdfpagemode = None, pdfstartview = FitH, colorlinks = true, urlcolor = blue,hyperfootnotes=false]{hyperref}
\usepackage{hypcap}
\usepackage{multirow}
\usepackage{listings}

\newcommand{\bra}[1]{\ensuremath{\langle #1 \vert}}
\newcommand{\ket}[1]{\ensuremath{\vert #1 \rangle}}
\newcommand{\braket}[2]{\ensuremath{\langle #1 \vert #2 \rangle}}
\newcommand{\be}{\begin{equation}}
\newcommand{\ee}{\end{equation}}
\newcommand{\bea}{\begin{eqnarray}}
\newcommand{\eea}{\end{eqnarray}}
\newcommand{\bF}{\begin{figure}}
\newcommand{\eF}{\end{figure}}

\newcommand{\bi}{\begin{itemize}}
\newcommand{\ei}{\end{itemize}}
\newcommand{\ud}{\mathrm{d}}
\newcommand{\mbf}[1]{\mathbf{#1}}
\newcommand{\rf}{\text{rf}}
\newcommand{\mw}{\mu\text{w}}
\newcommand{\cv}{\mathbf{c}}
\newcommand{\Tr}{\textrm{Tr}}

\newcommand{\stm}[1]{\mathbf{}}

\newtheorem{theorem}{Theorem}
\newtheorem{lemma}{Lemma}

\begin{document}

\frontmatter

\title{Quantum Control of d-Dimensional Quantum Systems with Application to Alkali Atomic Spins}

\author{Seth Merkel}

\degreesubject{Ph.D., Physics}

\degree{Doctorate of Philosophy \\ Physics}

\documenttype{Dissertation}

\previousdegrees{B.S., Worcester Polytechnic Institute, 2003}

\date{June, \thisyear}

\maketitle

\makecopyright

\begin{dedication}

To my parents, for teaching me to appreciate a challenge.
   
\end{dedication}

\begin{acknowledgments}

First and foremost I would like to thank my advisor Ivan Deutsch.  Throughout my tenure in his group he has provided not only knowledge and direction about physics, but also knowledge and direction on how to be a physicist.  

UNM has a fantastic quantum information community which is due in large part to the tireless efforts of Carlton Caves.  I would like to thank the people from UNM I have collaborated with including Andrew Silberfarb, Carlos Riofrio, Steve Flammia, Matt Elliot and Brian Mischuck.  Some of the other quantum info group members who have provided invaluable criticism and suggestions over the years are Andrew Landahl, JM Geremia, Collin Trail, Pat Rice, Iris Reichenbach, Aaron Denney, Brad Chase, Anil Shaji, David Hayes, Animesh Datta, Satyan Bhongale, Sergio Boixo, Rob Cook, Heather Partner, Nick Menicucci and Alex Tacla.

Additionally, I would like to thank some of the professors at UNM outside of the info physics group.  All the classes I've taken at UNM were of exceptional quality, in particular Colston Chandler's quantum mechanics courses and Dan Finley's general relativity class as well as his math seminars.  I'd also like to thank Gary Herling for explaining Monte Carlo sampling to me.

Finally, I would like to thank my collaborators outside UNM including Gavin Brennen, Dan Browne, Akimasa Miyake and Tony Short as well as Poul Jessen and his lab at the University of Arizona, Souma Chaudhury, Aaron Smith, and Brian Anderson.

 \end{acknowledgments}

\maketitleabstract

\begin{abstract}

In this dissertation I analyze Hamiltonian control of $d$-dimensional quantum systems as realized in alkali atomic spins.  Alkali atoms provide an ideal platform for studies of quantum control due to the extreme precision with which the control fields are characterized as well as their isolation from their environment.  In many cases, studies into the control of atomic spins restrict attention to a 2-dimesional subspace in order to consider qubit control.  The geometry of quantum 2-level systems is much simpler than for any larger dimensional Hilbert space, and so control techniques for qubits often are not applicable to larger systems.  In reality, atoms have many internal levels.  It seems a shame to throw away most of our Hilbert space when it could in principle be used for encoding information and performing error correction.  This work develops some of the tools necessary to control these large atomic spins.   

Quantum control theory has some very generic properties that have previously been explored in the literature, notably in the work from the Rabitz group.  I provide a review of this literature, showing that while the landscape topology of quantum control problems is relatively independent of physical platform, different optimization techniques are required to find optimal controls depending on the particular control task.  To this end I have developed two optimal control algorithms for finding unitary maps for the problems of: ``state preparation" where we require only that a single fiducial state us taken to a particular target state and ``unitary construction" where the entire map is specified.  State mapping turns out to be a simple problem to solve and is amenable to a gradient search method.  This protocol is not feasible for the task of finding full unitary maps, but I show how we can weave state mappings together to form full unitary maps.  This construction of unitary maps is efficient in the dimension of the Hilbert space.

The particular system I have used for demonstrating these control techniques is that of alkali atoms, specifically $^{133}$Cs.  The state preparation algorithm was used to create a broad range of target states in the 7-dimension $F=3$ hyperfine manifold in an experiment using a combination of time-dependent magnetic fields and a static tensor light shift.  The yields from this experiment were in the range of $0.8-0.9$.  I have developed another control system for the full hyperfine manifold in the ground-electronic state of $^{133}$Cs, a 16-dimensional Hilbert space, based on applied radio frequency and microwave fields.  Numerical studies of the state preparation algorithm find good operating points commensurate with modest laboratory requirements.  This system of microwave and rf control also admits a Hamiltonian structure than can be used by my protocol for unitary construction.  I demonstrate the performance of this algorithm by creating a standard set of qudit gates using physically realistic control fields, as well as by implementing a simple form of error correction.

\clearpage 
\end{abstract}

\tableofcontents
\listoffigures
\listoftables


\mainmatter

\chapter{Introduction}

In recent years, it has become increasingly clear that quantum dynamics allow us to perform certain tasks in ways that are fundamentally more powerful than their classical counterparts.  Some examples are quantum computing, quantum cryptography, quantum-limited measurement, and many others.  This presents a challenge, in that quantum systems also appear to be fundamentally more difficult to control.  This is in part due to the technological challenges of manipulating systems deep within the quantum regime, but even with extremely ``clean" systems we still have to worry about our control routines inadvertently destroying the coherences we are trying to protect.  Unlike in classical control systems, it is not possible to monitor a quantum system passively.  This has lead to the study of quantum control theory, the goal of which is to develop techniques for implementing non-trivial maps on quantum systems in spite of the fragility of quantum states.

In the work in this dissertation I will be considering what is essentially the ``easiest" classes of quantum control problems.  For all of this work, measurement will only be considered as a verifier of the control protocol, and as a source for feedback control.  In addition, for the most part all the states considered will be pure and all the dynamics will be Hamiltonian.  Without feedback and open-quantum-system dynamics that lead to mixed states the math required to model our control systems will be much simpler.  I will also only consider two control tasks: state preparation and unitary construction.  In the first, we would like to find dynamics that arise from a physical Hamiltonian that map some particular initial state to an arbitrary but fixed final state, and in unitary construction we would like the same dynamics to describe a unitary map in its entirety.  Even in these idealized conditions, we will find that designing optimal quantum control protocols for real physical Hamiltonians is a rich and subtle problem.

\section{Quantum control}

Quantum control theory comes in two main flavors: ``open-loop" and ``closed loop."  Generically, we have a Hamiltonian for control system of the form
\be	
H(t) = H_0 + \sum_j c_j(t) H_j,
\ee   
where we would like to choose the ``control waveforms," $c_j(t)$, to implement some control task.  In open-loop control we must choose the control waveforms \emph{without} the benefit of measurement and feedback.  Quantum open-loop control has its origins in the fields of physical chemistry and nuclear magnetic resonance (NMR) spectroscopy.  In physical chemistry the goal is to use laser interactions to drive chemical reactions or to excite molecular vibrations and rotations  \cite{shapiro86,judson92,rabitz03}.  For NMR imaging, pulses of rf magnetic fields are used to produce spin rotations, and by shaping pulses rotations can be enacted in a more optimal way \cite{khaneja01,cory05,vandersypen04}.

Attempts to build a scalable quantum computer have demonstrated the need for more accurate quantum control.  One of the famous DiVincenzo criteria for implementations of quantum computing is the ability to apply elements from a universal set of quantum gates \cite{divincenzo}.  While studies in error correction and fault tolerance have shown that there is an error threshold below which arbitrary length quantum computing is possible \cite{aharonov97}, the precision of control required to reach this threshold is daunting.  This has led to many applications of open-loop quantum control in order to combat loss of coherence in open quantum systems and to engineer robustness to errors in the applied controls \cite{vandersypen04, khaneja05a, grace07, chiara08}.  One prime example is dynamical decoupling where one engineers sequences of pulses that prevent loss of coherence between a qubit and a non-markovian environment to enable quantum memories \cite{viola99,khodjasteh05} or more recently for protected quantum gates \cite{khodjasteh09}.

Of more relevance to the work in this manuscript is the study of optimal quantum control.  ``Optimal" is a bit of a loaded term in the quantum control theory literature and should probably be interpreted according to the colloquial English definition of the word.  Quantum controls can be optimal with respect to a variety of measures, some common examples being the time of a pulse length, fidelity of the time-evolved state with the target, or purity of the end product of the control sequence.  Any, or all, of these measures can be enforced by some sort of cost or objective functional $J[c_j(t)]$ which must be optimized by a set of control waveforms $c_j(t)$ in order for them to be considered optimal.  There exist two primary techniques to solve optimal control problems.  One technique is to solve the  problem analytically using geometrical or Lie algebraic methods \cite{khaneja01, brennen05, zhang03}.  Alternatively, one can attempt to numerically solve these optimization problems either by using gradient search methods \cite{khaneja05, merkel08} or by learning algorithms, such as genetic algorithms, where the objective function is calculated by simply performing an experiment \cite{judson92}.  I will discuss these two methods in more detail in Ch.~\ref{sec:generating} and provide a comparison between the benefits and detriments of each method.  Framing the problem of quantum control in such a general language, optimizing a functional, allows for broad applicabiltiy.  Quantum control protocols often have elements that are system independent and so the design of new protocols for quantum control can impact a wide spectrum applications.     Optimal quantum control techniques have been explored on a wide variety of platforms ranging from optical \cite{wu08}, to semiconductors \cite{hohenester06}, and superconductors \cite{sporl07,motzoi09}.

\section{Atomic spins}

The main application of the control techniques of this dissertation will be towards the control of atomic spins.  Atomic spins are natural carriers of quantum coherence for use in various quantum information processing applications.   These systems have been of particular interest given their excellent isolation from the environment and the available techniques in the ``quantum optics toolbox".  Examples include ensembles of atomic spins as quantum information processing elements \cite{ lukin00, polzik04, kuzmich05, kimble08, molmer08}, ion-trap quantum computers \cite{wineland04, monroe05, blatt07}, and neutral-atom optical lattices \cite{jaksch04}.  The latter has attracted tremendous attention in recent years, as controllable spin lattices are seen as a platform in which to perform quantum simulations of condensed matter systems \cite{lewenstein07} and studies of topological quantum field theory \cite{brennen07}.  

Quantum optics is a mature technology.  Starting from the work of Glauber, Cohen-Tanoudji and others \cite{glauber65, dalibard85,glauber66}, our understanding of the interactions of atoms with lasers and other electromagnetic fields has reached an unprecedented level.  Of particular relevance are the developments in laser cooling and trapping \cite{wieman99}.  In particular, due to the cesium frequency standard, the atomic properties of $^{133}$Cs are extremely well-characterized.  The ability to model our control system to high precision enables us to start considering quantum optimal control.  In the same way that liquid state NMR has provided an excellent platform for exploring quantum control protocols \cite{khaneja01, vandersypen04, cory05}, atomic spins in cold atomic ensembles provide a test-bed with unique physical properties that allows for new investigations into control and measurement techniques.

In most theoretical discussions on quantum information theory, the fundamental systems considered are qubits, 2-level quantum systems.  This certainly makes sense from a theoretical computer science perspective where one can make transparent analogies between qubits and classical bits, as well as from an engineering perspective since all possible coherent manipulations of 2D quantum systems are simply geometric rotations and we would like to easily control our carriers of information.  From a physics perspective, however, it is not clear whether we should restrict our attention to qubits.  Atoms have large spins with a rich internal structure.  Instead of qubits, what if we consider  $d$-dimensional quantum systems or qudits?  This is significantly more complicated control problem, however, it allows for the possibility of qudits as the fundamental information carriers \cite{brennen05} as well as the embedding of logical qubits in a qudit, which may be advantageous for control or protection from errors \cite{gottesman01}.  Additionally, manipulating a nontrivial Hilbert space allows us to explore interesting dynamics such as quantum chaos \cite{haake, ghose}.  The ability to fully control the Hilbert space within the atoms for various applications is an important addition to our toolbox of atomic controls.

\section{Outline of document}

The theoretical work in this dissertation has, for the most part, been  conducted in collaboration with the experimental group of Poul Jessen at the College of Optical Science, University of Arizona.    Many of the results I discuss here have been published previously in three papers.  The first, \emph{Quantum Control of the Hyperfine Spin of a Cs Atom Ensemble} \cite{chaudhury07}, is a direct collaboration with Poul Jessen's lab, in a project headed by Souma Chaudhury.  For this work I developed a protocol for state preparation in atomic spin systems and provided theoretical support for an experimental implementation of said protocol.  The control system in this experiment was the magnetic field and AC-Stark shift system initially explored by Silberfarb and Smith \cite{smith04, silberfarb05, smith06} and discussed in detail in this thesis.  \emph{Quantum Control of the Hyperfine-Coupled Electron and Nuclear Spins in Alkali Atoms} \cite{merkel08} is a theoretical study in which my collaborators and I proposed a new atomic control system that uses microwave and rf magnetic fields as the controls.  This system should be more favorable to implement in the lab and numerical simulations suggest that we can perform state preparation on a space that is twice as big in a time that is about an order of magnitude shorter than in the previous control system.  Finally, in the last paper in this dissertation, \emph{Constructing General Unitary Maps from State Preparations} \cite{merkel09}, we developed a protocol to implement the task of unitary construction based on our knowledge of how to create good state preparation routines.  This construction is efficient in the dimension of the Hilbert space of interest and as an example we have used this technique to create unitary maps in the microwave and rf magnetic field control system.

I have also participated in several other projects that will not appear in this dissertation.  Of direct relevance to the contents of this manuscript, there are two projects that are nearing completion.  The first is a project collaboration with Brian Mischuck on the topic of robust control in the microwave and rf magnetic field system.  The control fields in this system have a geometry similar to the controls in liquid-state NMR systems.  We are currently working to directly port some of the robust control techniques in NMR to this cold atomic spin system.  Another project is in collaboration with Carlos Riofrio and Steve Flammia regarding state estimation.  In that project we are trying to understand the power of random unitary dynamics with regard to the information content of measurement outcomes states undergoing such evolution.  In the case where the dynamics describe some fixed orbit in $\mathfrak{su}(d)$, we have proven the system is not driven through an informationally complete set of observables.  Even though this means  there will be density matrices we cannot reconstruct perfectly, it appears that on average we can still use this measurement procedure to obtain extremely high fidelity estimates for typical quantum states.  The last project I'll mention here is somewhat farther afield and outside the Deutsch group.  In published work with Dan Browne, Matt Elliot, Steve Flammia, Akimasa Miyake and Anthony Short \cite{browne08}, we were able to show a phase transition in the computational power of the cluster state model of quantum computation.  This model requires a certain quantum state, a cluster state, as a resource for computation.  We demonstrated that with faulty resource states there is a sharp phase transition in the computational power, with respect to the error rate, that occurs at the percolation threshold.        

\begin{table}
\begin{center}
\begin{tabular}{|c|c|c|}
\hline
 Journal Reference & Coauthors &  Chapter\\
\hline 
PRL $\mbf{99}$, 163002 (2007)&S.~Chaudhury, T.~Herr,&Ch.~\ref{chp:control_magAC}, Ch.~\ref{ch:experiment}\\
 &A.~Silberfarb, I.~H.~Deutsch&\\
  &and P.~S.~Jessen&\\
\hline
PRA $\mbf{78}$, 023404 (2008) &P.~S.~Jessen and I.~H.~Deutsch&Ch.~\ref{chp:hamil_mwrf}, Ch.~\ref{ch:sp_mwrf}\\
\hline
eprint arxiv:0902.1969 (2009)&G.~K.~Brennen, P.~S.~Jessen&Ch.~\ref{ch:unitary}\\
to appear in PRA (2009)& and I.~H.~Deutsch&\\
\hline
NJP $\mbf{10}$, 023010 (2008)&D.~Browne, M.~Elliott,&\\
 &S.~Flammia, A.~Miyake &\\
& and A.~Short&\\
\hline
\end{tabular}
\caption[Table of published work.]{Table of published work with location in text.}
\end{center}
\end{table}

The remainder of this dissertation is as follows.  In Ch.~\ref{ch:control} I present a background review of some of the basics of open-loop quantum control theory.  The mains goals of this chapter are to understand: when Hamiltonian dynamics are controllable, the relative difficulty between unitary construction and state preparation (as explored by the Rabitz group \cite{rabitz04,rabitz06, shen06, rabitz05, hsieh08, moore08}), and some practical methods for finding optimal control waveforms.  In Ch.~\ref{ch:atomic} I describe the physics behind the two atomic control systems in this paper.  The first control system consists of an ``always on" nonlinear interaction derived from the AC-Stark effect combined with controllable quasi-static magnetic fields.  The second system has no laser interaction and instead utilizes magnetic fields oscillating at rf and microwave frequencies.  I explain the Hamiltonians dynamics of these two systems and rewrite the Hamiltonians in a form that is conducive to our quantum control techniques.  I will also show under what circumstance these systems are controllable.  Chapter \ref{chp:state_prep} describes the state preparation algorithm I helped to develop.  I describe the basic form of the algorithm and its application to both control systems,  experimentally in the case of the magnetic field and Stark shift systems and in numerical simulation for the microwave and rf system.  Finally, in Ch.~\ref{ch:unitary} I discuss the unitary construction protocol we proposed in \cite{merkel09} and show some examples of unitary matrix construction in the microwave and rf control system.

\chapter{Controlling Quantum Systems}\label{ch:control}

In this dissertation I will be looking exclusively at open-loop techniques for controlling quantum systems.  Open-loop control involves designing time-dependent fields to generate a dynamical map without using measurement and feedback.  This is nice in that we are not required to estimate system parameters in real time, but we instead can perform a more thorough modeling of our quantum system offline.  In the attempt to find feedback routines one is often forced to consider measurements that form a ``classical" commutative subalgebra in order to make the problem tractable.  This is not the case with open-loop control.  In some sense, open-loop control allows us to explore more of the ``quantum'" nature of our control protocols.  On the other hand, it can be argued that it is really the process of measurement, and in particular measurement backaction, that distances quantum control theory from classical control theory.  With open-loop control, we are really deriving classical control schemes, but control schemes for systems that live in complex manifolds such as $SU(d)$.  This allows us to directly port over some of the work in classical control theory regarding control over Lie groups.      

In this chapter I will provide an overview of some very general results from open-loop quantum control, in particular as they apply to the problems of state preparation and unitary construction.  In Sec.~\ref{Sec:Controllability}, I will review what it means to be controllable.  In the abstract, controllability determines whether a Hamiltonian system has the degrees of freedom necessary to perform a given control task.  Next, in Sec.~\ref{sec:topology}, I will look at some of the results on the control landscape topology of state preparation and unitary construction.  Analyzing the topology allows one to make some surprisingly general statements regarding the complexity of finding optimal controls for the two different problems.  Finally, I'll describe some of the methods we use to find control fields.  I will discuss two different classes of optimization algorithms and describe some illustrative examples.

\section{Controllability}\label{Sec:Controllability}
Before actually trying to control a quantum system, it is a worthwhile endeavor to determine whether the system is controllable in principle.  When we ignore physical constraints like bandwidth and decoherence, what types of control are possible at a later, finite time?  There are many different aspects of a quantum evolution that we might wish to control, and accordingly there are many different concepts of controllability in the quantum control theory literature.  Some common controllability questions are whether the available  dynamics allow: mappings between arbitrary states (pure or mixed), the construction of general unitary maps, or the simulation of arbitrary observables.  

In this dissertation, I will primarily consider ``unitary controllability", that is whether our dynamics allow us to construct any unitary map in a finite time.  The reasons for considering this type of controllability are twofold.  First, unitary controllability is sufficient for the types of tasks we will consider, i.e. state preparation and unitary construction, and indeed most of the pure-state control tasks in the literature.  Secondly, the conditions for controllability in the case of unitary control are by far the most intuitive and geometrical.

The conditions for unitary controllability have been studied in depth in the control theory literature,~\cite{jurdjevic72,Brockett73}, and more recently from a quantum information perspective~\cite{Schirmer01}.  Formally, we consider a quantum evolution that is governed by a general Hamiltonian evolution of the form
\be	
H(t) = H_0 + \sum_j c_j(t) H_j.\label{eq:generic_hamil}
\ee   
Here, the functions $c_j(t)$ are the control waveforms we are allowed to manipulate.  We take the operators $\{ H_0, H_1 \ldots H_n\}$ as traceless and Hermitian, which leads to unitary dynamics from the group $SU(d)$.  In principle, we could consider Hamiltonians with a nonzero trace leading to dynamics from $U(d)$, but for quantum system global phases are irrelevant.  For a Hamiltonian system of this form to be considered controllable we must show that, starting from the identity operator, we can generate any arbitrary unitary operator.  More formally, if we let $U(t)$ be the solution of the Schrodinger equation
\be
i \frac{\partial U(t)}{\partial t} = H(t) U(t)
\ee
with $U(0) = \mathbb{I}$, then for some $T < \infty$ there exist  control waveforms $c_j(t)$ such that $U(T) = U$ for any $U \in SU(d)$.

Requiring controllability places constraints on the structure of the independent terms in the Hamiltonian $\{ H_0, H_1 \ldots H_n\}$.  In order to generate any element of the Lie group $SU(d)$, it is both necessary and sufficient that the operators $\{ H_0, H_1 \ldots H_n \}$ be a generating set for the corresponding Lie algebra $\mathfrak{su}(d)$.  The Lie algebra generated by $\{ H_0, H_1 \ldots H_n \}$ is defined as the closure of the generating set with respect to general linear combinations and commutators.     

A Lie algebra is a linear vector space with an algebraic product defined by the commutator.  We can see that we can generate any linear combination of our initial set of generators by looking at very short square-pulses according to the Trotter formula, where
\be
 e^{-i H_1 \alpha \Delta} e^{-i H_2 \beta \Delta} \approx e^{-i (\alpha H_1 + \beta H_2) \Delta}.
\ee  
Such short pulses are allowed since we assume access to arbitrary control waveforms.  In addition to linear combinations it is also possible to generate the commutators by the approximation
\be
 e^{-i H_1 \Delta}e^{-i H_2 \Delta}e^{i H_1 \Delta}e^{i H_2 \Delta}\approx e^{-[H_1,H_2] \Delta^2}.
\ee     	
The ability to generate, in principle, any linear combination and any commutator means that one can simulate any element of the the Lie algebra generated by our initial independent Hamiltonians, $\{H_0, H_1,\ldots,H_n\}$. 

It is reasonably intuitive to see why $\{H_0, H_1,\ldots,H_n\}$ generating $\mathfrak{su}(d)$ will be necessary and sufficient for controllability.  We can treat the Lie group $SU(d)$ as a smooth manifold and $\mathfrak{su}(d)$ as its tangent space.  Since we are ignoring physical limitations on the control fields $c_j(t)$ we can create infinitesimal displacements along the directions described by $\{ H_0, H_1 \ldots H_n\}$.  To be controllable it is necessary that using a finite sequence of these displacements we can simulate a infinitesimal displacement along any arbitrary direction, since all infinitesimal displacements of the identity operator are elements of $SU(d)$.  Therefore, it is necessary that the operators $\{ H_0, H_1 \ldots H_n\}$ generate the Lie algebra $\mathfrak{su}(d)$ through linear combinations and commutators.  Sufficiency is a consequence of the fact that $SU(d)$ is compact and simply connected.  This implies that any two elements of $SU(d)$ are linked by a finite length geodesic.  Access to infinitesimal displacements along all directions in $\mathfrak{su}(d)$ allows us to create an arbitrary geodesic though the identity operator, and thus any element of $SU(d)$.

There are a number of ways to determine whether the independent terms in a Hamiltonian control system generate the Lie algebra $\mathfrak{su}(d)$.  The most general approach is to compute the iterated commutators numerically, see appendix \ref{appen:alg_size_code} for Mathmatica code.  We take our initial set of operators and form an orthonormal basis with respect to the standard trace inner product $\langle x_j, x_k \rangle = \Tr \left( x_j^{\dagger} x_k \right)$.  Then we compute the commutators of all pairs of these Hermitian basis operators and see if this results in any operators that have support outside of the initial set.  If so, we append these to our basis for the algebra.  We can look at the commutators of these new terms with our basis and iterate until either we span the entirety of $\mathfrak{su}(d)$, or we close on a sub-algebra.  

While this technique, in principle, can work for any set of control Hamiltonians, in practice, numerical errors start to become a problem for larger systems.  It can also, in the worst cases, require calculating something on the order of $d^4$ commutators.  For large systems it is easier if one can prove controllability analytically by exploiting the geometry of the Hamiltonians.  

To close this section I'll prove a simple theorem that we have been able to exploit to show controllability in many of the atomic systems I'll be considering in this dissertation.

\begin{theorem}\label{t:rank_2}
In an $d$-dimensional Hilbert space with $d>2$, if one has access to the irreducible generators of rotations, $J_x$ and $J_y$, then in order to fully control the space it is sufficient to add an operator $h$ that has a non-zero overlap (according to the trace inner product) with at least one rank-2 irreducible spherical tensor.  That is 
\be
 \exists ~q~ \textrm{s.t.} ~Tr\left( h T^{(2)}_q \right) \neq 0\quad \Rightarrow \quad\{J_x, J_y, h\}_{L.A.} = \mathfrak{su}(d).\nonumber
 \ee
\end{theorem}

Here we have introduced the orthonormal basis of irreducible spherical tensor operators, 
\be 
\label{irreducible}
T^{(k)}_q(J) = \sqrt{\frac{2k+1}{2J+1}}\sum_m \braket{J,m+q}{k,q;J,m} \ket{J,m+q}\bra{J,m},
\ee
satisfying the fundamental commutation rules,  
\bea
\left[J_z, T^{(k)}_q \right] &=& q T^{(k)}_q \\
\left[J_\pm, T^{(k)}_q \right] &=& \sqrt{k(k+1)-q(q\pm1)}T^{(k)}_{q \pm 1} \nonumber,
\eea
where $J_\pm = J_x \pm i J_y$.   It follows from these commutators that given the set $\{J_x, J_y,$ $ T^{(k)}_q\}$ one can simulate any rank-$k$ irreducible tensor, and since these are an operator basis, the generators of rotation can map any rank-$k$ operator to any other rank-$k$ operator.  With this property we are now prepared to prove a lemma.

\begin{lemma}\label{t:Z2}
 $\{J_x, J_y, T^{(2)}_0\}$ generates $\mathfrak{su}(d)$.
\end{lemma}
We prove this by first noting that
\be
\left[ T^{(2)}_0, T^{(k)}_q\right] = c_{k,q} T^{(k+1)}_q + d_{k,q} T^{(k-1)}_q.
\ee
The exact form of the constants is irrelevant except for the fact that there is always some rank-$k$ tensor for which $c_{k,q}$ is nonzero.  Given this, the proof follows by induction.  Suppose our library of simulatable operators contains all operators of ranks $k$ and $k-1$. By commuting some rank $k$ operator with $T^{(2)}_0$ we obtain an operator with support on operators of rank $k-1$ and $k+1$, thus containing a component in the space of rank $k+1$ operators that is linearly independent from the current set of Hamiltonians in our library.  Commutation with the generators of rotation allow us to simulate all other rank $k+1$ operators.  Since we can simulate all rank-1 from the generators $\{J_x,J_y\}$, and the rank-0 operator is the trivial identity operator, it follows by induction that we can simulate all rank-$k$ operators that are supported on the Hilbert space, $k \le d-1$.  Therefore $\{J_x, J_y, T^{(2)}_0\}$ generates $\mathfrak{su}(d)$.QED   

With this lemma, we see that in order to show theorem \ref{t:rank_2}, we need merely to show that the set $\{J_x, J_y, h\}$ can simulate the operator $T^{(2)}_0$.  We will do this in essentially three steps.  Before we start we expand the Hamiltonian $h$ in our spherical basis, $h = \sum_{k=1}^{d-1} \sum_{q=-k}^{k}  h^{(k)}_q T^{(k)}_q $.

\be
\mathbf{Step~ 1:} ~~\textrm{Simulate} \quad h_1 = T^{(2)}_0 +  \sum_{k=3}^{d-1}  \sum_{q=-k}^{k}  h'^{(k)}_q T^{(k)}_q \nonumber 
\ee

To simulate $h_1$ we note that $h$ is defined to have some nonzero rank-2 component.  With rotations we can transform the rank-2 component to $T^{(2)}_0$.  Additionally, since we have all the rank-1 tensors in our library already, we can remove the rank-1 piece of $h$ through linear combinations to yield $h_1$.
   
\be
\mathbf{Step~ 2:} ~~\textrm{Simulate} \quad h_2 = T^{(2)}_0 +  \sum_{k=3}^{d-1}  h''^{(k)}_0 T^{(k)}_0 \nonumber 
\ee

Consider the double commutator 
\be
\left[ J_z, \left[ J_z,h_1\right] \right]=\sum_{k=3}^{d-1}  \sum_{q=-k}^{k} q^2 h'^{(k)}_q T^{(k)}_q.
\ee
If we take a linear combination $h_1 -a \left[ J_z, \left[ J_z,h_1\right] \right] $ the resulting operator has the same coefficients for $q=0$. For $q_0 \neq 0$, choosing $a=1/q^2$, we can sequentially remove all rank-2 tensor components, and we are left with $h_2$.

\be
\mathbf{Step~ 3:} ~~\textrm{Simulate} \quad T^{(2)}_0 \nonumber 
\ee

Consider the double commutator
\bea
[J_x,[J_x,h_2]] = && \frac{3}{2}T^{(2)}_0+\frac{\sqrt{6}}{2}(T^{(k)}_2+T^{(k)}_{-2})  \nonumber\\
&+&\frac{1}{4} \sum_{k=3}^{d-1} h''^{(k)}_0  \bigg( 2 k (k+1) T^{(k)}_0  \nonumber\\
&+&\sqrt{ (k-1) k (k+1) (k+2)}(T^{(k)}_2 + T^{(k)}_{-2}) \bigg). \nonumber \\
\eea

We repeat the process in Step 2 to remove the components from $h_2$ with $q \neq 0$ to obtain
\be
h_2' =  \frac{3}{2} T^{(2)}_0  + \sum_{k=3}^{d-1}  a''^{(k)} \frac{ k (k+1)}{2} T^{(k)}_0.
\ee
If we now take the linear combination $h_2 - 2 h_2' /(k_0(k_0+1))$ we remove the $T^{(k_0)}_0$ component, but are left with a nonzero $T^{(2)}_0$ term.  Repeating this procedure for $k_0 = 3\ldots (d-1)$ yields an operator that is proportional to $T^{(2)}_0$.  This completes our proof of theorem \ref{t:rank_2}.

\section{Control landscape topology}\label{sec:topology}

In the last section we discussed how to determine whether a Hamiltonian system was controllable in principle, but for practical applications we need some way of finding the appropriate controls.  One would suspect that the relative difficulty of finding controls must be very system specific, however, it turns out that it is possible to make extremely general statements about the complexity of finding control waveforms.  This type of analysis derives from studies of the topology of the ``quantum control landscape".  

Finding optimal quantum controls always corresponds to maximizing some objective function $J[\cv]$ with respect to some control parameters $\cv$.  Traditionally, $J$ takes the form of a fidelity or distance measure and $\cv$ describes the control waveforms we use to drive the system.  The quantum control landscape is the multidimensional surface described by the value of the objective function as a function of the control parameters $\cv$.  Of particular interest are the critical points on this surface where the gradient $\nabla_{\cv} J = 0$ since some must describe the highest quality controls.       

The contents of this section follows from a sequence of papers from the Rabitz group on control landscape topology \cite{rabitz04,rabitz06, shen06, rabitz05, hsieh08, moore08}.  While the Rabitz group has studied a wide variety of control problems, the outcome appears to be the same --- the landscape topology depends on the dimension of the quantum system and the type of objective function, i.e., state preparation, unitary construction, etc., but not on any properties of the target or initial states or the particulars of the Hamiltonian, excepting controllability.  This is an incredibly powerful property since the landscape topology alone appears to set the complexity of finding good controls.  In this section I will paraphrase the arguments in the Rabitz papers in the language I have been using in this dissertation.          

The punch-line of this section will be that the problems of state preparation and unitary construction have very different control landscape topologies.  We will find that state preparation, mapping a single initial state to a single target state, has an extremely favorable topology that will allow for the construction of very efficient search routines for finding control fields.  In contrast, the landscape topology of unitary construction, mapping the identity to a target unitary operator, is much more complex.  Numerical surveys \cite{moore08} suggest that it takes exponentially more resources in the dimension of the Hilbert space to search for controls that generate unitary maps when compared to those required for state preparation. 

As an aside, in this chapter I will mostly consider the dynamics to be elements of $U(d)$, as opposed to $SU(d)$ like in the rest of this manuscript.  This assumption will greatly simplify some of the arguments of this section.  Adding a global phase is irrelevant to the physics of the problem and in no way diminishes the intuition gleaned from these studies.

\subsection{Landscape topology of state preparation}

In the problem of state preparation, we would like to map an initially known pure state of a $d$-dimensional quantum system, $\ket{i}$, to a fixed but arbitrary target pure state $\ket{f}$, up to a global phase.  The system evolves according to Hamiltonian of the form given in Eq.~\ref{eq:generic_hamil}, and we control this system by specifying the functions $c_j(t)$, which are defined from $t= [0, T]$.  For simplicity, instead of using continuous functions as our optimization variables, we will assume that the information content of the control waveforms $c_j(t)$ can be completely described by some finite length control vector $\cv$, e.g.~square pulses control waveforms or waveforms from cubic splines.  We can write the Hamiltonian as a function of this control vector, $H[\cv].$  From the Schrodinger equation, this Hamiltonian leads to a unitary propagator we can write as  
\be
U[\cv] = \mathcal{T}\left[ e^{-i \int_0^T  \ud t \, H[\cv]} \right], 
\ee
where $\mathcal{T}$ is the time-ordering operator.  Finding good controls amounts to optimizing the fidelity between the time-evolved quantum state and the target state, given by the objective function,
\be
J[\cv] = |\bra{f} U[\cv] \ket{i}|^2.
\ee

The first step to understanding the topology of the quantum control landscape is to determine the set of critical points where $\nabla_{\cv} J[\cv] = 0$.  A perfect state preparation, $J=1$, is an extremal point of the control landscape and thus must be a member of the set of critical points.  To simplify our calculation of the critical points we define the Hermitian matrix $A$ as the logarithm of $U$, $U[\cv] = e^{-i A[\cv]}$.   This will always exist since $U$ is unitary, however, $A$ is in general an extremely complicated functional of $\cv$.  For the following calculation I'll write the eigen-decomposition of $A$ as  
\be
A = \sum_j a_j \ket{a_j}\bra{a_j}.
\ee
An alternative description of $A$ is in terms of some orthonormal Hermitian basis, $E_{\eta}$, so that
\be
A = \sum_{\eta} A_{\eta} E_{\eta},
\ee
with $A_{\eta} = \Tr (A E_{\eta})$.  The canonical basis we will use consists of $d^2$ terms of the form $\ket{e_j}\bra{e_j}$,  $\left(\ket{e_j}\bra{e_k} + \ket{e_k}\bra{e_j} \right)/\sqrt{2}$ for $j<k$ and $\left(-i \ket{e_j}\bra{e_k} + i \ket{e_k}\bra{e_j} \right)/\sqrt{2}$, also for $j<k$.

The key insight from \cite{rabitz04} is that it is possible to remove essentially all of the particulars of the Hamiltonian dynamics from the condition $\nabla_{\cv} J = 0$ by a very simple argument from controllability.  We can use the chain rule to rewrite $\nabla_{\cv} J = 0$ as
\be
0 = \sum_k \frac{\partial J}{\partial c_k} \mbf{e}_k = \sum_k \sum_{\eta} \frac{\partial J}{\partial A_{\eta}} \frac{\partial A_{\eta}}{\partial c_k} \mbf{e}_k = \sum_{\eta} \frac{\partial J}{\partial A_{\eta}} \nabla_{\cv} A_{\eta} ,\label{eq:remove_controls}
\ee 
It follows from controllability that the vectors $\nabla_{\cv} A_{\eta}$ are linearly independent for different $\eta$, and so the  derivatives $\partial J / \partial A_{\eta}$ must each independently go to zero.  This leaves us with new constraint equations that take the form
 \be
\frac{\partial J}{\partial A_{\eta}} =\frac{\partial |\bra{f} U \ket{i}|^2}{\partial A_{\eta}} = \bra{i} U^{\dagger} \ket{f} \bra{f}\frac{\partial U}{\partial A_{\eta}}\ket{i} +  \textrm{c.c.}  = 0 \quad \forall ~ \eta,\label{eq:crit_constraint}
\ee
which has removed all the dependence on $H$ and $\cv$.  

To see that the vectors $\nabla_{\cv} A_{\eta}$ are linearly independent, consider the following argument.  One of the necessary implications of controllability is that we can construct any unitary matrix of the form $\textrm{exp} (-i \alpha E_{\mu})$ for all $\mu$ and $\alpha = [0,\infty).$  This means there must always be a control vector $\cv$ such that $A_{\mu}[\cv] = \alpha$ and $A_{\eta \neq \mu} [\cv] = 0.$  We can now show why the gradient vectors must be independent through a linearity argument.  Assume that there exist coefficients $\beta_{\eta}$ such that
\be
\nabla_{\cv}A_{\mu} = \sum_{\eta \neq \mu} \beta_{\eta}  \nabla_{\cv} A_{\eta}.
\ee  
This implies that 
\bea
0&=&\nabla_{\cv}A_{\mu} - \sum_{j \neq k} \beta_{\eta}  \nabla_{\cv} A_{\eta} \nonumber\\
&=& \nabla_{\cv}(A_{\mu} - \sum_{\eta \neq \mu} \beta_{\eta}   A_{\eta} )
\eea 
or 
\be
A_{\mu} = \sum_{\eta \neq \mu} \beta_{\eta}   A_{\eta} + C,
\ee
where $C$ is a constant with respect to $\cv$.  In this case the only unitary matrix we can construct of the form $\textrm{exp} (-i \alpha E_{\mu})$ is a single matrix, $\textrm{exp} (-i C E_{\mu})$.  This implies that any system where the vectors $\nabla_{\cv} A_{\eta}$ are linearly dependent is not controllable, and so by the contrapositive, if our system is controllable, the vectors $\nabla_{\cv} A_{\eta}$ must be linearly independent.

We can look at $\partial J / \partial A_{\eta}$ in more detail by first evaluating $\partial U / \partial A_{\eta}$.   We do this by explicitly differentiating the operator and expressing $A$ in its eigenbasis
\bea
&&\frac{\partial U}{\partial A_{\eta}} \nonumber\\
&=& -i  \int_0^1 \ud s \, e^{-i A (1-s)} \frac{\partial A}{\partial A_{\eta}} e^{-i A s}  \nonumber\\
&=& -i \sum _{m,n} \int_0^1 \ud s \,  \ket{a_m}  e^{-i a_m (1-s)} \bra{a_m} E_{\eta} \ket{a_n} e^{-i a_n s} \bra{a_n}\nonumber\\
&=&\sum_{m,n} \ket{a_m}  \bra{a_m} E_{\eta} \ket{a_n}  \bra{a_n} F(a_m,a_n).\label{eq:dUdA}
\eea
Here $F(a_m, a_n)$is the result of the integral and has the form
\be
F(a_m,a_n) = \left\{ \begin{array}{ll}
 (-i) e^{-i a_m} & \textrm{if} ~ a_m = a_n\\
 \frac{e^{-i a_m} - e^{-i a_n}}{a_m - a_n} &  \textrm{if} ~ a_m \neq a_n
  \end{array} \right.
\ee

We can plug this back into Eq.~\ref{eq:crit_constraint} to get the set of constraint equations
\bea
0 &=&\frac{\partial J}{\partial A_{\eta}}\nonumber\\
&=&\sum_{m,n} \bra{i} U^{\dagger} \ket{f} \braket{f}{a_m}\bra{a_m} E_{\eta} \ket{a_n}  \braket{a_n}{i} F(a_m,a_n) + \textrm{c.c.}\nonumber\\
\eea
Since these equations must be zero for all $\eta$, they must also be zero for any general linear combination.  In particular, by making a unitary transformation, we obtain that for all $r$ and $s$ 
\bea
0 &=&\sum_{\eta}\bra{a_s} E_{\eta} \ket{a_r} \frac{\partial J}{\partial A_{\eta}}\nonumber\\
 &=&\sum_{\eta}\sum_{m,n} \bra{i} U^{\dagger} \ket{f} \braket{f}{a_m}\bra{a_m} E_{\eta} \ket{a_n} \bra{a_s} E_{\eta} \ket{a_r}  \braket{a_n}{i} F(a_m,a_n)  +\textrm{c.c.}\nonumber\\
 \eea
 One of the consequences of our choice of basis is that it is easy to show that 
 \be
 \sum_{\eta} \bra{\psi_1} E_{\eta} \ket{\phi_1} \bra{\phi_2} E_{\eta} \ket{\psi_2} =  \braket{\psi_1}{\psi_2} \braket{\phi_2}{\phi_1},
 \ee
 which leaves us with
 \bea
 0 &=&\sum_{m,n} \bra{i} U^{\dagger} \ket{f} \braket{f}{a_m}\braket{a_m}{a_r} \braket{a_s}{a_n} \braket{a_n}{i} F(a_m,a_n)  +\textrm{c.c.}\nonumber\\
&=& \bra{i} U^{\dagger} \ket{f} \braket{f}{a_r}  \braket{a_s}{i} F(a_r,a_s) + \textrm{c.c.}  \label{eq:unit_basis}
\eea
To simplify this expression further by we write $\ket{\psi} = U \ket{i}$.  When we remove $\ket{i}$ we are left with
\bea
0&=&  \bra{a_s} U^{\dagger} \ket{\psi} \braket{\psi}{f} \braket{f}{a_r} F(a_r,a_s) + \textrm{c.c.}\nonumber\\
&=& \braket{a_s}{\psi} \braket{\psi}{f} \braket{f}{a_r} e^{i a_s} F(a_r,a_s) + \textrm{c.c.}
\eea

At this point it helps to look separately at the cases where $a_r = a_s$ and $a_r \neq a_s$ in order to see what restrictions are placed on the time-evolved state, $\ket{\psi}$, by these equations.  When $a_r = a_s$ the constraint equations reduce to 
\bea
0 &=& \braket{a_r}{\psi} \braket{\psi}{f} \braket{f}{a_r}e^{i a_r}(-i) e^{-i a_r}  + \textrm{c.c.}\nonumber\\
&=& -i \braket{a_r}{\psi} \braket{\psi}{f} \braket{f}{a_r} + \textrm{c.c.}\nonumber\\
&=& -i \left( \braket{a_r}{\psi} \braket{\psi}{f} \braket{f}{a_r}   - \braket{a_r}{f} \braket{f}{\psi} \braket{\psi}{a_r} \right) \nonumber\\
&=&-i\left( \bra{a_r}\left[ \ket{\psi} \bra{\psi}, \ket{f} \bra{f} \right] \ket{a_r} \right).\label{eq:constraint_diag}
\eea

The equations concerning $a_r \neq a_s$ are a bit more tricky to deal with.  We first define, only for $a_r \neq a_s$, the function 
\be
G(a_r,a_s)  = e^{i a_s}F(a_r,a_s) =\frac{e^{-i (a_r-a_s)} - 1}{a_r - a_s}. 
\ee
$G$ has two properties of note that one can easily show:  $\textrm{Re}(G) \neq 0$ and $G^*(a_r,a_s) = -G(a_s,a_r)$.  We can rewrite the constraint equations for $a_r \neq a_s$ in terms of $G$ as 
\bea
0&=& \braket{a_s}{\psi} \braket{\psi}{f} \braket{f}{a_r} G(a_r,a_s) + \textrm{c.c.}\nonumber\\
&=& \braket{a_s}{\psi} \braket{\psi}{f} \braket{f}{a_r} G(a_r,a_s) +\braket{a_r}{f}\braket{f}{\psi} \braket{\psi}{a_s}  G^*(a_r,a_s)\nonumber\\
&=& \braket{a_s}{\psi} \braket{\psi}{f} \braket{f}{a_r}  G(a_r,a_s) -\braket{a_r}{f}  \braket{f}{\psi}\braket{\psi}{a_s} G(a_s,a_r).\nonumber\\
\eea
While this doesn't immediately look to be an improvement we can look at the sum of the two constraint equations for $(r,s)$ and $(s,r)$ to obtain
\bea
0&=& \braket{a_s}{\psi} \braket{\psi}{f} \braket{f}{a_r}  G(a_r,a_s) -\braket{a_r}{f}  \braket{f}{\psi}\braket{\psi}{a_s} G(a_s,a_r)\nonumber\\
&+& \braket{a_r}{\psi} \braket{\psi}{f} \braket{f}{a_s}  G(a_s,a_r) -\braket{a_s}{f}  \braket{f}{\psi}\braket{\psi}{a_r} G(a_r,a_s)\nonumber\\
&=& \bra{a_s} [\ket{\psi} \bra{\psi}, \ket{f} \bra{f} ] \ket{a_r}  G(a_r,a_s) +  \bra{a_r} [\ket{\psi} \bra{\psi}, \ket{f} \bra{f} ] \ket{a_s} G(a_s,a_r)\nonumber\\
&=& \bra{a_s} [\ket{\psi} \bra{\psi}, \ket{f} \bra{f} ] \ket{a_r}  G(a_r,a_s) +  \bra{a_s} [\ket{\psi} \bra{\psi}, \ket{f} \bra{f} ] \ket{a_r} G^*(a_r,a_s)\nonumber\\
&=& \bra{a_s} [\ket{\psi} \bra{\psi}, \ket{f} \bra{f} ] \ket{a_r}  (G(a_r,a_s)+G^*(a_r,a_s)).
\eea
Remembering that the real part of $G$ is never zero, we can combine this result with the outcome of Eq.\ref{eq:constraint_diag} to  determine that if $\nabla_{\cv} J = 0$, 
\be
\left[\ket{\psi} \bra{\psi}, \ket{f} \bra{f}  \right] = 0\label{eq:state_com}.
\ee
That is,  the time-evolved state must commute with the target state.  The implication is that either, the time-evolved state is orthogonal to the target state, or, up to a global phase, it is equivalent to the target state.  Therefore, when $\nabla_{\cv} J = 0$, $J=0,1$.  This result is independent of the target and initial states as well as the details of the Hamiltonian evolution.        

This implication that $\nabla_{\cv} J = 0$ if and only if $J=0,1$ dramatically impacts the ease of search when looking for optimal controls.  The control landscape has no sub-optimal traps.  It isn't necessary to resort to complicated algorithms like genetic searches or annealing methods to find global optima.  Instead, local algorithms, like gradient searches, should converge on globally optimal controls.

By analyzing the topology of the set of critical points $J =1$, we find the structure of state preparation is even more favorable.  It turns out the set of good controls form a manifold.  We can see this by looking at the set of unitary operators $\mathcal{W} \subset SU(d)$ for which $|\bra{f} W \ket{i}|^2 = 1$.  Since $SU(d)$ is invariant to right multiplication we can re-express the  critical set as $\mathcal{W} = \mathcal{V}R$, where $R \in SU(d)$ and satisfies $R \ket{i} = \ket{f}$.  Now the condition for $J=1$ is  $|\bra{f} V \ket{f}|^2 = 1$, implying that the only requirement on $\mathcal{V}$ is that its elements have $\ket{f}$ as an eigenstate.  The elements of $\mathcal{V}$ are allowed to have any unitary action on the orthocomplement of $\ket{f}$, and so $\mathcal{V}$ is isomorphic to $U(d-1)$.  This is $U(d-1)$ and not $SU(d-1)$ due to the unconstrained phase associated with the eigenvalue of $\ket{f}$.   

The importance of the critical points, $J=1$, forming a smooth submanifold of $U(d)$ is that it lends the state preparation problem a certain amount of robustness to variations in the control fields.  The optimal control fields form a large plateau in the control landscape as opposed the case where high fidelity controls could have been represented by isolated points in the landscape.  When we perturb the control fields, only the resulting displacements in $SU(d)$ that have support outside of the tangent space of the critical submanifold will lead to a decrease in fidelity.  The dimension of $U(d-1)$ is $(d-1)^2$, which is a very large fraction of $d^2-1$, dimension of $SU(d)$.  The difference between these two dimensions is only $2d-2$, which should come as no surprise since it is the exact number of parameters necessary to describe a pure state.

\subsection{Landscape topology of unitary construction}

In the problem of unitary construction, instead of solely mapping one known state to some other state, we would like the final, time-evolved unitary map, $U[\cv]$, to be some particular, but arbitrary, unitary map $V \in SU(d)$.  We can quantify how close the time-evoloved unitary is to the target by the Hilbert-Schmidt distance 
\be
\|U[\cv]-V \|_{HS} = \sqrt{\Tr{|U[\cv]-V|^2}} = \sqrt{2d - 2 \textrm{Re} \left( \Tr{( V^{\dagger}U[\cv])} \right)},
\ee
from which we obtain 
\be
J[\cv] =2 \textrm{Re} \left( \Tr{( V^{\dagger}U[\cv])}  \right),
\ee   
as the objective function we would like to maximize for perfect unitary construction.         

The analysis of this problem proceeds very similarly to that of state preparation, even though the two objective functions are quite different.  We would like to determine the nature of the critical manifolds for which $\nabla_{\cv} J = 0$.  First, using the same decomposition and insights on the nature of controllability as we did in Eq.~\ref{eq:remove_controls}, we remove all dependence on the particulars of the evolution to get the independent constraints
\be
\Tr{ \left( V^{\dagger} \frac{\partial U}{\partial A_{\eta}}\right) } + \textrm{c.c.}= 0 \quad \forall ~ \eta.\label{eq:unit_obj}
\ee    

We can directly plug in the value of $\partial U[\cv] / \partial A_{\eta}$ from Eq.~\ref{eq:dUdA} into the set of constraint equations to obtain 
\be
\sum_{m,n} \bra{a_m} V^{\dagger} \ket{a_n} \bra{a_m} E_{\eta} \ket{a_n} F(a_m,a_n) +\textrm{c.c.} = 0.
\ee
To get this into a more manageable form we make the same change of basis as in Eq.~\ref{eq:unit_basis} yielding
\be
0 =\sum_{\eta}\bra{a_s} E_{\eta} \ket{a_r} \frac{\partial J}{\partial A_{\eta}} = \bra{a_s} V^{\dagger} \ket{a_r} F(a_r,a_s)+\textrm{c.c.}\label{eq:unit_con}
\ee

Again, we look separatly at the cases $a_r = a_s$ and $a_r \neq a_s$, but this time we will first look at the case $a_r \neq a_s$.  We can explicitly write out $F(a_r,a_s)$ and simplify to get    
\bea
0&=& \bra{a_s} V^{\dagger} \ket{a_r} F(a_r,a_s)+\textrm{c.c.}\nonumber\\
&=&\bra{a_s} V^{\dagger} \ket{a_r}  \frac{e^{-i a_r} - e^{-i a_s}}{a_r - a_s}+\textrm{c.c.}\nonumber\\
&=&\frac{1}{a_r - a_s}\left( \bra{a_s} V^{\dagger} U \ket{a_r}- \bra{a_s}U V^{\dagger} \ket{a_r} \right) +\textrm{c.c.}\nonumber\\
&=&\frac{1}{a_r - a_s} \bra{a_s}[ V^{\dagger}, U] \ket{a_r}+\textrm{c.c.}
\eea
Since these equations must be true for all $a_r \neq a_s$ we have that both the real and imaginary parts of the off-diagonal elements of $[V^{\dagger},U]$ must be zero.  The diagonal components of the commutator must be zero independently in this particular basis since $\ket{a_r}$ is an eigenstate of $U$ and so
\bea
\bra{a_r}[ V^{\dagger}, U] \ket{a_r} &=& \bra{a_r}V^{\dagger}U \ket{a_r} - \bra{a_r}UV^{\dagger}\ket{a_r} \nonumber\\
&=&  \bra{a_r}V^{\dagger}\ket{a_r} e^{-i a_r} -e^{-i a_r} \bra{a_r}V^{\dagger}\ket{a_r} =0. 
\eea   

These constraints lead to a similar commutator restriction as in Eq.\ref{eq:state_com}.  That is, when $\nabla_{\cv} J = 0$,
\be
[ V,U] = 0. 
\ee
Unlike in state preparation this is not the whole story.  In the state preparation problem both the evolved and target states were rank-1 projectors, and since global phases are irrelevant, the map was defined solely by its eigenvectors.  In order to construct a full unitary map we must not only consider the eigenvectors of the evolved operator, but also their eigenvalues. For this we need the equations corresponding to $a_r = a_s$.  We know that $U$ and $V$ have simultaneous eigenstates, and write the eigenvalues of $V$ as $\bra{a_s} V \ket{a_s} = e^{-i b_s}$.  This leads to the constraint equations 
\be
0= \bra{a_s} V^{\dagger} \ket{a_s} F(a_s,a_s)+\textrm{c.c.}=e^{i b_s} (-i)e^{-i a_s}+\textrm{c.c.} = 2 \sin{(b_s-a_s)}.
\ee
For this expression to be zero, $b_s - a_s = n_s \pi$, where $n_s$ is an integer.  

We can now consider what $\nabla_{\cv} J = 0$ implies about the value of $J$.  If the gradient of $J$ is zero then
\be
J[\cv] = 2\textrm{Re} \left( \Tr (V^{\dagger} U) \right) =  2\textrm{Re} \left( \sum_s e^{-i (a_s - b_s)}\right) = 2 \sum_s (-1)^{n_s}.
\ee 
This leads to $d$ different values of the objective function ranging from $-2d, -2d+4,\ldots,2d-4,2d$, or Hilbert-Schmidt distances $0,2,\ldots 2d.$  Unlike in the case of state preparation, when optimizing unitary maps there are $d+1$ critical manifolds for which the gradient is zero. 

In order to more fully understand the topology of the control landscape for this problem, we can look at the group structure of the critical manifolds exactly like in \cite{hsieh08}.  Since $U(d)$ is invariant under left multiplication we can make a transformation to some $W \in U(d)$ such that $W = V^{\dagger} U$.   Under this mapping, the subspace where $\textrm{Re} \Tr (W)$ is equal one of the critical values of $J$ is topologically equivalent to the subspace where $\textrm{Re} \Tr (V^{\dagger} U)$ is equal to the same critical value.  In one of these critical manifold, the matrix elements of $W$ have the form
\be
W = \sum_j e^{-i n_j \pi} \ket{a_j} \bra{a_j} = \sum_j (-1)^{n_j} \ket{a_j} \bra{a_j}. 
\ee
$W$ has a block structure of the form $\mathbb{I}_{d-n} \oplus -\mathbb{I}_n$, where $n$ is the number of eigenvalues with value $-1$.  In fact, the critical manifold is all such $W \in U(d)$ that have this eigenspectrum since $\textrm{Re} \Tr(TWT^{\dagger}) = \textrm{Re} \Tr(W)$ for all $T \in U(d)$.  From here on we will label the separate critical manifolds by a canonical representative $W_n$ that is diagonal and whose matrix values on the diagonal are arranged $(1,\ldots, 1, -1, \ldots,-1)$.  

More formally, the set $\textrm{Orb}(W_n) = \{T W_n T^{\dagger}: T \in U(d)\}$ is defined as the orbit of the group action of $U(d)$ with respect to $W_n$.  Since $U(d)$ is a compact Lie group, the orbits form smooth submanifolds of $U(d)$.  Additionally, while $\textrm{Orb}(W_n)$ is not necessarily a group, it is diffeomorphic to the quotient group $U(d)/\textrm{Stab}(W_n)$.  Here $\textrm{Stab}(W_n)$ is the stabilizer group of $W_n$ in $U(d)$, defined as $\textrm{Stab}(W_n) = \{R \in U(d): RW_nR^{\dagger} = W_n \}$.  Because $W_n$ has the block structure $\mathbb{I}_{d-n} \oplus -\mathbb{I}_n$, $\textrm{Stab}(W_n)$ is simply $\textrm{Stab}(W_n) = \{ U_{d-n}\oplus U_n: U_{d-n} \in U(d-n),U_n \in U(n)  \}$.  The critical submanifold has the structure of the Grassmannian manifold, that is the manifold of $U(n)$ subspaces of $U(d)$ or 
\be
G(n,d) = \frac{U(d)}{U(n) \times U(d-n)}.
\ee     
The dimensionality of these manifolds is
\bea
\textrm{dim}(G(n,d))&=&\textrm{dim}(U(d)) - \left(\textrm{dim}(U(d-n))+\textrm{dim}(U(n)) \right)\nonumber\\
&=& d^2 - ((d-n)^2+n^2 ) = 2n(d-n).
\eea

Unlike in state preparation, where the optimal critical submanifold had a relatively high dimension, for unitary construction the optimum is a single point.  In this control landscape, it is the suboptimal manifolds that have dimensions on the order of $d^2$.  If any of the suboptimal manifolds were traps, using local searches would become hopeless.  We can examine the curvature in the vicinity of the critical manifolds to determine whether they are saddles or local maxima by computing the Hessian.  This wasn't necessary in the case of state preparation since the only critical manifolds were at the extrema, and thus had to be either maxima or minima.

The Hessian is essentially the second derivative of the control landscape and has matrix elements defined by 
\be
\mathcal{H}_{j,k} = \frac{\partial^2 J}{\partial c_j \partial c_k}.
\ee 
The eigenvalues of the Hessian matrix describe the curvature of the control landscape.  The key quantity of interest is the sign of the eigenvalues, which determine whether the suboptimal manifolds are traps or saddles.  Like the rest of the analysis of the landscape topology we'll look at variations with respect to the manifold of unitary operators as opposed to variations in the control fields.   

The easiest way to understand the eigenspectrum of $\mathcal{H}$ is to look at the Hessian quadratic form.  We can rewrite our objective function $J[\cv]$ as a functional of the time evolved unitary map, $U$,
\be
J[U] = 2 \textrm{Re} \Tr (V^{\dagger}U).
\ee  
The Hessian quadratic form, $\mathcal{H}_h (U)$, is the second order term of the Taylor expansion about $h$ of $J[e^{-ih} U]$, where $h$ is an arbitrary infinitesimal Hermitian operator.  The Taylor expansion up to second order of $J$ is
\be
J[e^{-ih} U] = 2 \textrm{Re} \Tr \left( V^{\dagger}(\mathbb{I} -i h -h^2/2)U \right).
\ee 
Therefore, the Hessian quadratic form is 
\be
\mathcal{H}_h (U) = -\textrm{Re} \Tr \left(V^{\dagger}h^2 U \right).
\ee
We would like to evaluate this quantity when $U_n$ is a member of one of the critical submanifolds.  If we write the matrix values of $h$ in the eigenbasis of $V$ as $\bra{\beta_j} h \ket{\beta_j} = \gamma_{jj}$ and  $\bra{\beta_j} h \ket{\beta_k} = \gamma_{jk} + i\eta_{jk}$, where the $\gamma$'s and $\eta$'s are real, we are left with,
\bea
\mathcal{H}_h (U_n) &=& -\textrm{Re} \sum_s \bra{\beta_s}UV^{\dagger}h^2 \ket{\beta_s}\nonumber\\
&=& -\textrm{Re} \sum_s (-1)^{n_s}\bra{\beta_s}h^2 \ket{\beta_s}\nonumber\\
&=&- \sum_{s,t} (-1)^{n_s} \gamma_{ss}^2 - 2\sum_{s>t} \left((-1)^{n_s} + (-1)^{n_t} \right) (\gamma_{st}^2 + \eta_{st}^2).   \nonumber\\
\eea  
The independent terms in this sum give us the eigenvalues of $\mathcal{H}$.  We can enumerate the number of positive, $\mathcal{H}_+$, negative, $\mathcal{H}_-$, and zero, $\mathcal{H}_0$, terms in this sum to obtain
\be
\mathcal{H}_+ = n^2, \qquad \mathcal{H}_- = (d-n)^2, \qquad \mathcal{H}_0 = 2 n (d-n).
\ee
The size of the zero eigenspaces confirm our previous geometric arguments.

The eigenspectrum of the Hessian tells us that the topology has no traps, only saddles.  Ruling out the possibility of traps might give us hope that the same local searches that are efficient in the problem of state preparation should apply here.  That is not the case.  Numerical simulations have shown \cite{moore08} that the amount of computational resources necessary to optimize unitary maps grows exponentially with the dimension of the system.  It is not fully understood why the resources should scale exponentially with this topology.  One clue that is suggested from the numerical studies of the landscape is that the path traversed by the optimization increases linearly with problem size for optimizing full unitary matrices, while with state preparation this distance is roughly constant.  For the problem of state preparation, any arbitrary control vector is close to some optimal control.  This is impossible in the case of unitary construction when the optimal control is a solitary point.

\section{Generating optimal control waveforms}\label{sec:generating}

We have discussed how to determine whether a Hamiltonian system is controllable and the relative difficulty of the two types of control tasks in this dissertation.  In this section I'll review some of the techniques for the practical construction of control waveforms.  For the most part, the algorithms used to construct controls fall into one of two broad categories which I will label ``stochastic searches" and ``geometric constructions."  In this section I will discuss the relative strengths and weaknesses of these two approaches and describe some of the representative algorithms from each set.   

\subsection{Stochastic searches}

The algorithms that I will refer to as stochastic search algorithms all involve the same basic steps.  First, we select an arbitrary control field from some distribution to serve as a random seed.  We then use this seed to perform an optimization that attempts to maximize our objective function.  If this optimization yields controls that are insufficient for our needs, we simply draw a new random seed and repeat the process.  Eventually, this process will find control waveforms such that the value of the objective function is arbitrarily close to the global optima.  Some optimization routines such as simulated annealing or genetic algorithms incorporate the stochasticity in a more regular way, but the end result is the same.     

This kind of technique represents a brute force approach to finding optimal controls.  We essentially ignore everything we know about the underlying physics of the system and make random guesses that we hope are in the neighborhood of a global optima or at least a path to a global optima.  Ignoring the structure of the problem comes at a steep price.  For some problems the time required for these types of algorithms to converge on an acceptable answer may become prohibitive, e.g. the computational complexity scales exponentially in $d$.  

While it may seem silly to try to guess the answer, the fact that we can ignore all of the particulars of a problem is also a virtue.  These types of optimization procedures can be constructed for any type of control problem.  Stochastic searches always represent a possible avenue of last resort, and for small dimensional problems the asymptotic scaling can be insignificant.  Also, from a practical perspective, since these algorithms are all very similar, once one has implemented a stochastic search algorithm for one problem, it is almost trivial to retool it for use on a different physical system.  The ease of implementation is furthered by the availability of canned numerical solvers for these search problems for most computer algebra packages.    

Stochastic search algorithms become important when we consider the results from Ch.~\ref{sec:topology} regarding the landscape topology of state preparation.  State preparation has a topology that is extremely favorable towards stochastic searches since it has no suboptimal traps and the optimal points form a submanifold of reasonably high dimension.  With the problem of state preparation we can be sure that a random guess not only will always lead us to a global optima but also will be able to do so for local searches.  To solve a state preparation problem we do not need genetic or simulated annealing algorithms, but instead can make do with simpler gradient ascent techniques.  For this reason gradient searches have yielded some very powerful optimal control search routines.

Gradient searches are most simply explained in a couple lines of pseudocode.
\begin{tt}
\indent \indent $c$ = RANDOM \\
\indent \indent while $\| \nabla_c J[c] \| > \delta$ \\
\indent \indent \indent $c = c + \epsilon \nabla_c J[c] $ \\
\indent \indent end \\
\indent \indent output $c$\\
\end{tt}
We start from a random seed and calculate the gradient of $J$.  As long as we are not at a critical point already, the algorithm takes a small step in the direction of the gradient.  If $\epsilon$ is small enough the algorithm will converge on a critical point where $\nabla_c J[c] = 0$.  In the problem of state preparation this will always be a global optima.  There are extra bells and whistles one can add to the algorithm, e.g.~adaptively choosing $\epsilon$ or adding some stochasticity to the objective to help traverse saddles, but gradient searches will still find global optima reliably for only the most simple topologies.  Luckily for us, state preparation has such a topology.  For unitary construction, we must consider different methods for all but the smallest size problems.

\subsection{Geometric constructions}

The algorithms for generating quantum controls that I have described as geometric constructions are many and varied.  Depending on the structure of the Hamiltonian and the type of control problem one is considering, it is occasionally possible to find deterministic algorithms that create good control waveforms.  These constructions are particularly nice since they generally require only minimal computational resources, e.g.~solving a simple geodesic equation \cite{khaneja01}.  While we know that it is easy to construct state preparations using stochastic searches, geometric constructions have, until very recently, been the only way to construct unitary operators with a reasonable asymptotic scaling.  

The limitations with these approaches is that $SU(d)$ is a pretty complicated place.  Unlike the broad applicability of stochastic techniques, the set of problems for which we understand the geometry well enough to develop efficient unitary constructions is limited.  Additionally, geometric controls very often aren't optimal with respect to measures such as the total time of the control waveform or the robustness to errors.  When performing a stochastic search we could simply make adjustments to the objective function, but with a geometric construction, altering the objective can very easily destroy the geometric property one is exploiting.

Perhaps the simplest type of geometric construction for unitary matrices is that of the Euler angle construction for 2-level systems.  While a trivial example, but it does encapsulate some of the flavor of these techniques.  Our understanding on how to construct a $2$-level unitary matrix relies on the fact that $SU(2)$ is a double cover of $SO(3)$, the symmetry of the 2-sphere, which is geometry about which we understand well.  Given two Hamiltonians, $H_0$ and $H_1$, we can find a set $\{\alpha, \beta, \delta \}$ trivially, using only trigonometric functions, such that $U = e^{-i \alpha H_0 }e^{-i \beta H_1 }e^{-i \delta H_0 }$, for any $U \in SU(2)$.  There does, however, most likely exist some continuous control waveform $c_0(t)H_0 + c_1(t)H_1$ that creates this transformation with a smaller energy cost.

Of course, the main limitation of the Euler angle approach is that it fails for anything other than 2-dimensional systems.  The special unitary group is only isomorphic to a sphere for $d=2$.  Furthermore, while there exist some similar constructions in higher dimensions, e.g. the Cartan decomposition in $d=4$, these decompositions place requirements on the nature of the Hamiltonian beyond simple controllability.  It is more interesting to look at families of geometric constructions that are applicable to any dimension.  Since there is really no overarching algorithm that describes all geometric constructions, I will describe two particular examples from the literature that exemplify some of the powers and limitations of this approach.

\subsubsection{Unitary construction from a QR decomposition}

A procedure to exactly construct general unitary operators on a qudit was put forward in \cite{brennen05}. This construction requires some very specific Hamiltonian structure.  The Hamiltonians all come in pairs and these provide controllably on a 2d subspace of the form
\be
H_j^{(x)} = \ket{j}\bra{k} + \ket{k}\bra{j}, \qquad H_j^{(y)} = -i \ket{j}\bra{k} + i \ket{k}\bra{j}.
\ee
We can define a coupling graph for this system as a graph where the vertices are the basis states of our qudit and the edges connect the coupled 2d subspaces.  It is possible to show the system is controllable if and only if this coupling graph is connected.  

An arbitrary unitary map on this system can be implemented through a method that is derived from the QR decomposition.  All invertible matrices $V$ can be written in the form $V = QR$ where $Q$ is a sequence of Given's rotations, $Q = G_1G_2\ldots G_n$, and $R$ is upper triangular.  If $V$ is a unitary matrix, $R$ must additionally be diagonal.  A Given's rotation is rotation in a plane spanned by two coordinate axes, i.e.,
\be
G =\mathbb{I} + (\cos \theta -1) \ket{j}\bra{j} + .(\cos \theta -1) \ket{k}\bra{k} +(\sin \theta) \ket{j}\bra{k} +(\sin \theta) \ket{k}\bra{j}.
\ee  

\begin{figure}[t!]
\begin{center}
\begin{tabular}{cc}
\includegraphics[width=8cm,clip]{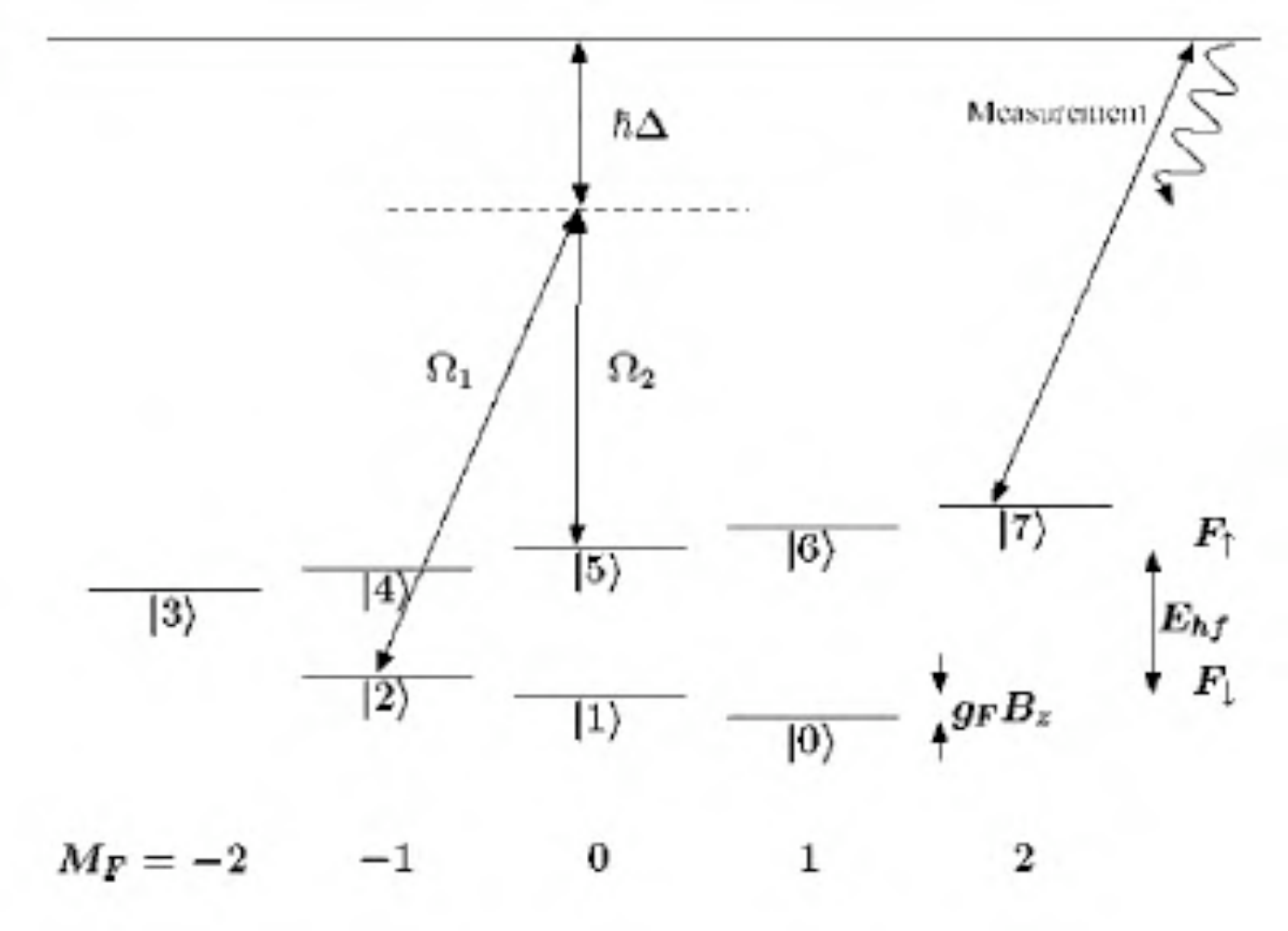} &
\includegraphics[width=5cm,clip]{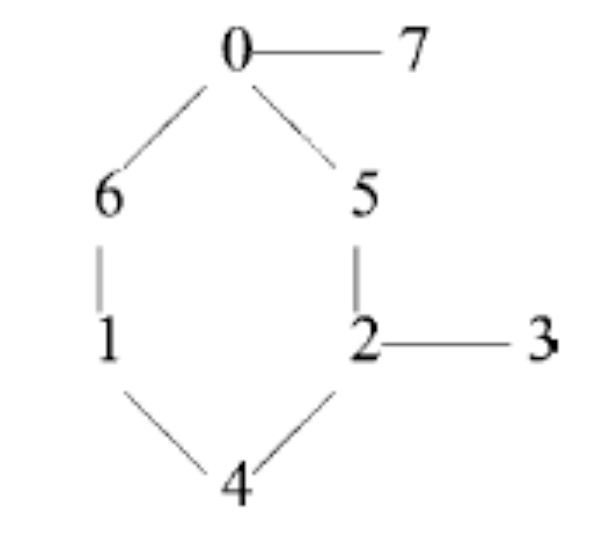}\\
(A) & (B)
\end{tabular}
\caption[System for use with QR decomposition unitary construction]{In (A) we have an example of a system that allows for the unitary construction technique in this section \cite{brennen05}.  The system of interest is the electronic ground state of $^{87}$Rb.  The coupling Hamiltonians are realized by two lasers driving Raman transitions.  These resonances can only couple hyperfine levels satisfying the selection rules $\Delta_{m_F} = 0,\pm 1$, which leads to the connected coupling graph in (B).}
\label{F:QRdecomp}
\end{center}
\end{figure}

This decomposition provides a method to construct a general unitary matrix by way of backwards-evolving the target to the identity.  We simply find a sequence of rotations in our 2d subspaces such that $G_n^{\dagger} \ldots G_1^{\dagger} V$ is diagonal.  We can do this by finding rotations that sequentially set the off-diagonal matrix elements of $V$ to zero.  There is a systematic way to set these elements to zero using spanning trees of the coupling graph.  Details can be found in \cite{brennen05}.  Once we have a diagonal matrix it is simple to remove the phases by considering rotations along $H_j^{(z)}$ in our 2-dimensional subspaces.  We can create these easily enough since the 2d subspaces are fully controllable.  Now that we have a construction for $Q^{\dagger}$ and $R^{\dagger}$ we can simply apply the time-reversed fields to map the identity to $V$.

It should be noted that not only does this technique only work for a very restricted class of control Hamiltonians such as the one in Fig.~\ref{F:QRdecomp}.  This construction does not make particularly efficient usage of the available resources. One can discard couplings terms and as long as the graph remains connected it turns out that total time of the construction remains constant.  This construction is more of the form of a proof of principle, similar to the Trotter expansion from Ch.~\ref{Sec:Controllability} in that the construction is a sequence of single-Hamiltonian propagators.  Unlike the Trotter expansion this has no infinitesimals and thus could be used in practice.  While this construction is not time-optimal, the length of the waveforms is still scales polynomially in $d$, and more importantly describes a deterministic algorithm.   
           
\subsubsection{Time-optimal control with Riemannian symmetric subspaces}

In \cite{khaneja01}, the authors developed a very clever way to find time-optimal controls for certain types of spin systems by solving a simple geodesic equation.  The geometric requirements for this scheme are that we have a standard control system on a $d$-dimensional Hilbert space,
given in Eq.~\ref{eq:generic_hamil}, where the constant term is much weaker than than the time-dependent pieces, $\| H_0\| \ll \| c_j H_j \|$.  Furthermore, the system must contain of Riemannian symmetric subspace which has the following form.  We will label the Lie algebra generated by just the time-dependent terms, $\{H_1,H_2,\ldots H_n \}$, as $\mathfrak{k}$, with corresponding Lie group $K$.  The (right) coset space of the respective Lie group, $SU(d)/K$, must be a Riemannian symmetric subspace.  More precisely, let $\mathfrak{m}$ denote the orthogonal complement of $\mathfrak{k}$ in $\mathfrak{su}(d)$.  The coset space $SU(d)/K$ is Riemannian symmetric if all elements in $\mathfrak{k}$ and $\mathfrak{m}$ satisfy the commutator relations 
\be
[\mathfrak{k},\mathfrak{k}] \subset \mathfrak{k} \quad [\mathfrak{k},\mathfrak{m}] \subset \mathfrak{m} \quad [\mathfrak{m},\mathfrak{m}] \subset \mathfrak{k}.
\ee  
This is obviously a fairly restrictive property. One common example however is in $SU(4)$ where $K = SU(2)\times SU(2)$.  That is, the drift term describes a coupling term between two qubits and we completely control the single qubit Hamiltonians.

\begin{figure}[t!]
\begin{center}
\includegraphics[width=15cm,clip]{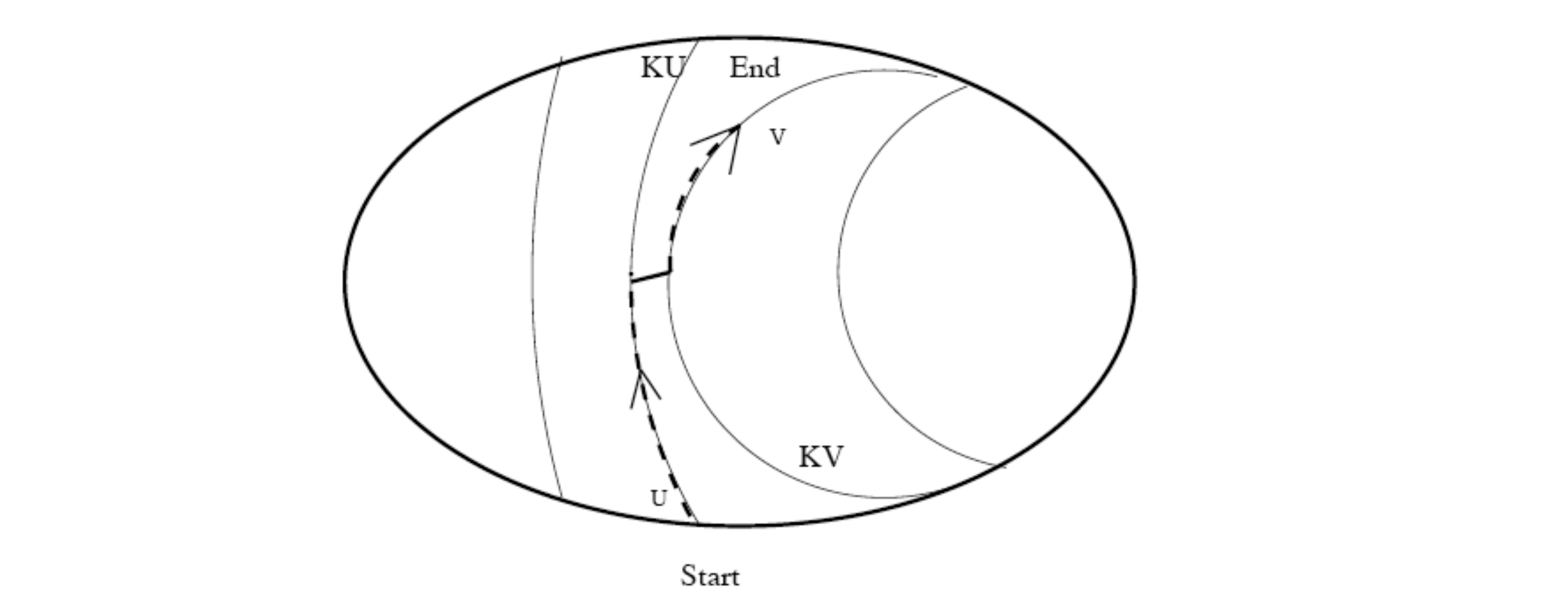}
\caption[Time optimal control with Riemannian symmetric subspaces]{Figure from \cite{khaneja01} that schematically describes the time optimal controls.  The dotted line denotes the optimal control to drive the system to $V$ in the shortest time.  Since movement along the cosets has essentially no cost the algorithm minimizes how long we must evolve according to the drift term, which is the only way to move between cosets. }
\label{F:geodesic}
\end{center}
\end{figure}

The importance of this type of system is that there is now an equivalence between finding controls that minimize the time $T$ such that
\be
 \dot{U}(t) = -i H(t) U(t) \qquad U(0) = \mathbb{I}, \quad U(T) = V
\ee 
and finding time-optimal controls $X$ such that
\be
\dot{P}(t) = X(t) P(t) \qquad P(0) = \mathbb{I} \quad P(T) = K V.
\ee
Here $X$ belongs not to the entire unitary group, but simply $X = \textrm{Ad}_K (H_0) = \{ k^{-1} H_0 k:k \in K \}$.  This second optimization is much easier because since the solution basically describes geodesic equation.  

Instead of moving on $SU(d)$ the second optimization moves through cosets, see Fig.~\ref{F:geodesic}.  Since the time-dependent terms are much stronger than the drift term, moving within a coset has essentially no cost.  Our optimizations simply needs to find the point on our current coset where $H_0$ describes the greatest rate of change.  This means we can use a simple greedy search to find time optimal controls since we optimize that rate of change independently at each point.   It is crucial that the coset space is Riemannian symetric since otherwise the optimal controls may involve backtracking, which makes a greedy search impossible.

This construction is very nice in that it provides not only the optimal controls with respect to the Hilbert-Schmidt norm, but also the optimal controls with respect to the duration of the control pulses.  The final algorithm for constructing controls is simple and deterministic.  Again, however, the restrictions on the character of the control Hamiltonians reduce its applicability to a small set of physical systems.

\chapter{Alkali Atomic Systems}\label{ch:atomic}

As stated in the introduction, atomic spins are a natural system to consider for storing and manipulating quantum information.  Because of the advances in laser cooling, ensembles of alkali atoms are a natural system to explore.  When the atoms are cold, their motion is negligible and they can be considered to be frozen in space over the time scale of interaction.  This vastly simplifies  the description and allows us to focus solely on the internal dynamics.  The internal state of alkali atoms is dependent only on a single valence electron plus nuclear spin, leading to a hydrogen-like level structure.  For many isotopes this leads to electronic ground states that have a non-trivial number of hyperfine states, e.g.~$^{133}$Cs has a nuclear spin $I = 7/2$ a thus $2(2I + 1) = 16$ sublevels.  Since these atoms are neutral and have no dipole in the ground state, they are extremely well-isolated from the environment.  Furthermore, we have easy access to the mature technology of diode lasers that can be tuned to the D1 and D2 resonance lines in alkalis, which lie in the near infrared.               

\begin{figure}[t!]
\begin{center}
\includegraphics[width=9cm,clip]{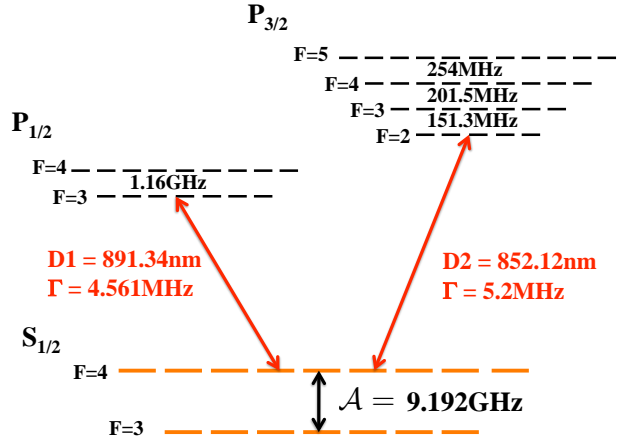}
\caption[Hyperfine levels for $^{133}$Cs]{The level structure of $^{133}$Cs for the D1 and D2 line.  Our control system of interest is the $S_{1/2}$ ground state, highlighted in orange.}
\label{F:levels_bare}
\end{center}
\end{figure}

We seek to control the quantum state of a multilevel atom.  Though single-atom addressing and measurement are possible \cite{schlosser01,dotsenko05,nelson07}, in practice we consider ensembles of uncorrelated particles.  To the degree that the atoms are identically prepared and uniformly addressed, with no interactions between them either from interatomic forces or through measurement backaction, we can take the joint state of the system as effectively $N$ identical copies, $\rho^{\otimes N }$.  More general many-body control is not considered here.  Restricting then to a single atom, the relevant Hilbert space of an alkali atom in its electronic ground state is the tensor product space of electronic spin $S$ and nuclear spin $I$ subsystems, $\mathcal{H}=\mathfrak{h}_S\otimes \mathfrak{h}_I$.  Given the single valence electron $S=1/2$, the Hilbert space is spanned by two irreducible subspaces of total angular momentum $F_{\pm}=I\pm1/2$, such that $\mathcal{H}=\mathfrak{h}_{+}\oplus\mathfrak{h}_{-}$.  With $^{133}$Cs, where the nuclear spin is $7/2$, these spin manifolds are $F_+ = 4$ and $F_- = 3$, see Fig.~\ref{F:levels_bare}.   

The Hamiltonian describing the atom and its interactions with external magnetic and electric fields in the electronic ground state is given by
\be
H = H_{\textrm{ATOM}} + H_{\textrm{MAG}} + H_{\textrm{LASER}} =  \mathcal{A}\mbf{I} \cdot \mbf{S} -\mbf{ \mu} \cdot \mbf{B}(t)-\frac{1}{4} E_i(t)^* E_j(t) \alpha_{ij}.\label{eq:full_hamil}
\ee    
Throughout this discussion I will set $\hbar=1$. For all the work in this dissertation the dominant term in this Hamiltonian will be the hyperfine interaction, $\mathcal{A}\mbf{I} \cdot \mbf{S}$.  In units of Plank's constant, in cesium the strength of the hyperfine coupling is $\mathcal{A}/\hbar = 9.2$~GHz.  The strength of the applied magnetic fields will at most be $2 \mu_B B_0/\hbar \approx 1$~MHz for a static bias field but will more typically be on the order of $2 \mu_B |\mbf{B}(t)|/\hbar \approx 10$~kHz for our time-dependent control fields.  The goal of this chapter is to rewrite this Hamiltonian in a way that is conducive to the types of control techniques we discussed in the previous chapter, as well as showing the resultant systems are controllable.  

\section{Quasi-static magnetic fields and light shift}\label{chp:control_magAC}

One approach to controlling atomic spins is with Zeeman and AC-Stark shift interactions \cite{smith04, silberfarb05, smith06}.   In this control system, the space of interest is restricted to the manifold $F_-$.   For the remainder of this section I will label the irreducible generators of angular momentum on this space as simply $\mbf{F}$.  Restricting to $F$, we can write $H = \mathbb{P}_F H \mathbb{P}_F$.  Since the hyperfine interaction 
\be
\mathcal{A} \mbf{I}\cdot \mbf{S} = \frac{\mathcal{A}}{2} \left(F^2 - I^2 - S^2 \right),
\ee      
 is a constant when reduced to one spin manifold we are left with two terms in our control Hamiltonian
\be
 H = -\mbf{\mu} \cdot \mbf{B}(t)-\frac{1}{4} E_i(t)^* E_j(t) \alpha_{ij}.
\ee

The nuclear magneton $\mu_I$ is about three orders of magnitude smaller than the Bohr magneton $\mu_B$.  We can thus, with high accuracy, write the magnetic field Hamiltonian as an operator purely on the electronic spin 
\be
H_{\textrm{MAG}} = 2 \mu_B \mathbf{B}(t) \cdot \mbf{S}.
\ee
In the linear Zeeman regime, with no resonant effects, this Hamiltonian approximately preserves $F$ and can be written according to the Land\'{e} projection theorem
\be
H_{\textrm{MAG}} \approx \mu_B  g_{f} \mbf{B}(t) \cdot \mbf{F}.
\ee
In the experiment I will discuss in Chapter \ref{chp:state_prep}, we controlled just the $x$ and $y$ components of the magnetic field.  We can combine constants to get the Larmor frequencies, $\Omega$, in the two  directions to write
\be
H_{\textrm{MAG}} = \Omega_x(t)F_x + \Omega_y(t)F_y
\ee
 
It should be clear that magnetic fields only generate rotations, and thus a representation of $\mathfrak{su}(2)$ and not the full algebra $\mathfrak{su}(2F+1)$.  To create a controllable system we need to consider the laser light shift interaction
\be
H_{\textrm{LASER}} = -\frac{1}{4} E_i(t)^* E_j(t) \alpha_{ij}.
\ee  
Here, $\alpha_{ij}$ is the polarizability tensor
\be
\mbf{\alpha} = - \sum_{g,e} \frac{\mbf{d}_{ge}\mbf{d}_{eg} }{\Delta_{eg}}
\ee  

We can reduce this to a more manageable form by expressing the light shift Hamiltonian in terms of its irreducible spherical components
\be
H_{\textrm{LASER}} =-\frac{1}{4} \left( \alpha^{(0)} |\mbf{E}|^2 + \mbf{\alpha}^{(1)}\cdot (\mbf{E}^* \times \mbf{E}) +\alpha^{(2)}_{ij} (E^*_iE_j - \frac{1}{3} |\mbf{E}|^2 \delta_{if}) \right).
\ee
We can rewrite this Hamiltonian as an effective operator on the atomic spin by expressing $\alpha_{ij}$ in terms of the generators of angular momentum on $F$ like 
 \be
 H_{\textrm{LASER}} =  c^{(0)} |\mbf{E}|^2 + c^{(1)} (\frac{\mbf{E}^* \times \mbf{E}}{i})\cdot \mbf{F} +c^{(2)} (|\mbf{E} \cdot \mbf{F}|^2 - \frac{1}{3} |\mbf{E}|^2 |\mbf{F}|^2 ).
 \ee
The constants, $c^{(j)}$, can be found through the Wigner-Eckart as in \cite{deutschjessen}.

For our control system we use monochromatic light with polarization along the $x$-direction.  In this case the light shift Hamiltonian, dropping constant terms, reduces simply to 
\be
H_{\textrm{LASER}} = c^{(2)} |\mbf{E}|^2 F_x^2 =   \beta \gamma_s F_x^2.
\ee
Here we can rewrite the constants in terms if the photon scattering rate, $\gamma_s$, and a dimensionless parameter $\beta$ which is a measure of the timescales for coherent versus incoherent evolution. Its value depends on the atomic structure and the frequency of the driving field and for Cs takes on a maximum value $\beta=8.2$ when tuned between the hyperfine transitions of the $D_1$  line at $894$nm. This is enough to allow considerable coherent manipulation.  Due to technical concerns the laser was an ``always on" interaction leading to a final control Hamiltonian 
\be
H = \beta \gamma_s F_x^2 + \Omega_x(t) F_x + \Omega_y (t) F_y.
\ee  
$F_x^2$ is itself a rank-2 operator of angular momentum, and so we see that this system is controllable by direct application of Thm.~\ref{t:rank_2}.  In the experiment, the photon scattering rate is typically around $\gamma_s / 2 \pi \approx 0.77$~kHz and the amplitudes of the applied magnetic fields are about $B \approx 40$mG which leads to $\beta \gamma_s/ 2 \pi \approx 0.5$~kHz and $\Omega / 2 \pi \approx 15$~kHz.   

In addition to the nonlinear light-shift, the laser interaction also leads to spontaneous photon scattering.  This is important since, in the large detuning limit, the photon scattering rate has the same scaling with respect to the intensity and detuning of the laser as the nonlinear contribution to the Hamiltonian.  By choosing the optimal parameters we can get some nontrivial evolution before we lose too much coherence to spontaneous emission, but since the incoherent and coherent rates are intrinsically related there is an upper bound on the length of the coherent control fields it is possible to consider.

\section{Microwave and rf magnetic fields}\label{chp:hamil_mwrf}

An alternative route to controlling the atomic spins is to employ solely magnetic interactions, and remove the necessity of the laser-induced AC-Stark shift.  This approach has the advantage that we can perform control on the entire electronic ground state rather than one irreducible manifold, a 16-dimensional Hilbert space.  Additionally, none of the control fields are intrinsically tied to decoherence, with spontaneous scattering of rf or microwave photons completely negligible, in principle allowing for much richer landscape of possible controls.

The Hamiltonian describing the atom and its interaction with external magnetic fields takes the form given in Eq.~\ref{eq:full_hamil}, with laser coupling set to zero.  In this control scheme we consider the application of three fields, $\mbf{B}(t)=B_0 \mbf{e}_z + \mbf{B}_{\rf}(t) + \mbf{B}_{\mw}(t)$.  The static bias field $B_0$ defines the quantization axis and Zeeman splittings between the magnetic sublevels.  The terms $\mbf{B}_{\rf}(t)$ and $\mbf{B}_{\mw}(t)$ describe magnetic fields oscillating at radio and microwave frequencies, respectively.  The hyperfine coupling between spins provides an effective nonlinearity that will allow full controllability of the Hilbert space for appropriate choices of external fields.  

In the linear Zeeman regime, $\mu_B B_0 \ll A$, the static field acts separately in the two irreducible subspaces, and according to the Land\'{e} projection theorem, the Hamiltonian is approximately,
\be
H_{B_0} \approx \mu_B \sum_{f=\pm} g_f \mbf{B}_0 \cdot \mbf{F} ^{(f)}. \label{eq:bfield}
\ee
Here $\mbf{F}^{(\pm)} \equiv P_\pm \mbf{F}P_\pm$ refers to the total angular momentum operator projected onto the subspaces with quantum number $F_\pm$. Neglecting the nuclear magneton contribution, the g-factors for the two manifolds have equal magnitude but opposite sign, i.e. $g_+ = -g_-=1/F_+$.  The hyperfine coupling plus bias magnetic field thus determine the static Hamiltonian,
\be
H_0  = \frac{\Delta E_{HF}}{2}\left(P_+ - P_- \right)+ \Omega_0 (F^{(+)}_z - F^{(-)}_z),
\ee
where $\Delta E_{HF}=A F_{+}$ is the hyperfine splitting and $\Omega_0= \mu_B B_0/ F_+$ is the Zeeman splitting between neighboring magnetic sublevels.

\begin{figure}[t!]
\begin{center}
\includegraphics[width=9cm,clip]{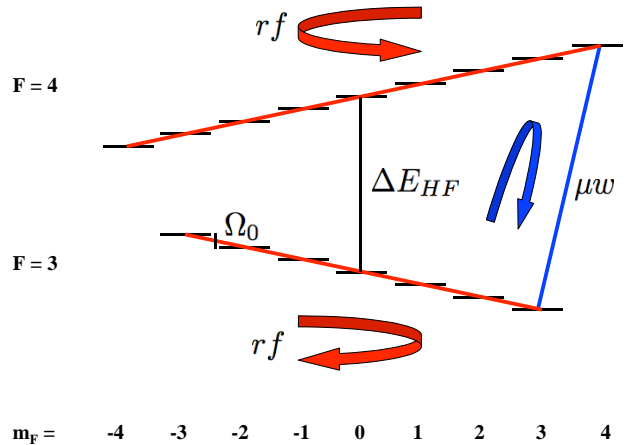}
\caption[Microwave and Rf magnetic fields]{The ground state hyperfine manifold of $^{133}$Cs.  Rf-magnetic fields (in red) lead to independent rotations in the two manifolds.  Microwaves (in blue) are the generators of rotation in a two-dimensional subspace between states in the two manifolds, here the stretched state transition $\ket{4,4} \rightarrow \ket{3,3}$. }
\label{F:hyperfine}
\end{center}
\end{figure}

As our first control field, we consider rf-magnetic fields oscillating near the frequency of the Zeeman splitting, $\omega_{\rf} \approx \Omega_0$, realized by Helmholtz coils driven with the appropriate current. We take two sets of coils that produce fields with $x$ and $y$ polarization, independent amplitude and phase control, but equal carrier frequency, $\omega_{\rf}$.  Again, for a moderate current such that the amplitude of the magnetic field is in the linear Zeeman regime, the rf-Hamiltonian takes a form equivalent to the interaction with the static field
\bea
H_{\rf}(t)& =& \Omega_x (t) \cos{\big(\omega_{\rf} t - \phi_x(t)\big)}  \big(F^{(+)}_x - F^{(-)}_x\big)\nonumber\\
&+&\Omega_y (t) \cos{\big(\omega_{\rf} t - \phi_y(t)\big)}  \big(F^{(+)}_y - F^{(-)}_y\big) .
\eea
The time dependent amplitudes $(\Omega_x(t), \Omega_y(t))$ and phases $(\phi_x(t),\phi_y(t))$ of the two sets of  rf coils will be used to control the system. 

To better understand the effect of the rf field, consider a resonant interaction, $\omega_{\rf}=\Omega_0$.  In the rotating frame, $H_{\rf}(t) \rightarrow H_{\rf}'(t) =U_{\rf}^{\dagger}H_{\rf}(t) U_{\rf}$, where $U_{\rf} =\exp \left\{-i \omega_{\rf}t (F^{(+)}_z - F^{(-)}_z ) \right\}$ is a rotation of the two manifolds about the $z$-axis in opposite directions, $F_x^{(\pm)} \rightarrow F_x^{(\pm)} \cos(\omega_{\rf} t) \pm  F_y^{(\pm)} \sin(\omega_{\rf} t)$, $F_y^{(\pm)} \rightarrow F_y^{(\pm)} \cos(\omega_{\rf} t) \mp  F_x^{(\pm)} \sin(\omega_{\rf} t)$.  Performing this unitary transformation and averaging over a cycle, the rf-Hamiltonian in the rotating wave approximation is,
\bea
H_{\rf}'(t) &=&\frac{\Omega_x (t)}{2} \cos{\big( \phi_x(t)\big)} \big(F^{(+)}_x -F^{(-)}_x \big)  \nonumber\\
&+& \frac{\Omega_x (t)}{2} \sin{\big( \phi_x(t)\big)} \big(F^{(+)}_y + F^{(-)}_y\big)\nonumber\\
&+&\frac{\Omega_y (t)}{2} \cos{ \big( \phi_y(t)\big)}   \big(F^{(+)}_y -F^{(-)}_y \big)\nonumber\\
 &-& \frac{\Omega_y (t)}{2} \sin{\big( \phi_y(t)\big)} \big(F^{(+)}_x+F^{(-)}_x \big). 
\label{eq:rf_hamils}
\eea
Rf-control of the two spin manifolds differs from the familiar spin resonance problem.  In the latter, a single magnetic field in either the $x$ or $y$-direction would be sufficient to generate the entire $SU(2)$ algebra for rotations.   With two irreducible manifolds there is an added freedom -- the two angular momenta $F_+$ and $F_-$ can rotate in the same or opposite directions.  Amplitude and phase control of two rf-magnetic field polarizations allows us to perform arbitrary and independent rotations on the two hyperfine manifolds.  With only a single direction of $\mbf{B}_{\rf}$ we would be restricted to either co-rotating or counter-rotating in the two subspaces.  

The weak rf-magnetic fields alone will not be sufficient to fully control our atomic system; they don't couple the $F_+$ and $F_-$ manifolds, nor do they provide a nonlinear Hamiltonian within these subspaces.  In order to make our system fully controllable, we look to resonant microwaves.  While the fundamental Hamiltonian governing the microwaves is exactly of the same form as the quasistatic magnetic fields, the resonant behavior leads to very different dynamics than the previous interactions.  Depending on the polarization and frequency, the microwave couples a Zeeman sublevel in $F_+$ manifold with one in the $F_-$ manifold whose magnetic quantum number differs by $\Delta m =  0,\pm 1$.  For a sufficiently strong bias $B_0$ we can ignore any off-resonant excitation, and restrict the Hamiltonian to act only on a 2D subspace spanned by the states we are trying to couple.  In that case the microwave Hamiltonian has the form
\be
 H_{\mw}(t) = \Omega_{\mw}(t) \cos{\big( \omega_{\mw} t - \phi_{\mw}(t) \big)}\sigma_x,
\ee
where $\sigma_x$ is the Pauli sigma-$x$ matrix for this pseudospin, $\sigma_x = \ket{F_+,m_+}\bra{F_-,m_-} + \ket{F_-,m_-}\bra{F_+,m_+}$ and $\Omega_{\mw}(t)$ is the (time-dependent) Rabi frequency depending on the microwave power and the transition matrix element.  Again, the amplitude and phase of the microwave fields are control parameters.  In this subspace, the problem takes the form of the standard two-level resonance problem.  We must take care in going to the rotating frame to account for the simultaneous transformation we perform due to the rf-fields.  The complete frame transformation is achieved by the unitary matrix
\be
U=U_{\rf} \exp\left\{-i \frac{\alpha t}{2} \left(P_+ -P_-\right) \right\},
\ee
where $\alpha = \omega_{\mw} -(m_+ + m_-)\omega_{\rf}$. Under this transformation, the Hamiltonian in the rotating wave approximation for resonant microwaves is 
 \bea
 H'_{\mw}(t) &= &\frac{\Omega_{\mw}(t)}{2} \cos{\big(  \phi_{\mw}(t) \big)}\sigma_x \nonumber\\
 &+& \frac{\Omega_{\mw}(t)}{2} \sin{\big(  \phi_{\mw}(t) \big)}\sigma_y,\label{eq:mw_hamil}
 \eea
generating rotations of this pseudo-spin on the Bloch sphere.

Combining the static, rf, and microwave interactions the final Hamiltonian in the rotating frame is
\be
H'(t) = H'_0+H_{\rf}'(t)+H_{\mw}'(t).
\ee
Allowing for a finite detuning of the oscillating fields from resonance, the static Hamiltonian in the rotating frame becomes, 
\be
H'_0 =  \frac{\Delta_{\mu w }}{2}\left(P_+ - P_- \right)+ \Delta_{\rf}(F^{(+)}_z - F^{(-)}_z),
\ee
where $\Delta_{\mw}=\omega_{\mw}-\Delta E_{HF} -(m+m')\omega_{\rf}$ is the effective detuning of the microwaves from the two-level transition of interest, $\ket{F_-,m_-}\rightarrow\ket{F_+,m_+}$, and $\Delta_{\rf}=\omega_{\rf}-\Omega_0$ is the rf detuning.  This, together with Eqs. (\ref{eq:rf_hamils},\ref{eq:mw_hamil}), defines the Hamiltonian we employ for control, and which we will analyze for use in arbitrary state preparation.

For this Hamiltonian system, with arbitrary control of the amplitude and phase of the two orthogonal sets of rf-coils and a single microwave field, the control algebra generated by the six operators $\{ F_x^{(+)}, F_y^{(+)}, F_x^{(-)}, F_y^{(-)}, \sigma_x, \sigma_y \}$ is $\mathfrak{su}(d)$ in its entirety.   In this case, it is possible to prove controllability analytically for an arbitrary alkali, with an arbitrary nuclear spin $I$.  

The proof is as follows, first we would like to show that with our Hamiltonian the subspaces $F_+$ and $F_-$ are independently controllable.  To show controllability of the $F_+$ manifold we require an operator that has a nonzero overlap with a rank-2 tensor on that space.  Restricted to the $F_+$ subspace, the $\sigma_z$ operator looks like a projector onto some particular sublevel, $\ket{F_+,m_+}\bra{F+,m_+}$.  The overlap of this projector with $T^{(2)}_0$ is $\textrm{Tr}\left( \ket{F_+,m_+}\bra{F_+,m_+} T^{(2)}_0\right) = \sqrt{5/11} \braket{F_+,m_+}{2,0;F_+,m_+}$, which is nonzero for all values of $m_+$.  Of course, $\sigma_z$ also has support in the $F_-$ manifold, however, $\left[ F^{(+)}_x,\sigma_z\right]$ does not.  Since commuting by $F^{(+)}_x$ can't change the rank of a tensor, we are left with an operator confined to the $F_+$ manifold that has a nonzero overlap with some rank-2 tensor, and so according to theorem \ref{t:rank_2}, we have complete control of the $F_+$ manifold.  This proof directly carries over to the $F_-$ manifold. 

At this point we have shown that we have full controllability over both the $F_+$ and the $F_-$ subspaces, as well as the 2-dimensional subspace coupled by the resonant microwaves.  We can write this in matrix form   
\be
\setlength{\unitlength}{0.25cm}   
\left(
\begin{array}{c}
\begin{picture}(16,16)(0,0)
\put(0,16){\line(1,0){9}}
\put(0,16){\line(0,-1){9}}
\put(9,16){\line(0,-1){9}}
\put(0,7){\line(1,0){9}}
\put(9,7){\line(1,0){7}}
\put(9,7){\line(0,-1){7}}
\put(16,7){\line(0,-1){7}}
\put(9,0){\line(1,0){7}}
\put(8,8){\line(1,0){2}}
\put(8,8){\line(0,-1){2}}
\put(10,8){\line(0,-1){2}}
\put(8,6){\line(1,0){2}}
\put(3.75,11.5){$s_1$}
\put(11.75,3.25){$s_2$}
\put(10.5, 9){$\sigma$'s}
\end{picture}
\end{array} 
\right),\label{e:block}
\ee
where we have ordered the basis vectors so that the states coupled by the microwaves are adjacent to each other.  We have shown that we can simulate any operator that only has matrix elements within the three boxes in Eq.~\ref{e:block}, i.e.~all operators that have support only on the diagonal and super diagonal matrix elements.  The irreducible representations of angular momentum, $J_x$ and $J_y$, on the entire space have support only on the super-diagonals.  Therefore, we can simulate $J_x$ and $J_y$.  According to theorem \ref{t:rank_2} all we need to show for controllability is that we can simulate some operator with a nonzero overlap with some rank 2 operator.  Since we can simulate any diagonal operator, we can simulate $T^{(2)}_0$.  It thus follows that $\{F^{(+)}_x,F^{(+)}_y,F^{(-)}_x,F^{(-)}_y, \sigma_x, \sigma_y \}$ generates $\mathfrak{su}(d)$.

Though sufficient, the entire available set is not necessary to achieve controllability.  In practice, one can reduce the number generators in the control algebra and still implement an arbitrary unitary.  For an experiment, it is important to understand which components are really necessary so that we can chart the tradeoffs between ease of implementation and controllability.  In order to study the capability of various reduced sets of controls, we resort to numerical approach discussed in Ch.~\ref{Sec:Controllability}.  

\begin{table}[t!]
\begin{center}
\includegraphics[width=15cm,clip]{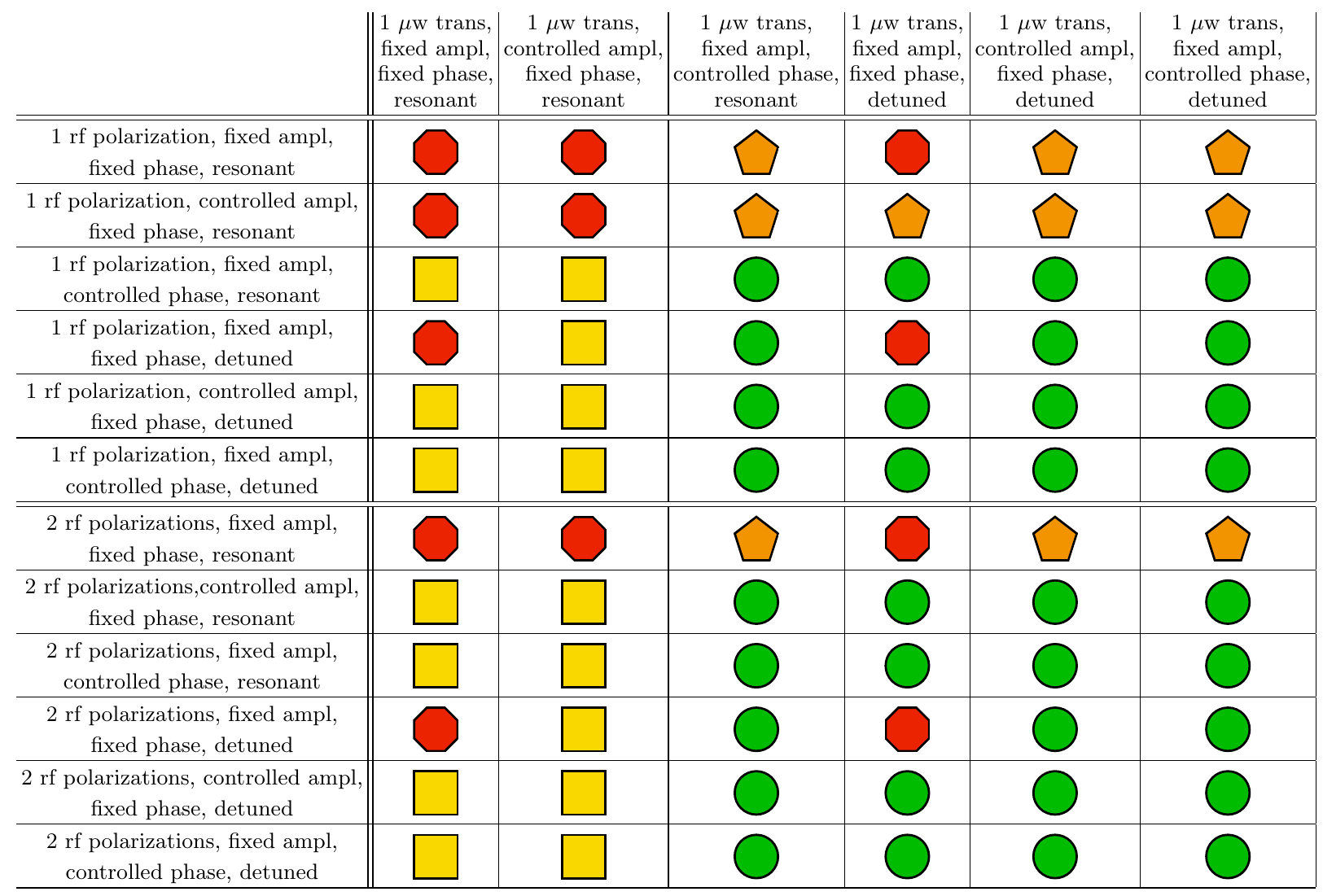}
\caption[Controllability with microwave and rf magnetic fields]{Table exploring controllability of the system for a variety of configurations: one microwave field driven on different two-level transitions, $\ket{F=3,M} \rightarrow \ket{F=4,M'}$, amplitude and/or phase control, one or two sets of orthogonal rf coils (rf polarizations), and resonant vs.~detuned fields. The different configurations yield one of four different outcomes: (green circle) all microwave transitions provide full controllability, (yellow square) all transitions but the clock transition $\ket{3,0} \rightarrow \ket{4,0}$  provide full controllability, (orange pentagon) only the transitions of the form $\ket{3,\pm 3} \rightarrow \ket{4,\pm 4}$ and $\ket{3,\pm 3} \rightarrow \ket{4,\pm 2}$  provide full controllability, and (red octagon) no transitions yield controllable Hamiltonian dynamics.  In this calculation we consider all valid microwave transitions, not only the energy non-degenerate ones.}
\label{F:control_table}
\end{center}
\end{table}

We carried out this procedure for the specific example of $^{133}$Cs with nuclear spin $I=7/2$ to study the capability of a variety of control sets to generate the entire $\mathfrak{su}(16)$ algebra.  We considered 8 different microwave configurations:  controlling or fixing the amplitude and the phase of the fields, and whether or not we are detuned from resonance.  The two cases where both the amplitude and phase are controlled and where the amplitude is fixed but the phase is controlled can be shown to be equivalent.  In the rf configurations we also allow for one or two orthogonal sets of magnetic coils.  The last free parameter is the choice of which microwave transition we excite.  We assume arbitrary frequency and polarization selectivity of the desired transition for this purpose.  The results are summarized in summarized in table \ref{F:control_table}.  In each box we enumerate the set of microwave transitions that yield controllable dynamics.   We find that our system is controllable in a wide number of regimes, though there are some configurations in which it is not.  For example, out of all the choices for microwave transitions, the clock transition, $\ket{F_+,0} \rightarrow \ket{F_-,0}$, is controllable in the least number of scenarios.  This shouldn't come as much of a surprise since we are controlling the system with rf magnetic fields and this transition's insensitivity to magnetic fields is what makes it useful for precision metrology.   

It is interesting to note that there exist configurations that are controllable in which there is one time-dependent control waveform and some fixed time-independent interactions.  This is the simplest system one could expect to find, and allows for bang-bang control, a well-studied protocol.  In the next chatper, however, we look at the control systems that utilize more parameters, decreasing the time needed for state preparation.

 \chapter{State Preparation with Alkali Atoms}\label{chp:state_prep}
 
In this chapter I will discuss the application of the theoretical methods for state preparation discussed in Ch.~\ref{ch:control} to the the physical systems of alkali atomic spins discussed in Ch.~\ref{ch:atomic}.  In the context of control via AC-Stark and quasi-static magnetic fields, this protocol was carried out in the laboratory and shown to yield good results as described below.  In the context of microwave/rf control, I have devised new protocols for control which have been studied numerically.  Experimental test should be forthcoming in the near future.

\section{A state preparation algorithm}\label{sec:stateprep_alg}
  
We seek to design Hamiltonian evolutions that take an initial known pure quantum state to an arbitrary pure state in the Hilbert space of interest.   We would like to maximize fidelity as a functional of the control waveform given by
\be
F[\mbf{c}(t)]  = |\bra{\psi_{target}}U[\mbf{c}(t)] \ket{\psi_0} |^2.
\ee
As we explored in chapter \ref{sec:topology}, this problem has an extremely favorable topology devoid of local maxima, and therefore, a local search of the space of control fields, starting from any random initial guess, will find a global maximum of the fidelity.  For this problem, gradient searches perform about as well as more computationally intensive searches like genetic or simulated annealing algorithms.  

In a real system, we will violate some of the assumptions required for the proof that there are no local maxima.  There will always be some decoherence and one does not have infinite time to perform the control.  In fact, we would like to perform state preparation as fast as possible in order to combat decoherence and various inhomogeneities that lead to accumulated errors.  Additionally, we need to consider control fields that have a limited bandwidth and slew rate constraints.  For these realistic conditions, not every gradient search from an arbitrary starting point yields a global maxima.  Nonetheless, we have found empirically that the results of the theorem are approximately true with moderate decoherence and after a sufficient time.  We still find excellent protocols after making only a small handful of searches, and these can be further filtered to find control waveforms that perform well under realistic operating conditions.  

As we are dealing with the optimization of waveforms that are functions of continuous time, the first step is to transform the problem into a search for a finite number of values at discrete times.  The physical constraints of bandwidths and slew rates of the controllers provide a natural scale. There is a minimum interval during which a field can vary over a maximum range.  A discretized version of a control waveform is thus specified as a vector of values within this range at these fixed intervals.  The continuous control waveforms are then found by interpolation using cubic splines, consistent with the bandwidth constraints, at least on a fine enough grid for use in our numerical integration of the Schr\"{o}dinger equation. 
  
We create optimal control waveforms by first fixing the total time of the state preparation procedure.  Due to our discretization technique, fixing the total time fixes the number of optimization variables.  Starting from a randomly chosen initial vector of control waveform values, $\mbf{b}_0$, we perform a gradient ascent search by taking small steps in the direction of steepest ascent, i.e. 
\be
\mbf{b}_{n+1} = \mbf{b}_n + \epsilon \nabla F(\mbf{b}_n).
\ee   
An optimal value corresponds to the maximum, where the gradient approaches zero.  We performed this search numerically on a Matlab cluster by optimizing waveforms from a handful of random seeds in parallel, and then chose the one that gave the highest fidelity.  The actual gradient search itself was performed using a canned algorithm from Matlab's ``Optimization Toolbox."  An alternative approach would have been to use the Gradient Ascent Pulse Engineering (GRAPE) algorithm developed in \cite{khaneja05}.  While this algorithm has been used to great affect in a number of quantum control protocols, it was of no use in our scheme for controlling atomic spins. For completeness, I give a brief summary of the GRAPE algorithm here and its limitations. 

The GRAPE algorithm is a gradient search algorithm whose outstanding feature is that the gradient is computed in a way that is much more efficient than a standard numerical differentiation routine.  In the simplest incarnation of the GRAPE algorithm we imagine the controls $\cv$ to describe the amplitude of square pulses with one time-varying control field.  We can write the fidelity of state preparation as 
\be
J[\cv] = |\bra{f} e^{-i \Delta t (H_0 + c_N H_1)}e^{-i \Delta t (H_0 + c_{N-1} H_1)} \ldots e^{-i \Delta t (H_0 + c_1 H_1)}\ket{i}|^2.
\ee  
When we numerically evaluate $J$ we need to perform $N$ matrix multiplications.  If we were to try to we calculate the $N$-dimensional gradient numerically we would need to evaluate $J$, approximately $N$ times.  This leads to a scaling of $O(N^2)$ for computing the gradient numerically.  

The GRAPE algorithm allows us to compute the gradient with a cost of $O(N)$.  The trick is to simultaneously forward evolve the initial state while backwards evolving the final state.  We define two sets of vectors
\bea
\ket{i(j)}&=&e^{-i \Delta t (H_0 + c_j H_1)}e^{-i \Delta t (H_0 + c_{j-1} H_1)} \ldots e^{-i \Delta t (H_0 + c_1 H_1)}\ket{i},\\
\bra{f(j)} &=& \bra{f} e^{-i \Delta t (H_0 + c_N H_1)}e^{-i \Delta t (H_0 + c_{N-1} H_1)} \ldots e^{-i \Delta t (H_0 + c_{j+1} H_1)},
\eea  
with $\ket{i(0)} = \ket{i}$ and $\bra{f(T)} = \bra{f}$.  Clearly 
\be
J = |\braket{f(j)}{i(j)} |^2 = |\bra{f(j)}  e^{-i \Delta t (H_0 + c_j H_1)} \ket{i(j-1)} |^2, \quad \forall ~~0\leq j \leq T.
\ee
In the GRAPE algorithm we first compute $\bra{f(1)}$ and $\ket{i(1)}$, at a cost of $O(N)$ matrix multiplications.  If $\Delta t$ is sufficiently small we can compute
\bea
\frac{\partial J}{\partial c_1} &=& |\bra{f(1)}  \frac{\partial e^{-i \Delta t (H_0 + c_1 H_1)}}{\partial c_1} \ket{i(0)} |^2\nonumber\\
&\approx& |\bra{f(1)}  (-i H_1) e^{-i \Delta t (H_0 + c_1 H_1)} \ket{i(0)} |^2\nonumber\\
&=& |\bra{f(1)}  (-i H_1)  \ket{i(1)} |^2
\eea
at a cost that is constant in $N$.  To compute $\partial J/ \partial c_2$ we require the vectors $\bra{f(2)}$ and  \ket{i(2)} which we obtain by evolving 
\be
\bra{f(2)} = \bra{f(1)} e^{i \Delta t (H_0 + c_1 H_1)}, \qquad \ket{i(2)} = e^{-i \Delta t (H_0 + c_2 H_2)}\ket{i(1)}.
\ee
This only takes  two matrix multiplications.  We can repeat this $N$ times in order to get all the components of $\nabla_{\cv} J$ with only $O(N)$ matrix multiplications.

The assumption I have made in this simple description of the GRAPE algorithm is that the number of optimization variables corresponds to the ratio between the total time of the state preparation and the sampling time for the Schrodinger integrator.  This assumption is valid in the case where the maximal slew rates for the applied fields are much larger than the Rabi or Larmor frequencies of the fields, as is the case in liquid state NMR systems.  In more concrete terms, the restriction for convergence in the GRAPE algorithm is that the integration step length of the Schrodinger integrator $\Delta t$ must satisfy
\be
\Delta t \ll \| H(t) \|^{-1}
\ee  
where, $\| H(t) \|$ is the seminorm of $H(t)$, or the largest frequency component of the Hamiltonian.  This constraint arises from the requirement that the derivative of $\textrm{exp} \left( -i \Delta t (H_0 + c_1 H_1\right)$ is approximately  $(-i H_1) \textrm{exp} \left(-i \Delta t (H_0 + c_1 H_1)\right)$.  For our state preparation routine the constraint is that
\be
\Delta t ' \ll \| \frac{\partial H(t)}{\partial t} \|^{-1},
\ee 
requiring that the Hamiltonian must be relatively uniform over a time step.    This leads to a computational cost of $O(T/\Delta t)$ for the GRAPE algorithm and a cost of $O((T/\Delta t')^2)$ for ours.  For instances of the control systems in this dissertation it turns out that the second term is smaller and thus the GRAPE algorithm is not an efficient approach for our system..

For the experimental implementation, we additionally require that the state preparation protocol is at least somewhat robust to inhomogeneities and noise.  To enforce this we use a two round optimization.  First, we find a set of state preparation protocols using the above technique.  For the parameters we considered this would typically yield fidelities of greater than 0.99.  At this point we switch to a more realistic estimate of control performance by modeling the evolution with a full master equation that incorporates decoherence from light scattering and inhomogeneity of the nonlinear strength across the atomic ensemble. This allows a second stage of optimization starting from the waveform generated in round one and using the more complete but computationally intensive model to predict the yield, which is now defined in terms of the overlap $\mathcal{Y} = \Tr \sqrt {\rho_P^{\phantom{}1/2} \rho_{T}^{} \rho_P^{\phantom{P}1/2}}$ between the target density matrix $\rho_T^{}$ and the predicted density matrix $\rho_P^{}$.

 \section{Quasi-static magnetic fields and light shift }\label{ch:experiment}
 
We demonstrate, in \cite{chaudhury07}, the quantum control of the spin-angular momentum associated with the $F=3$  hyperfine ground state of individual $^{133}Cs$ atoms, a $2F+1=7$ dimensional Hilbert space. Starting from an easily prepared fiducial state we use time-dependent magnetic fields and a fixed AC Stark shift to design and implement near-optimal controls and produce a range of target states. We evaluate our control performance by experimentally reconstructing the entire spin density matrix \cite{silberfarb05} and computing the overlap between the measured and target states. In most cases the estimated yield is in the $0.8-0.9$ range, limited by errors in the control fields and to a lesser extent by decoherence from light scattering.  The measured states can be compared also to the predictions of a full model that includes the effects of errors and decoherence. Typical fidelities between measured and predicted states are around $0.9$, which is close to the resolution limit of our procedure for quantum state estimation. We further use this universal approach to generate spin-squeezed states and compare against a method based on adiabatic evolution \cite{molmersorensen}. The latter is more robust against errors in the control fields, but also slower and thus more sensitive to light scattering and decoherence. Large spins provide a testing ground for the design of accurate and robust controls in a system where the Hamiltonian is well known and where errors and dissipation are well understood and can be accurately modeled. From a practical perspective, quantum control of hyperfine states has direct relevance to proposals for neutral atom quantum computing \cite{jessenQIP2004} wherein qubits (or higher dimensional qudits \cite{brennen05}) are encoded in the ground-state manifold, and may provide a simple route to modest spin squeezing and accompanying gains in precision atomic magnetometry \cite{geremia2005}.

\begin{figure*}
[t]\resizebox{15cm}{!}
{\includegraphics{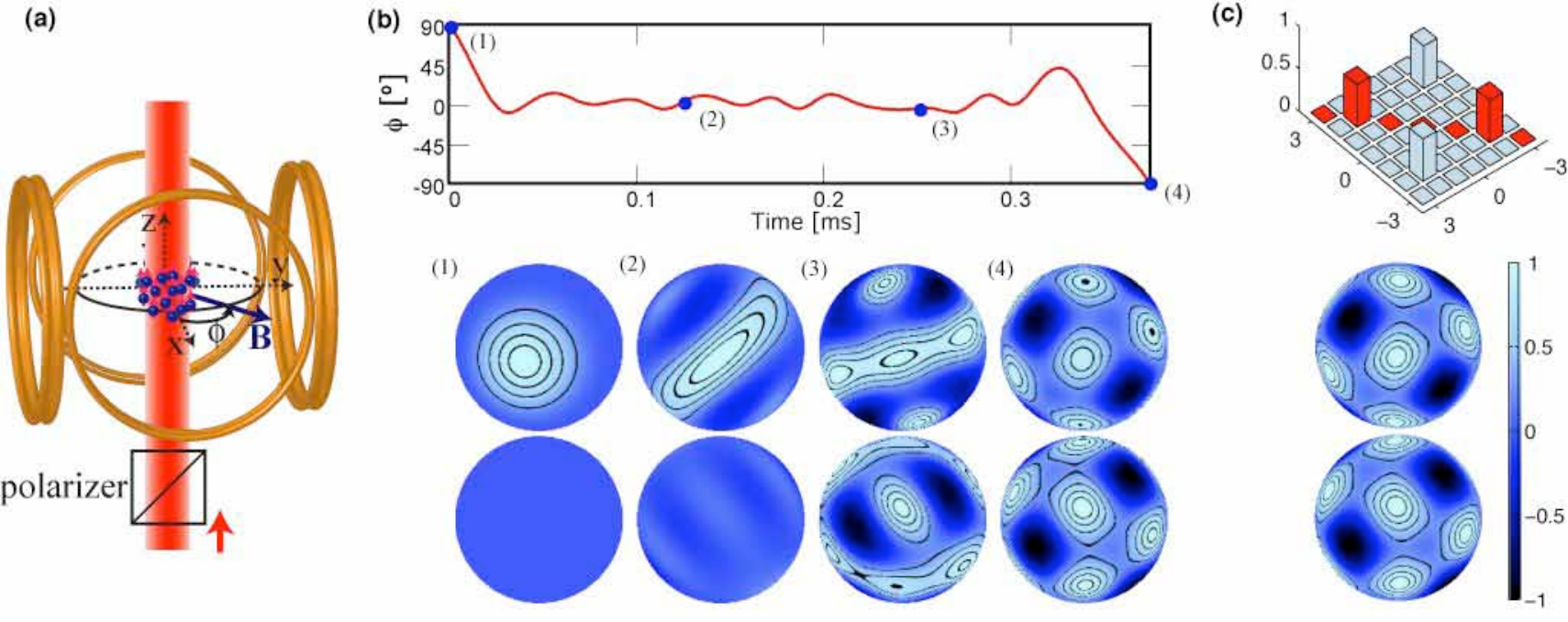}}
\caption[Schematic of AC-Stark shift experiment with time series of evolved states]{Quantum control of a large atomic spin.  (a) Schematic of the experiment.  (b) Example of a control waveform $\phi(t)$. (1-4) Wigner functions at four stages during the control sequence. Both sides of the sphere are shown.  The final result is close to the target state $ \ket{\psi_{target}} = (\ket{ m_z=2} +\ket{ m_z=-2 })/\sqrt{2}$. (c) Density matrix (absolute values) and  Winger function for $ \ket{ \psi_{target}}$}\label{fig:setup}
\end{figure*}

The combination of a time-dependent magnetic field and a constant x-polarized light field, discussed in chapter \ref{chp:control_magAC}, results in a control Hamiltonian \cite{smith04},
\begin{equation}
\hat{H}_C(t) = 
g_{f}\mu_{B} \mathbf{B}(t) \mathbf{\cdot \hat{F}} +
\beta \gamma_s F_x^2
\label{eq:controlH}
\end{equation}
A schematic of our setup for spin quantum control is shown in Fig.~\ref{fig:setup}(a). We begin with a sample of a few million Cs atoms, captured and laser cooled to $\sim2 \mu K$ in a magneto-optical trap and optical molasses. Once the atoms are released from the optical molasses their spin state is initialized by optical pumping into a state of maximum projection along the $y$-axis, $\ket{\psi_0} = \ket{F=3,m_y=3}$. We drive the spins by applying a time-dependent magnetic field from a set of low-inductance coils driven by arbitrary waveform generators, and by applying a static light shift from an optical probe beam. Using an all-glass vacuum cell, avoiding nearby conductive or magnetizable materials, and synchronizing our $\sim0.5 ms$ duration experiment to a fixed point during the AC line cycle allows us to null the background magnetic field to a few tens of $\mu$Gauss without the use of shielding or active compensation. The applied magnetic field can be controlled with an accuracy better than one percent in a bandwidth of more than $100$~kHz. Immediately following a period of quantum control we estimate the resulting quantum state as described in \cite{silberfarb05}. In this procedure the control magnetic and optical fields are applied to drive the spins for an additional 1.5 ms, while continually and weakly measuring a spin observable through its coupling to the probe polarization. To reduce the effect of noise, the measurement signal is averaged over 16 repetitions of the experiment and the full density matrix determined from the measurement record and the known evolution.

The objective is to start from the state $\ket{\psi_0}$ and to produce a specified target state $\ket{\psi_{target}}$ by modulating the field $\mathbf{B}(t)$ for a fixed time $\tau$. With readily available magnetic fields the timescale for geometric rotations is much shorter than for nonlinear evolution driven by the light shift, and the latter therefore becomes the time-limiting element of most transformations. In our experiment the maximum available Larmor frequency is $15$~kHz and the nonlinear strength is $\beta \gamma_S \approx 2\pi \times 500$~Hz. Under these conditions there is no significant sacrifice in control performance when the set of available rotations is somewhat restricted. We therefore choose the magnetic field to have constant magnitude and time-dependent direction in the $x$-$y$ plane. With this simplification the control Hamiltonian is completely determined by the time dependent angle $\phi (t)$ between $\mathbf{B}(t)$  and the $x$-axis. 
\be
B(t) = B_1 \left( \cos \left( \phi(t) \right) \mbf{e}_x+\sin\left( \phi(t)\right) \mbf{e}_y \right)
\ee 
The state $\ket{\psi_{target}}$ and the transformation $\ket{\psi_0}  \to \ket{\psi_{target}}$ belong to a $d=7$ dimensional Hilbert space and can be specified by a set of $2d-2=12$ real numbers, and full control therefore requires at least that many free parameters in the control Hamiltonian. To ensure sufficient flexibility we specify the control waveform $\phi(t)$ by its values $\{ \phi_i\}$ at $N=30$ discrete time steps.

\begin{figure}[t!]
\begin{center}
\includegraphics[width=8.75cm,clip]{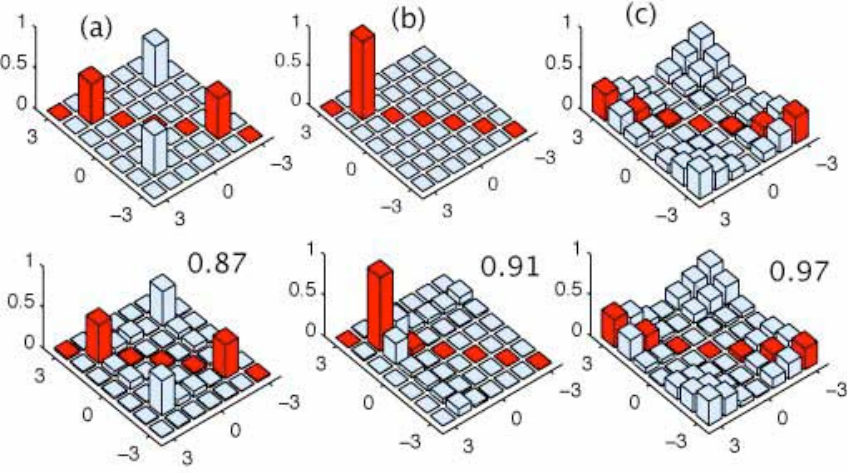}
\caption[Examples of target and time-evolved density matrices]{Examples of target and measured density matrices (absolute values). The target states are (a) $(\ket{ m_z=2 } + \ket{ m_z=-2 })/\sqrt{2}$ , (b)$\ket{ m_x=2}$, (c) $\mathbf{\Sigma}_{y}m_y \ket{ m_y}$. The experimental yield is indicated for each case.}\label{fig:estimates}
\end{center}
\end{figure}

An example of an optimized control waveform is shown in Fig.~\ref{fig:setup}(b), along with Wigner function representations of the spin ÒwavepacketÓ \cite{agarwal81} (see appendix \ref{appen:wigner}) at a few steps during the transformation as calculated using the complete master equation. Note that the nonlinear evolution initially produces a squeezing ellipsoid which later wraps around the sphere so that interference effects can be manipulated to create the desired state. The end product is very close to the target state shown in Fig.~\ref{fig:setup}(c). According to our model this and a wide variety of other control waveforms all produce yields near 0.95. Taking into account imperfect optical pumping in our experiment (the initial population in $|\psi_0\rangle$ is $\sim0.96$) reduces the expected yields to around 0.90.

\begin{figure}[t!]
\begin{center}
\includegraphics[width=8.75cm,clip]{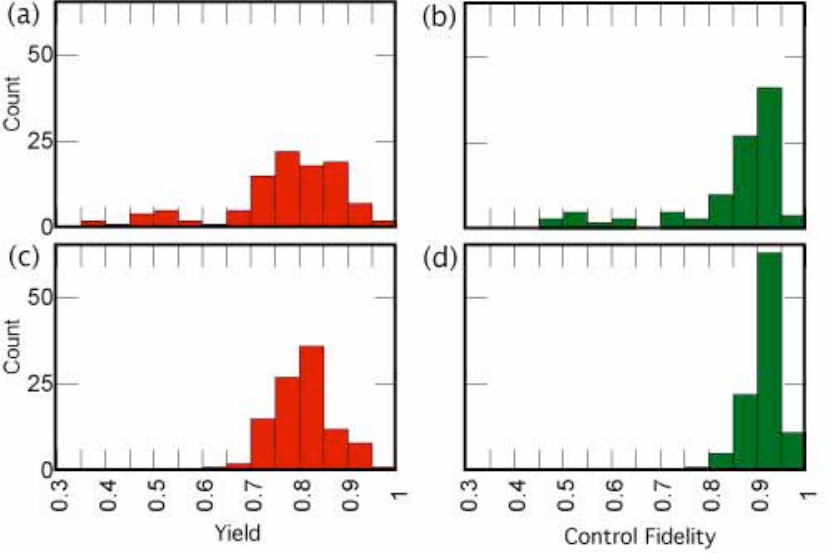}
\caption[Yields and fidelities of experimentally reconstructed states]{\label{fig:histogram} Histograms of (a) yields and (b) fidelities of measured vs. predicted states. (c) \& (d) Yields and fidelities when each measured state is geometrically rotated to optimize overlap with the predicted state.}
\end{center}
\end{figure}

We have generated and tested a sample of control waveforms designed to produce 21 different pure spin states. Fig.~\ref{fig:estimates} shows three examples of target and measured density matrices, with yields falling in the range 0.87-0.97. A more complete statistics of yields for over a hundred experimental realizations of control is compiled in the form of a histogram in Fig.~\ref{fig:histogram}(a), showing a fairly broad distribution centered on respectable value of 0.8. It is also informative to compare the experimentally measured density matrices $\rho_M^{}$ against the density matrices $\rho_P^{}$ predicted by our model, as quantified by the fidelity $\mathcal{F} = Tr\sqrt {\rho_P^{\phantom{P} 1/2} \rho_M^{} \rho_P^{\phantom{P}1/2}}$. Fig.~\ref{fig:histogram}(b) shows a histogram of fidelities for our data set. Note that both yield and fidelity can be affected by control errors (the real state is different from $\rho_P^{}$) as well as state estimation errors (the real state is different from $\rho_M^{}$), and that there is no way to distinguish between these possibilities. Numerical modeling shows that small background magnetic fields or miscalibration of the control fields will lead to apparent geometric rotations of the final state, but such errors are too small in our experiment to significantly affect the outcome. The obvious outliers in the yield and fidelity distributions are associated with two specific control waveforms, and closer examination shows that the estimated states are rotated relative to the predicted states. The axis of rotation corresponds the direction of the magnetic field at the transition between the control and state estimation phases, which suggests a problem with the way the corresponding control waveforms were joined together. We can numerically rotate a given $\rho_M^{}$ to maximize its fidelity relative to $\rho_P^{}$ and obtain new values for yield and fidelity. Carrying out this procedure for all data points takes care of the outliers without otherwise changing the yield distribution significantly, as shown in Fig.~\ref{fig:histogram}(c). This distribution can reasonably be interpreted as a measure of our ability to control the spins in a well designed experiment. The fidelity distribution (Fig.~\ref{fig:histogram}(d)) remains peaked at ~0.9, which we know from experience to reflect the accuracy of our state estimation algorithm. Finally we note that random errors in state estimation are far more likely to decrease than increase the apparent yield. A simple error model based on Gaussian random displacements in state space indicate that the yields are probably $10\%$ larger on average than indicated by Fig.~\ref{fig:histogram}(d). This puts most yields in the range 0.8-0.9, in good agreement with the $\sim 0.9$ predicted by the model used to design the control waveforms in the first place.

\begin{figure}[t!]
\begin{center}
\includegraphics[width=8.75cm,clip]{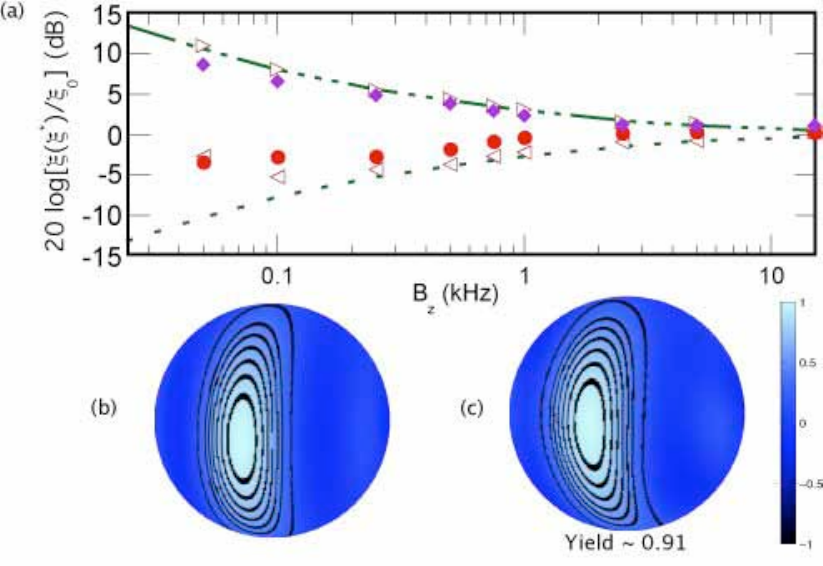}
\caption[Spin squeezing with state preparation]{\label{fig:squeezing} Spin squeezing by adiabatic control. (a)  Normalized squeezing parameter vs. final magnetic field for the squeezed and anti-squeezed components. Dashed lines: perfect squeezing. Open symbols: predictions from our theoretical model. Filled symbols: experimental results. (b) Target and (c) measured Wigner functions corresponding to the smallest observed $\xi$.}
\end{center}
\end{figure}

To further explore quantum control in our system we have studied the generation of spin squeezing both by optimal control as outlined above and by the adiabatic scheme described in \cite{molmersorensen}. The latter begins with an initial state, $|\psi_0\rangle = |F=3,m_y=-3\rangle$, which has equal uncertainties for the components $\Delta F_x$ and $\Delta F_z$ and is often referred to as a spin-coherent state. This state is a good approximation to the ground state of the control Hamiltonian $\hat{H}_C(t)$ when the magnetic field is of the form $\mathbf{B}(t)=B(t)\mathbf{y}$ and $B(t)$ is large. As the field magnitude is slowly reduced the state adiabatically evolves so as to minimize the squeezing parameter $\xi = \Delta F_x/|\langle F_y \rangle|$ of relevance for metrology \cite{wineland}. Fig.~\ref{fig:squeezing}(a) shows the progression of squeezing and anti-squeezing relative to a spin-coherent state with the same $|\langle F_y \rangle|$. Up to $\sim 4$~dB of squeezing is seen in the experiment, in good agreement with the predictions of our model. For the small spin magnitude used here the squeezing is quickly limited by the decrease in $|\langle F_y \rangle|$ as the squeezing ellipse wraps around the sphere. Fig.~\ref{fig:squeezing}(b)-(c) shows Wigner functions for the target and actual state for the smallest $\xi$ achieved in our experiment ($\sim 80\%$ of the coherent state value). We have produced the same spin squeezed states via optimal control, with small but significant reductions in both squeezing and yield. This suggests that gains from reduced decoherence (optimal control is as much as five times faster) is offset by increased sensitivity to control errors.

 \section{Microwave and rf magnetic fields}\label{ch:sp_mwrf}

In \cite{merkel08}, we developed the microwave and rf control system in chapter \ref{chp:hamil_mwrf}, and applied our state preparation technique to the complete 16-dimensional ground state manifold of $^{133}$Cs.  We take a static bias field to produce a Zeeman splitting of $\Omega_0= 1.0$~MHz, sufficient to give excellent resolution of the magnetic sublevels, but well within the linear Zeeman regime.  The rf field power is chosen so that on resonance the rotation rate is characterized by $\Omega_{\rf} = 15$~kHz.  As a generic case, we take one microwave field, resonant on one of the stretched transitions $\ket{F=3,M=\pm -3} \rightarrow \ket{F=4,M= \pm -4}$, where the microwave Rabi frequency is largest, and the system is controllable in a wide variety of scenarios.  The microwave power is chosen to give a Rabi frequency $\Omega_{\mw} = 40$~kHz.  The slew rates constrain the maximum rate of change of amplitude and phase of the control fields.  In the case of the rf-magnetic field, a ``slew time" of $\tau_{\rf} =10 \mu$s fixes the slew rates on the amplitude to 1.5~kHz$/\mu$s and phase to 0.2 $\pi/\mu$s.  In the case of microwaves, faster control is possible, with a slew time of $\tau_{\mw} = 1.0$ $\mu$s, or amplitude and phase slew rates of 40~kHz$/\mu$s and 2.0 $\pi/\mu$s respectively.

\begin{figure}[t!]
\begin{center}
\includegraphics[width=12.5cm,clip]{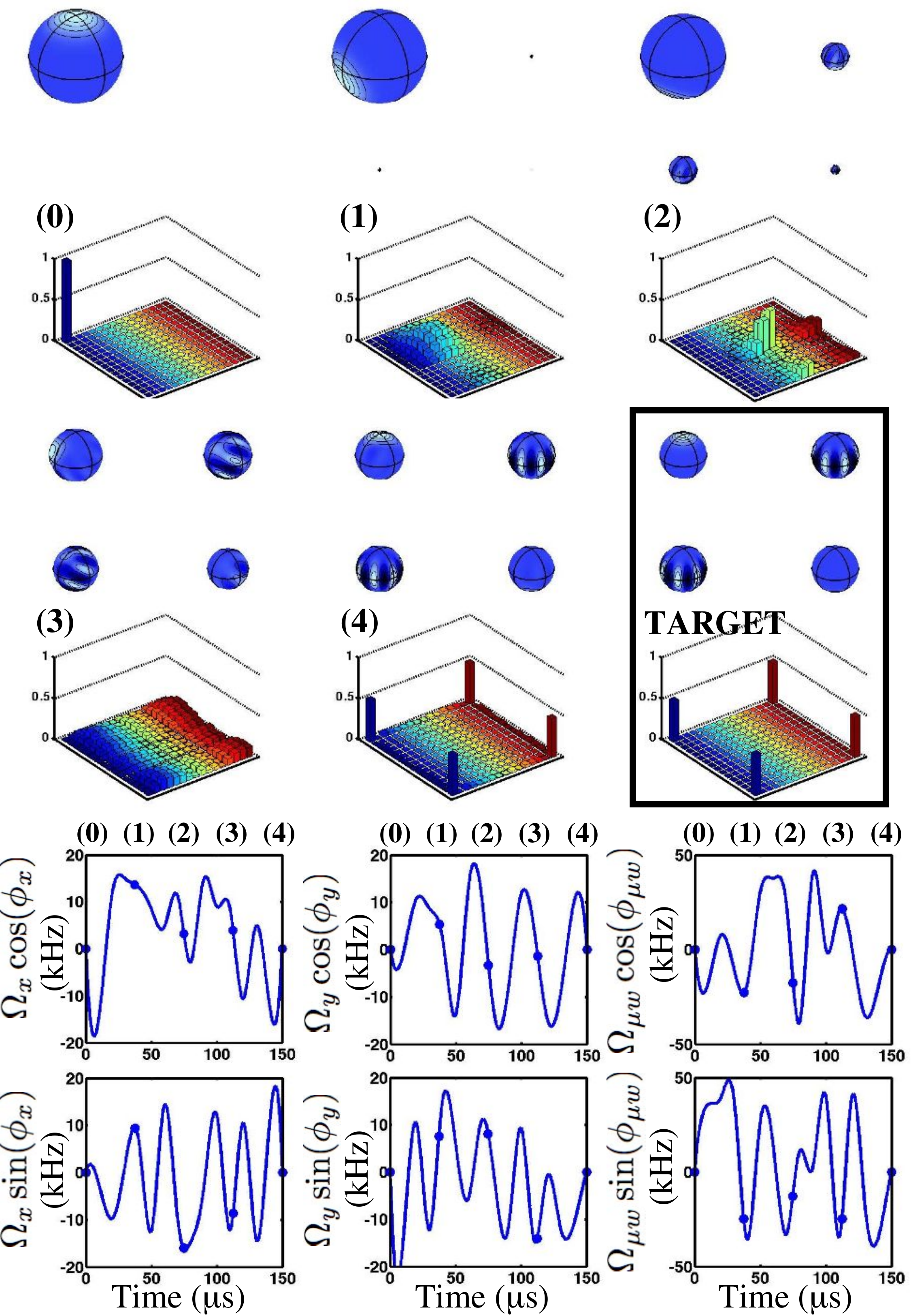}
\caption{$\frac{1}{\sqrt{2}}(\ket{4,4}+\ket{3,-3})$ prepared with fidelity 0.993.}
\label{F:time_series1}
\end{center}
\end{figure}

\begin{figure}[t!]
\begin{center}
\includegraphics[width=12.5cm,clip]{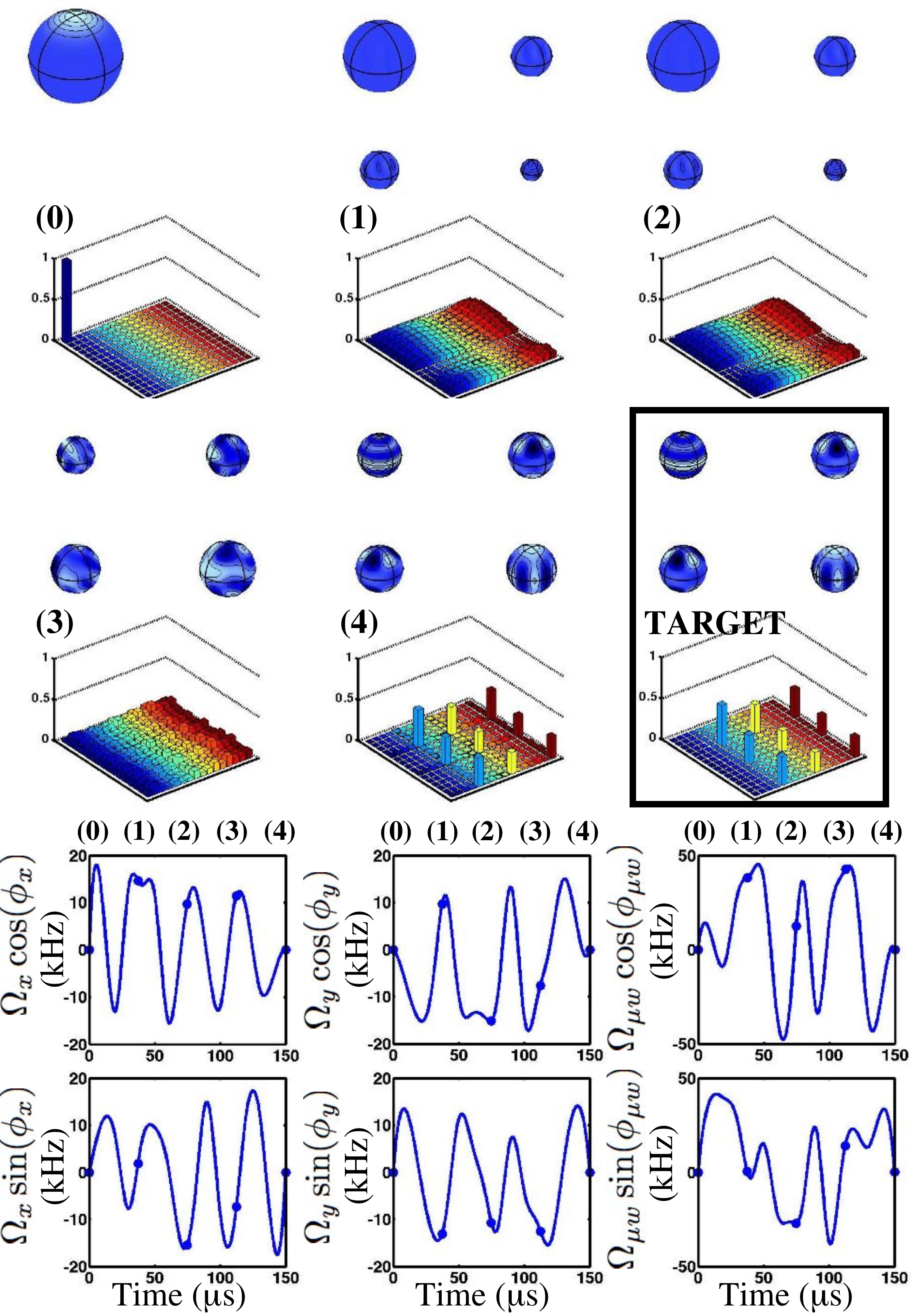}
\caption{$\frac{1}{\sqrt{2}}\ket{4,4}+\frac{1}{2}(\ket{3,3}+\ket{3,-3})$ prepared with fidelity 0.995.}
\label{F:time_series2}
\end{center}
\end{figure}

Two examples of the end product of this optimization are shown in Figs.~(\ref{F:time_series1},\ref{F:time_series2}) for target states $\frac{1}{\sqrt{2}}(\ket{4,4}+\ket{3,-3})$ and $\frac{1}{\sqrt{2}}\ket{4,4}+\frac{1}{2}(\ket{3,3}+\ket{3,-3})$ respectively.  The initial state for these examples is the stretched state $\ket{4,4}$, a state easily reached by optical pumping.  We control the amplitudes and phases of rf coils in both the $x$ and $y$ directions, as well as the amplitude and phase of a resonant microwave that couples the states $\ket{4,-4}$ and $\ket{3,-3}$.  In Figs.~(\ref{F:time_series1},\ref{F:time_series2}) we show the Cartesian components of the three control fields ($\Omega \cos \phi$ and $\Omega \sin \phi$) over the entire state preparation time of 150$\mu$s.   The figures show snapshots of the evolved state at five different times, identified as times (0)-(4).  Two different representations of the state are shown: bar charts of the absolute values of the density matrix elements, and a generalized spherical Wigner function.  The spheres on the diagonal represent the Wigner functions in the irreducible subspaces $F^{\pm}$ and the off-diagonal spheres represent the coherences between the manifolds.  For details see Appendix \ref{appen:wigner}.  The fidelities of preparation in both cases are greater than 99\%.  With a state preparations time of 150$\mu$s moderate searches yield high-fidelity waveforms.  More intensive optimizations can yield faster control waveforms.  

Our gradient search algorithm leads to waveforms that cause the system to undergo quite complex dynamics, as evidenced by the intermediate states seen in the course of the evolutions,  Figs.~(\ref{F:time_series1},\ref{F:time_series2}).  One may wonder whether there are simpler choices, since given a fixed initial state, there are many different waveforms that lead to same target state.  While our method does lead to waveforms that are hard to intuitively understand, some recent studies \cite{schirmer08} suggest that the waveforms derived from gradient searches may be more robust than those that come from more geometric algorithms.

We discussed the mathematical conditions necessary for our Hamiltonian dynamics to be controllable.  These conditions, while useful for ruling out large classes of Hamiltonians as unsuitable for our purposes, tell us nothing about the relative performance of different control scenarios.  Our figure of merit is the time after which we can be reasonably sure that our optimization will find a high fidelity waveform for any target state.  To determine this time for a given control protocol, we run our optimization up to a given final control time over a large collection of randomly chosen states and determine the average fidelity.  In this section we examine these results and discuss some of the tradeoffs and bottlenecks that might be encountered in the lab.

There are many parameters in this system that we can manipulate, including the number of independently controlled rf polarizations, the number of resonant microwave frequencies, the types of controls (amplitude vs. phase), detuning, slew rates, and the strengths of the different fields.  Based on some of our previous experiments we set as a baseline: one microwave frequency, two orthogonal rf polarizations,  rf power giving $\Omega_{\rf}$ = 15~kHz, a microwave Rabi frequency of $\Omega_{\mw}$ = 40~kHz, a rf slew time of 10 $\mu$s, and a microwave slew time of 1.0 $\mu$s.  While we could independently vary all these parameters, this would be an unwieldy computation.  Here we fix some of the parameters that are unlikely to differ in the future experiments we are considering.  In particular, we fix the rf slew time to be 10 $\mu$s and consider control with two sets of rf coils.  For simplicity we also consider all fields to be resonant, and the microwaves to couple the stretched states.       

\begin{figure}[t!]
\begin{center}
\includegraphics[width=15cm,clip]{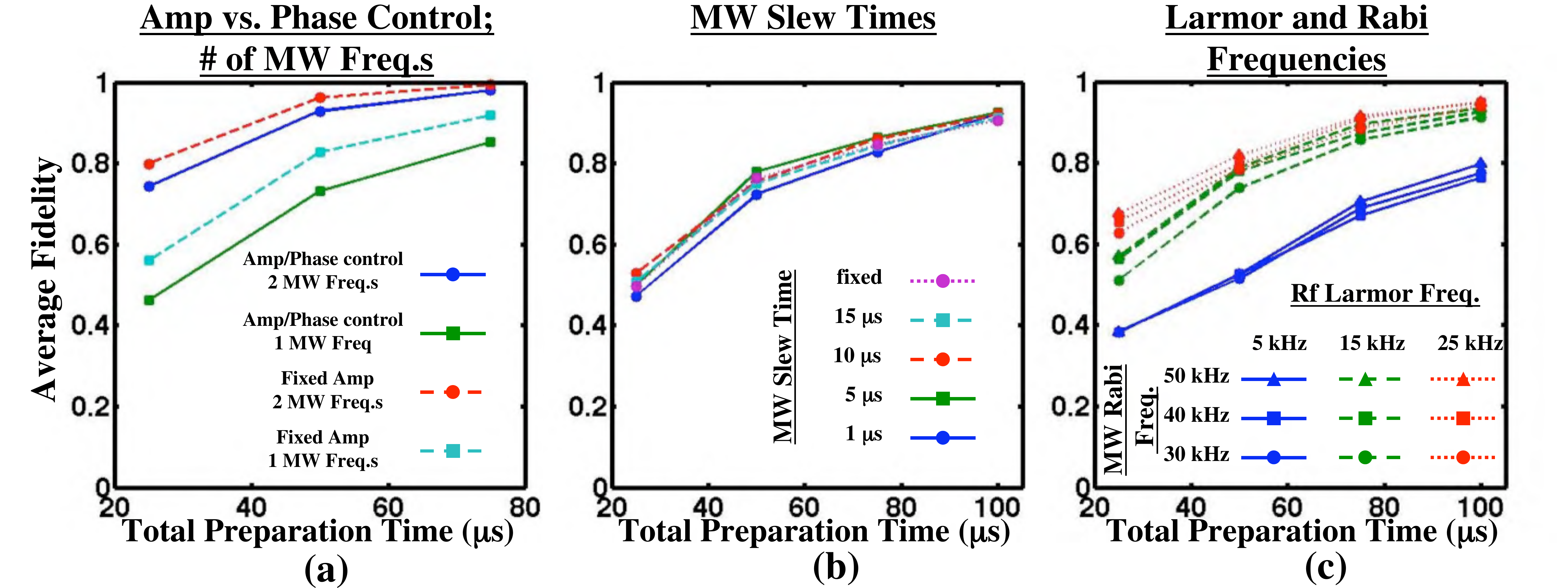}
\caption[Comparison of average fidelities for different control configuration in microwave and rf magnetic field system]{ Plots of the average fidelity of state preparation for different control configurations and total preparation times.  Each point represents the fidelity averaged over a set of 10 states randomly chosen from the Harr measure.  For each state and configuration, the gradient search was performed with 20 random seeds and we chose the protocol that generated the highest fidelity.}
\label{F:opt_stats}
\end{center}
\end{figure} 

Statistics were collected by running the state preparation algorithm for 10 different random states found by sampling using the Harr measure on $SU(16)$ \cite{Pozniak98}.  In all cases the initial state was the $\ket{4,4}$ state.  For each combination of total time, target state, and system configuration, we run the optimization 20 times starting from different random seeds of the vector that defines the control waveform.  Out of this set of 20, we choose the highest fidelity preparation.  The fidelities from the 10 random states are averaged to produce the data points shown in Fig.~\ref{F:opt_stats}.  In principle, more iterations would yield higher fidelity waveforms, but it is useful to understand which types of high-fidelity controls can be found after only modest searches.  

In Fig.~\ref{F:opt_stats}a, we study the effect of varying the characteristics of the microwave field. We compare the performance of one vs. two resonant microwave frequencies on one or both of the stretched transitions, $\ket{3,3} \rightarrow \ket{4,4}$ and  $\ket{3,-3} \rightarrow \ket{4,-4}$.  In addition, we examine the effect of removing control of the microwave amplitude (a scenario that still allows for full controllability of the system, as discussed in Ch.~\ref{chp:hamil_mwrf}).  As expected, since the microwave Rabi frequency is larger than the rf Larmor frequency, increasing the number of microwave fields has a large effect.  On the other hand, it was surprising that fixing the microwave amplitude, thereby substantially decreasing the number of control parameters, yielded higher fidelity waveforms.  We suspect that while there most likely exist higher fidelity waveforms with control of both amplitude and phase, increasing the number of microwave control parameters rapidly increases the dimension of the search space, requiring many more iterations of our algorithm to find a superior waveforms, on average. This suspicion is reinforced by Fig.~\ref{F:opt_stats}b, where we consider the effect of microwave slew time.  With our baseline parameters, it would appear that increasing the microwave slew time doesn't really limit the optimized control performance.  In fact, the smallest slew time we considered, 1.0 $\mu$s, performed slightly worse than the other slew times, including the case where the microwave amplitudes are fixed.  As a reminder, the slew time determines the information content of our waveforms, and thus the number of optimization variables.  Again, we see that for the modest searches we are performing, decreasing the dimension of the search space counterbalances the loss of control.

In Fig.~\ref{F:opt_stats}c, we study the effect of the power in the rf and microwave fields.  For these simulations we fixed the amplitudes of the fields and solely control their phases.  We find that varying the microwave power around our baseline makes little difference.  The rf power is slightly more important, but increasing the Larmor frequency above the baseline has a fairly  small effect.  These results indicate that the slew rate and bandwidth constraints we have imposed on the rf magnetic fields are the bottleneck for controlling the system, and limit the ability to more rapidly control the system through increases in power.  It would appear that the microwave parameters we employ as our baseline are also well above the limits imposed by this bottleneck and we can safely reduce the microwave power and slew rates without sacrificing performance. The rf Larmor frequency we employ is commensurate with the slew rate constraint.           

By optimizing many state preparations for a variety of control configurations we find state preparation protocols with this system that take between $50-150 \mu$s.  We can compare this to the types of control waveforms that were implemented in our previous work that employed a nonlinear AC-Stark shift to achieve controllability~\cite{chaudhury07}.  The waveforms we find here are about an order of magnitude faster, control a Hilbert space that is double the dimension, and have negligible decoherence as compared to the intrinsic decoherence that arises from spontaneous emission.

\chapter{Efficiently Constructing Arbitrary Unitary Maps on Qudits}\label{ch:unitary}

In chapter \ref{sec:topology}, I discussed the difference between the problems of unitary construction and state preparation.  The stochastic search, state preparation techniques used in the last chapter are ill-suited to the problem of designing general unitary maps.  Most known techniques for constructing arbitrary unitary maps fall under the category of geometric constructions (Ch.~\ref{sec:generating}) which, while powerful, lack broad applicability.  In \cite{merkel09}, my coauthors and I developed a new type of unitary construction protocol which is a hybrid of stochastic/geometric construction, similar to the protocol in \cite{luy05}.  Essentially, we leverage off of our ability to efficiently generate state preparations, and then splice state preparations together in a geometric way to create a general unitary map.  The types of Hamiltonian dynamics that this construction applies to have some restrictions beyond controllability.  These restrictions are, however, much less stringent than those in most geometric techniques.    

In chapter \ref{sec:construct} I present our hybrid protocol for constructing general unitary maps by combining efficient numerical searches with a deterministic algorithm.   In addition to unitary maps on the full Hilbert space, this scheme allows us to construct maps on a subspace with a complexity that scales as the dimension of that space.  In chapter \ref{sec:examples},  our unitary matrix construction is applied to control the large manifold of magnetic sublevels in the ground electric states of an alkali atom (e.g. $^{133}$Cs) \cite{merkel08}.   We show how to construct a set of unitary matrices on $SU(d)$ that are often considered as qudit logic gates in a fault-tolerant protocol.  In addition, we apply our construction for subspace mapping to encode logical qubits in our qudit, and simulate an error correcting code that protects against magnetic field fluctuations.

\section{Unitary construction}\label{sec:construct}

In this section we define an efficient protocol for constructing arbitrary unitary maps based on state preparation.  Any unitary matrix has an eigen-decomposition, 
\be
U = \sum_j e^{-i \lambda_j} \ket{\phi_j}\bra{\phi_j}= \prod_j e^{-i \lambda_j  \ket{\phi_j}\bra{\phi_j}},  
\ee 
where in the second form we expressed $U$ as a product of commuting unitary evolutions by moving the projectors into the exponential.  A general unitary map can be thus be constructed from $d$ propagators of the form $\exp\{-i \lambda_j  \ket{\phi_j}\bra{\phi_j}\}$, one for each eigenvalue/eigenvector pair.   These unitary propagators can now be constructed using state mappings.  We begin by noting that there exists some $V_j \in SU(d)$ that satisfies 
\be
e^{-i \lambda_j  \ket{\phi_j}\bra{\phi_j}} = e^{-i \lambda_j  V_j^{\dagger} \ket{0}\bra{0} V_j} = V_j^{\dagger} e^{-i \lambda_j  \ket{0}\bra{0}} V_j,
\ee 
where $\ket{0}$ is a fixed ``fiducial state".  The sole requirement on $V_j$ is that $|\bra{0} V_j \ket{\phi_j}|^2 = 1$, i.e., it must be a mapping from $\ket{\phi_j}$ to $\ket{0}$.  Therefore, we can create the unitary propagator $\exp\{-i \lambda_j  \ket{\phi_j}\bra{\phi_j}\}$ by using  a state preparation to map the eigenvector of $U$, $\ket{\phi_j}$, onto the fiducial state $\ket{0}$, applying the correct phase shift, and finally mapping the fiducial state back to the eigenvector with the time-reversed state preparation.  A general unitary map is thus constructed via the sequence,
\be
U=V_d^{\dagger} e^{-i \lambda_d  \ket{0}\bra{0}} V_d \ldots V_{2}^{\dagger} e^{-i \lambda_{2}  \ket{0}\bra{0}} V_{2}V_1^{\dagger} e^{-i \lambda_1  \ket{0}\bra{0}} V_1  .
\ee  
Each of the propagators $V_j$ is specified by a control waveform that generates a desired state mapping.  One can efficiently find such control fields based on a numerical search that employs a simple gradient search algorithm, as described above.  To generate an arbitrary element of $SU(d)$, we require at most $d$ such searches. Moreover, the full construction consists of $2d$ state preparations interleaved with $d$ applications of the phase Hamiltonian, leading to an evolution that is only of order $d$ times longer than a state mapping evolution.              

This construction places only two requirements on the Hamiltonian in addition to controllability.  Firstly, the dynamics must be reversible such that if we can generate the unitary evolution $V_j$, we can trivially generate the unitary $V_j^{\dagger}$ by time-reversing the control fields.  Note that this is not the same as finding a state preparation that goes in the opposite direction, $\ket{0} \rightarrow \ket{\phi_j}$; there are many unitary propagators that map $\ket{0}\rightarrow \ket{\phi_j}$, so it is unlikely to find the unique operator $V_j^{\dagger}$ from a stochastic search.  Generally, we can easily time reverse our controls if the Hamiltonian dynamics have no drift term.  In some cases with a non-zero drift term it is still possible to time reverse controls, however, in this case we need to be able to find a rotating frame that removes the drift term while leaving the remaining Hamiltonian terms reversible.  Secondly, we require access to a control Hamiltonian that applies an arbitrary phase to one particular fiducial state $\ket{0}$ relative to all of the remaining states in the Hilbert space, $\exp\{-i \lambda_j  \ket{0}\bra{0}\}$.  This latter requirement is the most restrictive, but can be implemented in a wide variety of systems.  An example is discussed in \ref{sec:examples}.

\subsection{Subspace maps}\label{sec:subspace}
We have so far considered two kinds of maps on our $d$-dimensional Hilbert space $\mathcal{H}$: $d \times d$ unitary matrices and state-to-state maps.  The former corresponds to a map $U : \mathcal{H} \rightarrow \mathcal{H}$, while the latter specifies a map on a one-dimensional space.  Intermediate cases are also important.  In particular, we are often interested in unitary maps that take subspace $\mathcal{A}$ of arbitrary dimension $n$ to subspace $\mathcal{B}$,  according to $T: \mathcal{A} \rightarrow \mathcal{B}$.  Examples include the encoding of a logical qubit into a large dimensional Hilbert space $( \mathcal{A} \neq \mathcal{B})$ and a logical gate on encoded quantum information $( \mathcal{A} =\mathcal{B})$. Above we showed that the design of a fully-specified unitary matrix required search for $d$ waveforms that define $d$ state preparations (trivially a state mapping requires one such search).  We show here how unitary maps on subspaces of dimension $n$ can be constructed from exactly $n$ such numerical solutions.

Formally, a unitary map between two subspaces  $ \mathcal{A}$ and $\mathcal{B}$  of dimension $n$ is defined as a map between between their orthonormal bases $\{\ket{a_i}\}$ and $\{\ket{b_i}\}$,
\be
T_n\left(\mathcal{A} \rightarrow \mathcal{B}\right) =  \sum_{i=1}^n \ket{b_i}\bra{a_i} \oplus U_{\perp},
\label{eq:desiredmap}
\ee
where $U_{\perp}$ is an arbitrary map that preserves unitarity on the orthogonal complement $ \mathcal{A}_{\perp}$ whose dimension is $d-n$.  State preparation is the case $n=1$; a full unitary matrix is specified when $n=d$.  Clearly for $n \neq d$ the map is not unique, with implications for the control landscape and the simplicity of numerical searches described above.  As a first na\"{i}ve construction of $T(\mathcal{A} \rightarrow \mathcal{B})$, one might consider a sequence of one-dimensional state mappings, 
\be
T_n\left(\mathcal{A} \rightarrow \mathcal{B}\right) \stackrel{?}{=} \prod_{i=1}^n T_1\left(\ket{a_i} \rightarrow \ket{b_i}\right). 
\label{eq:bad_map}
\ee
This does not, however, yield the desired subspace map because each state mapping acts also on the orthogonal complement, so, e.g. $\ket{b_1}$ is affected by $T_1\left(\ket{a_2} \rightarrow \ket{b_2}\right)$ and subsequent maps will move formerly correct basis vectors to arbitrary vectors in the orthogonal component.  We can resolve this problem by instead constructing subspace maps as a series all well-chosen rotations that maintain proper orthogonality conditions.

To construct the necessary unitary operators, we make use of the tools described above: arbitrary state mapping based on an efficient waveform optimization and phase imprinting on a fiducial state.  With these, we define the unitary map between unit vectors $\ket{a}$ and $\ket{b}$,
\be
S\left( \ket{a} ,\ket{b} \right) \equiv e^{-i \pi \ket{\phi}\bra{\phi}} = \hat{I} - 2 \ket{\phi}\bra{\phi}.
\ee
Here $\ket{\phi}=N(\ket{a} - \ket{b})$, where we have chosen the phases such that $\braket{b}{a}$ is real and positive, and $1/N^{2} \equiv 2\left(1-\braket{b}{a}\right)$ is the normalization.  This unitary operator has the following interpretation.  In the two-dimensional subspace spanned by $\ket{a}$ and $\ket{b}$, $S$ is a $\pi$-rotation that maps $S\ket{a} = \ket{b}$.  In contrast to the state preparation map, Eq.~(\ref{eq:desiredmap}) with $n=1$, this map acts as the {\em identity} on the orthogonal complement to the space.  This property is critical for the desired application.

With these 2D primitives in hand, we can construct the subspace map according to the prescription,
\be
T_n(\mathcal{A} \rightarrow \mathcal{B}) =s_n\ldots s_2 s_1,
\ee
where $s_k \equiv S\left( \ket{\tilde{a}_k} ,\ket{b_k} \right)$ and
\be
\ket{\tilde{a}_j}  \equiv s_{j-1}\ldots s_2 s_1 \ket{a_j}. 
\ee
This sequence does the job because each successive rotation leaves previously mapped basis vectors unchanged.  To see this, we must show that at step $j$, the basis vectors $\{ \ket{b_1},\ket{b_2}, \ldots, \ket{b_{j-1}} \}$ are unchanged by $s_j$.  This will be true when this set is orthogonal to the vectors $\ket{\tilde{a}_j}$ and $\ket{b_j}$.  Orthogonality to $\ket{b_j}$ is trivial since the basis vectors of $\mathcal{B}$ are orthonormal.  We must thus prove, $\braket{\tilde{a}_j}{b_k}=0$,  $\forall j>k$.  We can do this by induction.  For an arbitrary $k$, assume the conjecture is true for all $j$ such that $j_0 \geq j>k$, and thus $s_j \ket{b_k} = \ket{b_k}$ up to $j=j_0$.  This implies that $\braket{\tilde{a}_{j_0+1}}{b_k} = 0$ since,
\bea
\braket{\tilde{a}_{j_0+1}}{b_k} &=& \bra{a_{j_0+1}} s^{\dagger}_1 \ldots s^{\dagger}_k s^{\dagger}_{k+1} \ldots s^{\dagger}_{j_0} \ket{b_k} \nonumber\\ 
&=& \bra{a_{j_0+1}} s^{\dagger}_1 \ldots s^{\dagger}_k  \ket{b_k} \nonumber\\ 
&=& \braket{a_{j_0+1}}{a_k}=0.
\eea
To complete our proof by induction, we must show that for any $k$, the conjecture is true for $j=k+1$.  This follows since, 
\bea
\braket{\tilde{a}_{k+1}}{b_k} &=& \bra{a_k+1}s^{\dagger}_1 s^{\dagger}_2 \ldots s^{\dagger}_k \ket{b_k} \nonumber\\
&=& \braket{a_{k+1}}{a_k} =0.
\eea
With this protocol we can construct unitary maps on a subspace of dimension $n$ with optimized waveforms that corresponded to exactly $n$  prescribed state preparations.  In the following section we apply these tools to qudit manipulations in atomic systems.

\section{Applications to the microwave rf system}\label{sec:examples}

In this section, we apply our results to the control of the ground-electronic manifold of magnetic sublevels in alkali atoms discussed in chapter \ref{chp:hamil_mwrf}.  In addition to an efficient method for designing and implementing state-to-state mappings, our protocol places certain requirements on the available control tools. Firstly, the system dynamics must be reversible so that we can trivially invert a state mapping.  This is easily achieved through phase control.  Secondly, we require phase imprinting on a single fiducial state.  While this cannot be accomplished using solely microwave and rf-control, by introducing an excited electronic manifold, an off-resonant laser-induced light-shift can achieve this goal.  We restrict our system to one spin manifold (here the $F=3$, but in principle either will do) and a single state from $F=4$ manifold, e.g. $\ket{F=4,m=4}$, which acts as the fiducial state.  By detuning far compared to the excited state line width of  ~5~MHz, but close compared to the ground-state hyperfine splitting of ~10~GHz, we imprint a light shift solely on the $\ket{F=4,m=4}$ state with negligible decoherence. Using rf-magnetic fields to perform rotations in the $F=3$ manifold, and microwaves to couple to the fiducial state, we obtain  controllable and reversible dynamics.  Note that we may include the fiducial state in our Hilbert space, for a total of 8 sublevels, or treat it solely as an auxiliary state and restrict the Hilbert space to the 7-dimensional $F=3$ manifold.

\begin{figure}[t]
\begin{center}
\includegraphics[width=9cm,clip]{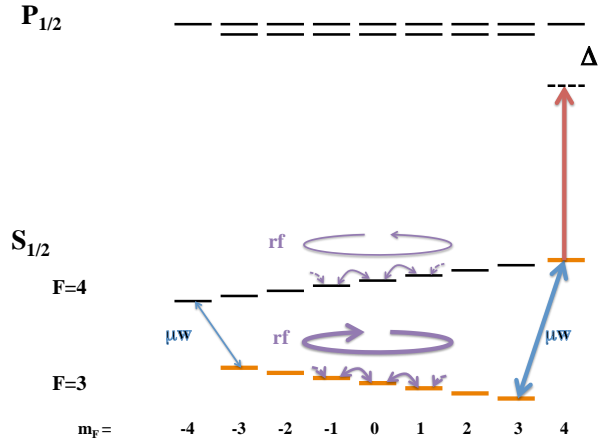}
\caption[Control system for unitary construction in $^{133}$Cs]{The hyperfine structure of $^{133}$Cs in  the 6S$_{1/2}$ ground state.  Microwaves (blue) and rf magnetic fields (purple) provide controllable dynamics on the 16-dimensional Hilbert space.  A detuned laser light shift (red) can be used to create a relative phase between the $F=4$ and $F=3$ manifolds. By considering controls on the subspace of the orange states we recover a system that satisfies the criteria proposed in chapter\ref{sec:construct}. }
\label{F:levels}
\end{center}
\end{figure}

\subsection{Constructing qudit unitary gates}
The standard paradigm for quantum information employs two-level systems -- qubits -- in order to implement binary quantum-logic based on $SU(2)$ transformations.  Extensions beyond binary encodings in $d>2$ system -- qudits -- based of $SU(d)$ transformations have also been studied and may yield advantages in some circumstances \cite{gottesman99,brennen05, grace06}.  Of particular importance for fault-tolerant operation is implementation of these transformations through a finite set of ``universal gates".  Our goal here is to show how important members of the universal gate set can be implemented using our protocol.

In choosing a universal gate set appropriate for error correction, it is natural to consider generalizations of the Pauli matrices $X$ and $Z$ which generate $SU(2)$.  The generalized discrete Pauli operators for $SU(d)$ are defined
\bea
X \ket{j} &=& \ket{j \oplus 1}\nonumber\\
Z \ket{j} &=& \omega^j \ket{j}.
\eea
Here $\oplus$ refers to addition modulo $d$ and $\omega$ is the primitive $d$th root of unity, $\omega =\exp \{i 2 \pi /d\}$.  By considering the commutation relation of $X$ and $Z$, the remaining generalized Pauli operators have the form $\omega^l X^j Z^k$, defining the elements of Pauli group for one qudit (up to a phase).  This group is a discrete (finite dimensional) generalization of the Weyl-Heisenberg group of displacements on phase space.  

Another important group of unitary matrices in the theory of quantum error correction is the single qudit Clifford group, given its relationship to stabilizer codes \cite{gottesman99}.  These group elements map the Pauli group back to itself under conjugation.  Expressed in terms of their conjugacy action on $X$ and $Z$, the generators of the Clifford group for single qudits are
\bea
HXH^{\dagger} = Z, &\quad &HZH^{\dagger} = X^{-1} \\ 
SXS^{\dagger} = XZ, &\quad &SZS^{\dagger} = Z \\ 
G_a X G_a^{\dagger} = X^a, &\quad &G_a ZG_a^{\dagger} = Z^{a^{-1}}\nonumber\\ &&\textrm{when gcd}(a, d)=1
\eea
$H$ and $S$ are direct generalization of the Hadamard and phase-gates familiar for qubits \cite{nielsen2000}.  The $d$-dimensional $H$  is the discrete Fourier transform
\be
H\ket{j} = \frac{1}{\sqrt{d}}\sum_k \omega^{jk}\ket{k}
\ee           
and $S$ is a nonlinear phase gate
\bea
S\ket{j} = \omega^{j(j-1)/2}\ket{j} \quad j~\textrm{odd},\\
S\ket{j} = \omega^{j^2/2}\ket{j} \quad j~\textrm{even}.
\eea
The operator $G_a$ is a scalar multiplication operator with no analog in the standard Clifford group on qubits, defined by
\be
G_a \ket{j} = \ket{a j},
\ee
where the multiplication is modulo $d$.  The only such multiplication operator for 2-level systems is the identity operator.

While both the generalized Pauli and Clifford groups have utility in quantum computing, it is clear from their descriptions that unlike their qubit $SU(2)$ counterparts, these unitary matrices do not arise naturally as the time evolution operators governed by typical Hamiltonians.  This fact is not relevant to our unitary construction, which requires only knowledge of the operators' eigenvectors and eigenvalues.  Using the time-dependent Hamiltonian dynamics with couplings illustrated in Fig.\ref{F:levels} we have engineered control fields to create the generators of both the Pauli and Clifford groups acting on the 7-dimension $F=3$ hyperfine manifold.  The duration of  waveforms is approximately 1.5 ms, which is significantly shorter than the decoherence time of the system.  In principle, the durations of these waveforms could be decreased by an order of magnitude or more by using more powerful control fields.  Our objective function for creating a desired unitary $W$ is the trace distance $J[W] = Tr\left(W^{\dagger} U \right)$, where $U$ is the unitary matrix generated by our control waveforms.  Based on our protocol, employing state mappings that have fidelities of ~0.99, our construction yields unitary maps that reach their targets with fidelities of $J[Z] = 0.9866$, $J[X] = 0.9872$, $J[H] = 0.9854$, $J[S] = 0.9892$ and $J[G_3] = 0.9801$.

\begin{figure}[t!]
\begin{center}
\begin{tabular}{cc}
\includegraphics[width=7cm,clip]{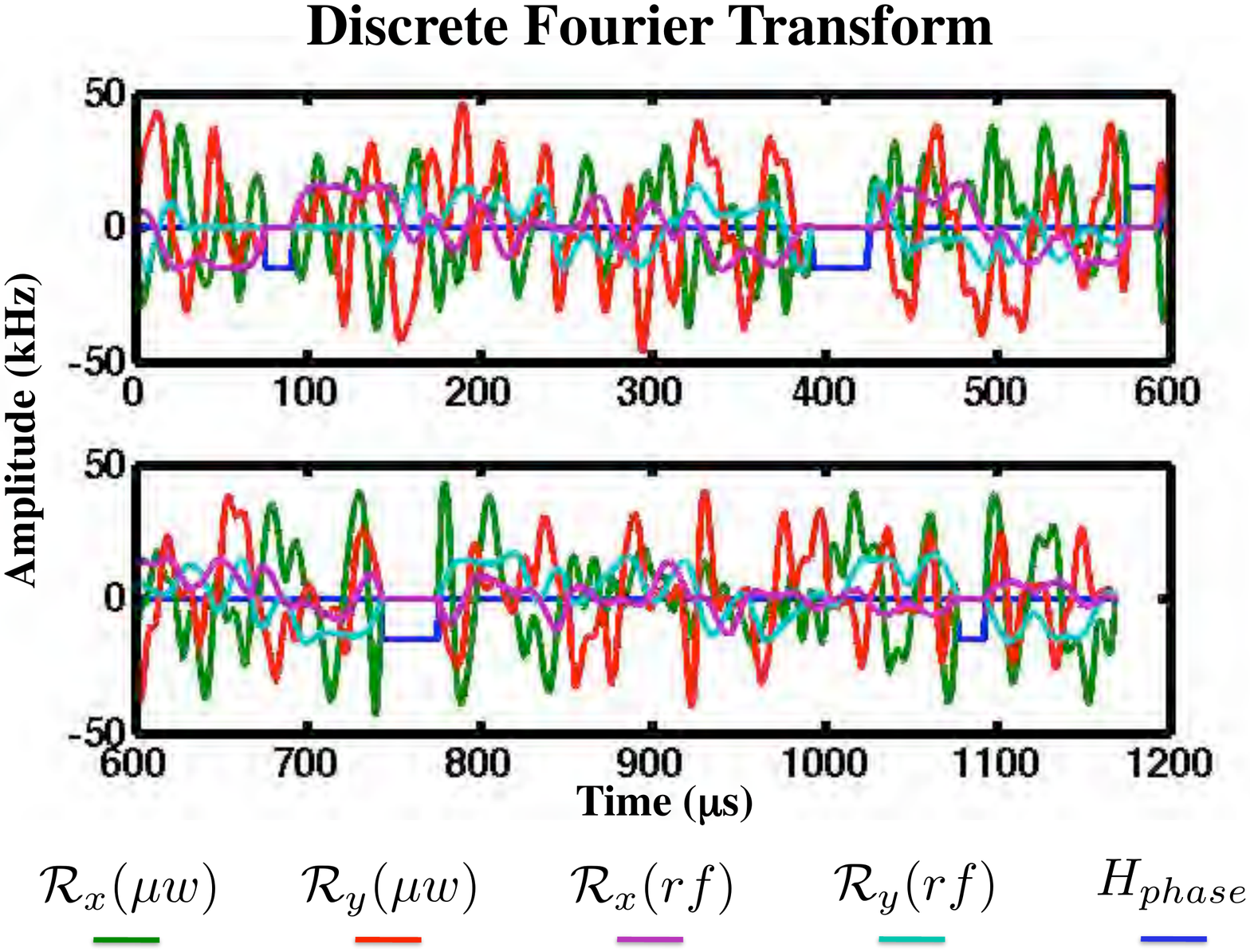}&
\includegraphics[width=7cm,clip]{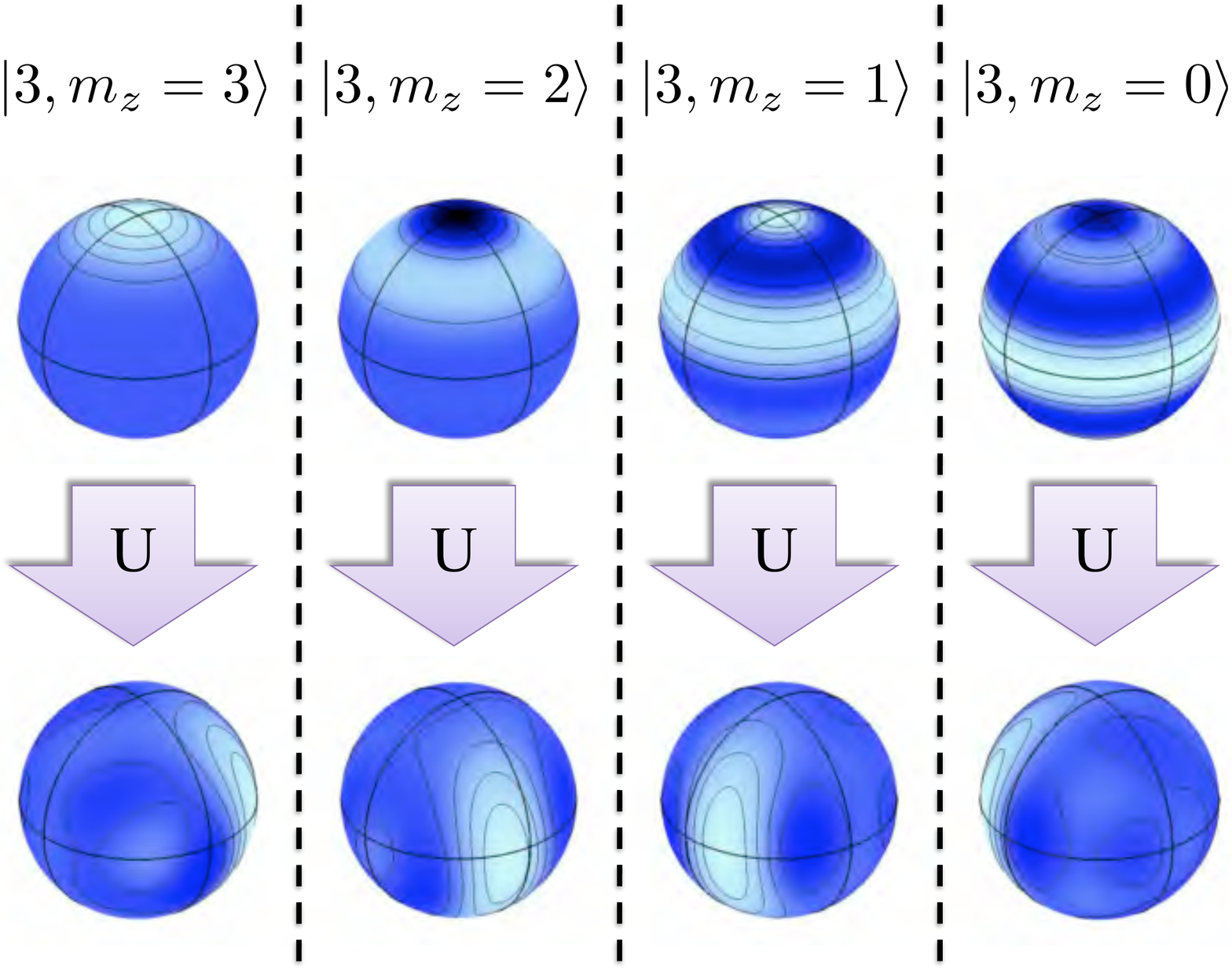}
\end{tabular}
\caption[Implementation of the discrete Fourier transform unitary operator]{Optimized control fields for implementing the 7-dimensional Fourier transform on the $F=3$ hyperfine manifold in $^{133}$Cs.  The duration of the pulse is less than 1.2 ms and yields a unitary map that has an overlap of 0.9854 with the desired target.  As an example, we show the action of the resulting unitary on the $Z$-eigenstates of angular momentum.  The conjugate variable of $F_z$ is the azimuthal angle $\phi$.  If we Fourier transform a $Z$-eigenstate, a state with a well defined value of $F_z$, we obtain a state that has a well defined value of $\phi$, a squeezed state.}
\label{F:wig}
\end{center}
\end{figure}

As an example, in Fig.~\ref{F:wig} we show the control sequence for the discrete Fourier transform.  The unitary map generated by this sequence should act to transform eigenstates of $Z$ into eigenstates of $X$ and vice versa.  We illustrate this through a Wigner function representation on sphere \cite{agarwal81}.  The $Z$ eigenstates are the standard basis of magnetic sublevels, whose Wigner functions are concentrated at discrete latitudes on the sphere, Fig.~\ref{F:wig}a.  Applying the control fields to each of these states yields the conjugate states, with Wigner functions shown in Fig.~\ref{F:wig}b.  These have the expected form.  They are spin squeezed states concentrated at discrete longitudes conjugate to the $Z$ eigenstates. The $Z$ and $X$ eigenstates are analogous to the number and phase eigenstates of the harmonic oscillator in infinite dimensions.

\subsection{Error-correcting a qubit embedded in a qudit}

The ability to generate unitary transformations on two-dimensional subspaces allows us to encode and manipulate a qubit in a higher dimensional Hilbert space in order to protect it from errors.  Such protection can take a passive form through the choice of a decoherence-free subspace \cite{lidar98,bacon00}, or active error correction through an encoding in a logical subspace chosen to allow for syndrome diagnosis and reversal \cite{calderbank96,aharonov97}.  Typically,  error protection schemes involve multiple subsystems (e.g. multiple physical qubits) to provide the logical subspace.  While tensor product Hilbert spaces are generally necessary to correct for all errors under reasonable noise models, for a limited error model, one can protect a qubit by encoding it an a higher dimensional qudit  \cite{gottesman01}. We consider such a protocol as an illustration of our subspace-mapping procedure.

As an example, we consider encoding a qubit in the ground-electronic hyperfine manifold of $^{133}$Cs and protecting it from dephasing due to fluctuations in external magnetic fields.  In the presence of a strong bias in the $z$-direction, the spins are most sensitive to fluctuations along that axis.  For hyperfine qubits, one solution is to choose the bias such that two magnetic sublevels see no Zeeman shift to first order in the field strength (a ``clock transition").  An alternative is to employ an active error correction protocol analogous to the familiar phase-flip code \cite{nielsen2000}.  

\begin{figure*}[t!]
\begin{center}
\begin{minipage}{6.5cm}
\includegraphics[width=6.4cm,clip]{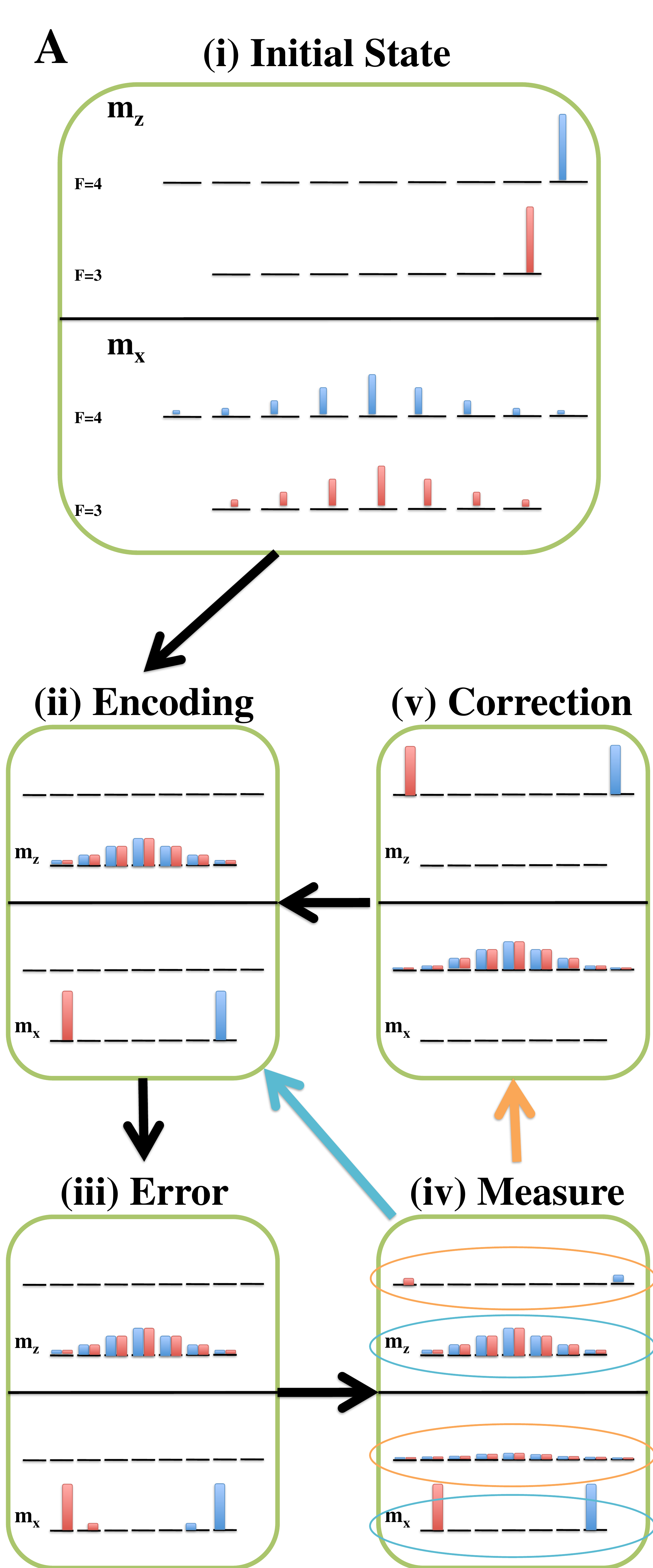}
\end{minipage}
\qquad
\begin{minipage}{7.5cm}
\includegraphics[width=7.4cm,clip]{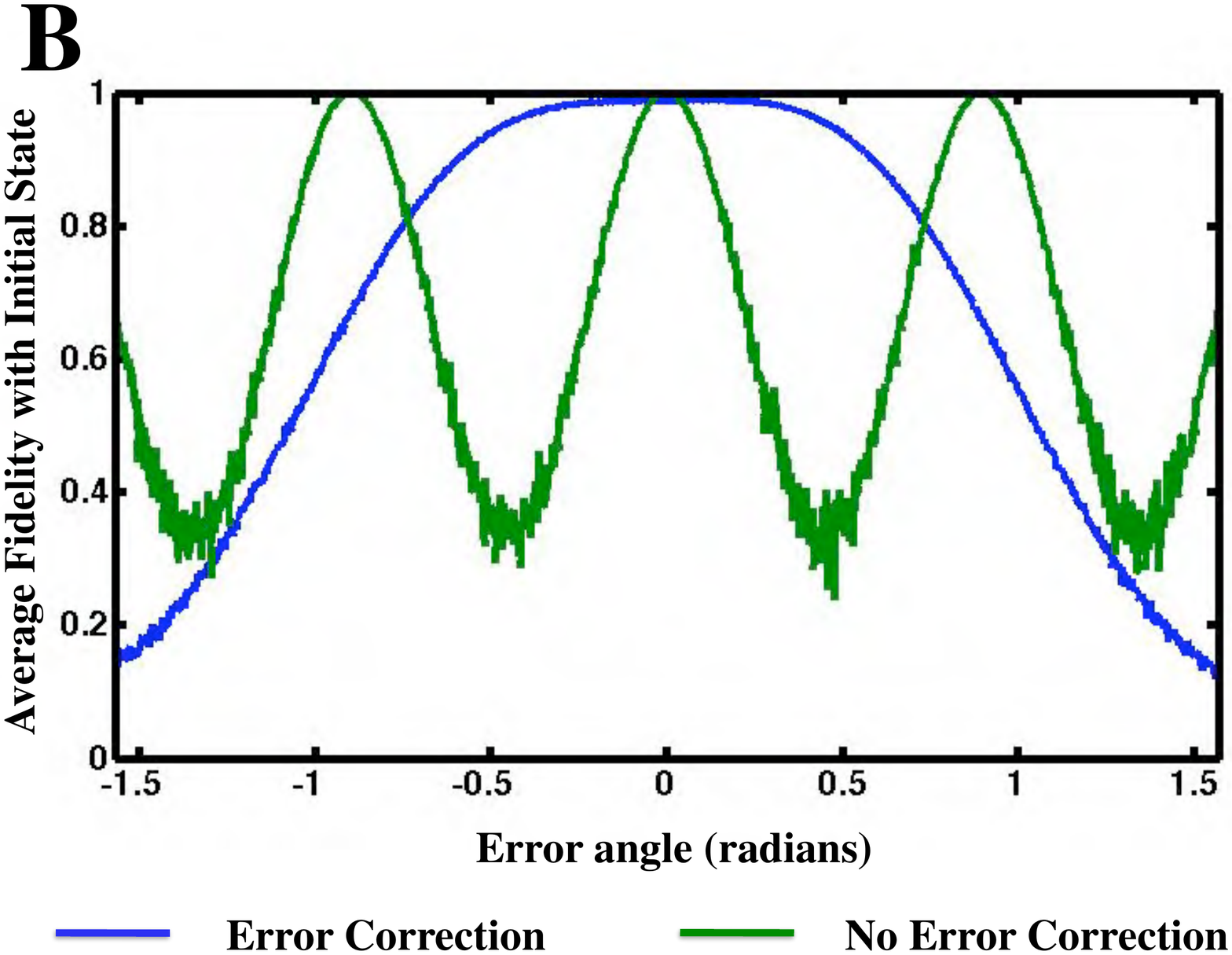}
\caption[A schematic for phase error correction in $^{133}$Cs]{(A)  A schematic of the error correction protocol we have designed using subspace maps.  We track the basis elements of our encoded subspace, here $\ket{0}$ is red and $\ket{1}$ is blue,  via their populations in the $x$ and $z$ bases.  The different configurations are explained in the text.  In (B) we examine the performance of the error correction.  On the $x$-axis we have the angle of rotation in the $z$-direction due to the magnetic field error.  On the $y$-axis is the fidelity between the initial and post error states, as average over pure states drawn from the Harr measure.  The blue line shows the fidelity of the error corrected states and the green the fidelity if the state had simply stayed in the subspace $\ket{4,4_z}$,  $\ket{3,3_z}$.}
\label{F:schem}
\end{minipage}
\end{center}
\end{figure*}

We take our ``physical qubit" computational basis to be the stretched states, $\ket{0}=\ket{3, 3_z}$ and $\ket{1}=\ket{4, 4_z}$, states easily prepared via optical pumping and controlled via microwave-drive rotations on the Bloch sphere Fig.~\ref{F:schem}(i).  Here we have used the shorthand labeling the two quantum numbers $\ket{F,m_z}$, and have denoted the relevant quantization axis by the subscript on the magnetic sublevel.  Such states, however, are very sensitive to dephasing by fluctuations along the bias magnetic field, and such errors are not correctable.  We choose as our encoded qubit basis stretched states along a quantization axis perpendicular to the bias ($x$-axis),  $\{\ket{\bar{0}}= \ket{3, 3_x}, \ket{\bar{1}}=\ket{3,-3_x} \}$, Fig.~\ref{F:schem}(ii).  Choosing this basis, a dephasing error in the $z$-direction acts to transfer probability amplitude into an orthogonal subspace.  Such errors that can be detected and reversed without loss of coherence.  

Our error correction protocol works as follows. Consider an encoded qubit $\ket{\bar{\psi}} = \alpha \ket{\bar{0}} + \beta \ket{\bar{1}}$.  The error operator due to B-field fluctuations is the generator of rotations, $F_z$. Assuming a small rotation angle $2 \epsilon$, when such an error occurs, the state of our encoded qubit is mapped to
\be
e^{-2i \epsilon F_z} \ket{\bar{\psi}} \approx \ket{\bar{\psi}} + \epsilon \left( \alpha \ket{3, 2_x}+ \beta \ket{3, -2_x} \right ) . 
\ee
The error acts to spread our qubit between two orthogonal subspaces, $|m_x|=3$ and $|m_x|=2$, Fig.~\ref{F:schem}(iii).  To diagnose the syndrome we must measure the subspace without measuring qubit.  We can achieve this by coherently mapping the error subspace to the upper hyperfine manifold, Fig.~\ref{F:schem}(iv), followed by a measurement that distinguishes the two hyperfine manifolds, $F=3$ and $F=4$.   Such a coherent mapping cannot be achieved through simple microwave-driven transitions since the bias field is along the $z$-direction while the encoded states are magnetic sublevels along the $x$-direction.  We can instead use the construction of unitary operators on a subspace described in chapter \ref{sec:construct} to design $\pi$-rotations that take the error states to the upper manifold.  This is tricky for our implementation because our protocol only included one magnetic sublevel in the $F=4$ manifold so as to ensure proper phase imprinting.  The solution is to switch the auxiliary state in the upper manifold between two different subspace maps.  First, we consider the control system where $\ket{4,4_z}$ is our auxiliary state and perform a $\pi$-rotation that maps $\ket{3,2_x }$ to $\ket{4,4_z}$, leaving the rest of the space invariant.  Then employ control on the system where $\ket{4, -4_z}$ is the auxiliary state and map  $\ket{3,-2_x }$ to $\ket{4, -4_z}$, with the identity on the remaining space.  A QND measurement of $F$ collapses the state to the initially encoded state when the measurement result is $F=3$, or to the state $ \alpha \ket{4,4_x}+ \beta \ket{4, -4_x}$,  if we find $F=4$, Fig.~\ref{F:schem}(v).  In the final step of the protocol, if an error occurred, we conditionally move the error subspace back to the encoded subspace, which can be achieved through reverse maps of the sort described above.

We simulate here the coherent steps in the error correction protocol.  These are implemented through our efficient search technique to construct subspace maps for the sequences 
\bea
\{ \ket{4,4_z}, \ket{3,3_z} \} &\rightarrow& \{ \ket{3,3_x}, \ket{3,-3_x} \}\nonumber \\
\{ \ket{3,2_x}, \ket{3,-2_x} \} &\rightarrow& \{\ket{4,4_z}, \ket{4,-4_z}\}  \nonumber \\
\{ \ket{4,4_z}, \ket{4,-4_z}\} &\rightarrow& \{ \ket{3,3_x} \ket{3,-3_x} \} \nonumber 
\eea
Each of these maps are achieved through a sequence of $SU(2)$ $\pi$-rotations on a two-dimensional subspace that leave the orthogonal subspaces invariant.  Starting from numerical searches for state preparation maps that have fidelity greater than 0.99, we obtain subspace maps with comparable fidelities.  The performance of this error correction procedure is shown in Fig.~\ref{F:schem}B.  We plot the fidelity between the initial state and the post-error-corrected state, averaged over random initial pure states of the physical qubit, versus the magnitude of the error as described by the rotation angle induced the stray magnetic field.  Even with imperfect subspace transformations the error correction protocol is significantly more robust than free evolution. Of course, like all quantum error correction protocols, we assume here that the time necessary for diagnosing the syndrome and correcting an error is sufficiently shorter than the dephasing time, so that the implementation of error correction does not increase the error probability.
                     
In practice, the most challenging step in the error correction protocol in this atomic physics example is measurement of the syndrome.  This requires addressing of individual atoms and measuring the $F$ quantum number in a manner that preserves coherence between magnetic sublevels.  In principle, this can be achieved through a QND dispersive coupling between an atom and cavity mode that induces an $F$-dependent phase shift on the field that could be detected \cite{khudaverdyan09}.  Alternatively, $F$-dependent fluorescence from a given atom would allow this code to perform ``error detection", without correction.

\chapter{Summary and Outlook}

In this thesis I have studied the general principles of quantum control of finite dimensional quantum systems and their application to the control of alkali atomic spins.  In Ch.~\ref{ch:control} I discussed the general complexity of open-loop quantum control, focusing on the two control tasks: state preparation and unitary construction.  In particular, I provided a pedagogical review of a sequence of papers from the Rabitz group,  \cite{rabitz04,rabitz06, shen06, rabitz05, hsieh08, moore08}, on the topic of control landscape topology.  The conclusion from these papers is that the problem of state preparation has a landscape that is very favorable with respect to local searches for optimal controls, while the landscape for the problem of unitary construction is more much complex, and makes finding optimal controls with the same types of searches unfeasible.  This implies that we must utilize smarter algorithms to find optimal controls when construct full unitary maps.       

The platform with which we explored these control protocols was the alkali atomic spin system discussed in Ch.~\ref{ch:atomic}.  I described two independent control systems developed for these atomic spins.  In the first, \ref{chp:control_magAC}, the control was achieved through applied time-dependent magnetic fields that give rise to a Zeeman interaction, which together with an ``always-on" nonlinear light shift, provided for full controllability.  The Hilbert space controlled with this system was the 7-dimensional $F=3$ ground-state hyperfine manifold in $^{133}$Cs.  The second control system contained no optical fields and instead used oscillating magnetic fields at both rf and microwave frequencies to control the entire 16-dimensional electronic ground state \ref{chp:hamil_mwrf}.  This system is favorable for the types of control we considered in later sections due to both the large number of tunable parameters, which led to many controllable configurations and the simple geometric description of the constituent Hamiltonians as representations of $\mathfrak{su}(2)$.  Additionally, this systems dynamics are essentially coherent, which was not the case with the previous system where the laser light shift induces decoherence in the form of spontaneous photon scattering.            

In Ch.~\ref{chp:state_prep} I discussed open-loop state preparation.  In this control task, the goal is to map a known fiducial state to an arbitrary target with unit fidelity.  I developed an algorithm for finding good control waveforms in Ch.~\ref{sec:stateprep_alg}, which is based on simple gradient search techniques.  We utilized this algorithm to construct state preparations for both of the control systems in Ch.~\ref{ch:atomic}.  By employing a nonlinear light shift in conjunction with time-varying magnetic fields, an experimental implementation of the state-preparation for  waveforms of duration of about 0.5ms yielded the target with a fidelity on the order of 0.8 - 0.9.  The difference between the fidelity in the optimization and in the experiment can be traced to known quantities, such as the precision in the density-matrix reconstruction protocol , inhomogeneities in the laser field or a rotation of the final state due to a mismatch between the state preparation and reconstruction waveforms.  In addition, we looked at preparing squeezed states using our state preparation algorithm in comparison to the adiabatic technique proposed in \cite{molmersorensen}.  Our method for state preparation was about five times faster, which meant that there was less decoherence due to photon scattering.  The imprecision in our application of these complicated control waveforms, however, produced squeezing that was slightly smaller than what was seen using the adiabatic technique.            

With the microwave and rf control fields we were able to numerically construct state preparation protocols from our algorithm that were about an order of magnitude faster (50$\mu$s - 150$\mu$s) even though they acted on a space about twice as large (see Ch.~\ref{ch:sp_mwrf}).  The main reason for the speedup is that we can increase the strength of the effective nonlinearity without fear of increased decoherence rates.  We looked at the performance of a variety of scenarios, restricting some control parameters by, e.g., fixing the amplitudes of the fields or the number of resonant microwaves frequencies.  Under certain conditions, restricted control yielded better performance.  We suspect this is a numerical issue related to the complexity of searching a large dimensional control space. These unrestricted control system should contain higher fidelity waveforms, but for realistic parameters also contain more local optima.  There is simply a larger space to sample from and the moderate length of our searches was fixed independently of the size of the control space.    

With regard to unitary construction, in Ch.~\ref{ch:unitary} I described a protocol that utilizes stochastic searches to construct state preparations, as opposed to stochastically searching for full unitary maps.  The computational resources for this algorithm scale only polynomially with the dimension of the system's Hilbert space and the duration of the control waveforms also scale polynomially with $d$.  This hybrid search technique places only very mild restrictions on the types of Hamiltonians with which our protocol is applicable as opposed to the constraints from other geometric constructions.  The conditions for applicability are that the system dynamics are controllable, we can time-reverse our control fields, and that we can imprint an arbitrary phase on a single fiducial state $\ket{0}$.  With this system we can construct not only unitary matrices on the full Hilbert space, but also maps on subspaces.  With a subspace of dimension $n \leq d$, the difficulty of the numerical search as well as the duration of the control scales like $n/d$ with respect that of the full unitary construction.  As an example we looked at constructing unitary maps in the microwave and rf control system \ref{sec:examples}.  The most restrictive constraint on our system is the necessity of imprinting a phase on a single state.  To achieve this we considered control restricted to the $F=3$ manifold with $\ket{4,4}$ as an auxiliary state.  We used a laser light shift to apply a phase to the $F=4$ manifold relative to the $F=3$, which in this restricted control space acts to imprint a phase on the state $\ket{4,4}$.  With this control system we constructed generators for the generalized Pauli and Clifford groups on the $F=3$ manifold, e.g.~the discrete Fourier transform.  Starting from physically reasonable state preparations of $0.99$ we obtained unitary constructions with maps with a target fidelity of around $0.98$.  We have not performed any detailed error analysis, but this scaling suggests that the fidelity of full unitary maps may go as the square of the state preparation fidelity.  Additionally we looked at constructing subspace maps to enable correction of errors due to unknown $z$-magnetic fields for a qubit encoded in this larger spin.

There are some obvious extensions to the control systems in this thesis that are already being pursued by myself and others.  The microwave and rf control system has the nice property that the independent terms in the Hamiltonian look like the generators of $SU(2)$ in overlapping subspaces described by the two hyperfine spin manifolds, $F=3$ and $F=4$, as well as the 2D subspace spanning these manifolds that is coupled by the microwave interaction.  Much of the power of qubit control comes from our understanding of rotations on the 2-sphere.  In work with Brian Mischuck, we have shown that it is possible to create general unitary operators from $SU(16)$ in this control system out of sequences of $SU(2)$ rotations.  We are currently working to make these individual rotations robust to detuning and amplitude errors using robust composite pulse design techniques from NMR control \cite{vandersypen04,kobzar05} in order to create robust $SU(16)$ evolutions.  

From a more general perspective, in Ch.~\ref{sec:topology}, we discussed the relative complexity for stochastically searching for control waveforms that generate state preparations versus a full dimensional unitary map.  It takes resources polynomial in $d$ to find good state preparation waveforms but resources exponential in $d$ to find unitary constructions.  While it is possible to determine some of the properties of the landscape analytically, to get the scaling, numerics are required.  It would be nice to understand the transition point between polynomial and exponential scaling in these problems.  A state preparation map put constraints on one row of a unitary matrix as opposed to constructing unitary maps where the entire matrix is specified.  Subspace mappings lie somewhere in between.  A reasonable hypothesis would be that the dimension of the subspace must have some intermediate scaling with the Hilbert space dimension in order to require exponential resources.  Confirming this conjecture would be useful since it would imply that stochastic search techniques would scale efficiently with the problem of 2D subspace mapping.  

Beyond the uses for 2D subspace for encoding a qubit into a qudit, the ability to search for 2D subspace maps directly would greatly improve the applicability of the unitary construction technique in Ch.~\ref{ch:unitary}. Finding natural Hamiltonians of the form $\ket{0}\bra{0}$, to imprint a phase on the fiducial state, is fairly difficult.  In most cases one has to proceed by coupling to an ancilla subspace that is not part of the Hilbert space being controlled, and then producing the effective Hamiltonian $\ket{0}\bra{0}$ by tracing out the ancilla.  If the primitive for the phase gate was of the form $\ket{0}\bra{0}-\ket{1}\bra{1}$ no ancilla subspace is required.   In fact, in the examples we have discussed there is already such a term in our Hamiltonian in the form of the microwave coupling.  We have been able to show that it is trivial to extend our unitary protocol to work with this phase primitive and so all that remains is to determine whether one can efficiently search for 2D subspace mappings.  

An important tool in developing these sorts of quantum control protocols that I haven't discussed in much detail in this manuscript is measurement.  Prior work on this subject was carried out in the collaboration between the Deutsch and Jessen groups in the PhD works of Andrew Silberfarb \cite{silberfarb05} and Greg Smith \cite{smith06}.  An important goal for the near future is to extend their work, applied in the context of magnetic field and nonlinear light shift, to the microwave and rf control system. I have collaborated with Carlos Riofrio, who has recently begun to work on this extension.  In the density matrix reconstruction procedure in \cite{silberfarb05}, the observables must be driven through dynamics, in the Heisenberg picture, so as to span an informationally complete set of measurements.  Without an informationally complete set there will always exist some density matrices that cannot be reconstructed with unit fidelity.  In a real physical system, there will additionally be errors in the measurement record due to a sensitivity to external fields as well as Gaussian noise associated with finite measurement statistics.  In this case it is important not only to sample from an informationally complete set, but to sample in some unbiased manner.  Optimizing the dynamics so as to drive the observables uniformly through an informationally complete set has proved exceptionally difficult, and is just barely possible in a seven dimensional Hilbert space with reasonable computational resources.  This is a problem when we would like to consider a 16-dimensional system.  In this case we have been looking at dynamics that correspond to sampling from pseudorandom unitary matrix distributions, such as those that arise from quantum chaotic maps \cite{haake}, in the hopes that random unitary evolution will provide measurement records that are sufficient for reconstruction without requiring a huge computational overhead to optimize the dynamics.      
       
With the tools in this thesis, state preparation and unitary construction, as well as the ability to perform density matrix reconstruction, we reach a level of control that allows us to explore new and interesting physics which, as a physicist, is a primary goal of developing quantum control techniques.  With the AC-Stark shift control system, in \cite{ghose}, the authors were able to use the state preparation techniques developed in \cite{chaudhury07} and the density matrix reconstructions methods from \cite{silberfarb05, smith06} to explore quantum chaos in the quantum-kicked top.  In a $7$-dimensional Hilbert space, which is deep within the quantum regime, it is possible to see features of the classical phase space if one can prepare and measure atomic spins.      

The AC-Stark shift system has truly been a workhorse for exploring quantum control techniques \cite{smith04,smith06,chaudhury07,ghose}, and in the future, I expect that the microwave and rf magnetic field system should be able to fill a similar role.  Currently, this control system is being constructed in Poul Jessen's lab at the University of Arizona.  With state preparation and unitary construction we can hope to see more explorations of quantum chaos in this system, as well as perhaps studies of into many-body physics.  The techniques in this dissertation should also be applicable to single atoms, such as atoms trapped in an optical lattice which is a well-known paradigm for quantum computing.

\chapter*{Appendices}

  

\appendix

\chapter{Generalized Wigner function representation}\label{appen:wigner}

\begin{figure}[t!]
\begin{center}
\includegraphics[width=14cm,clip]{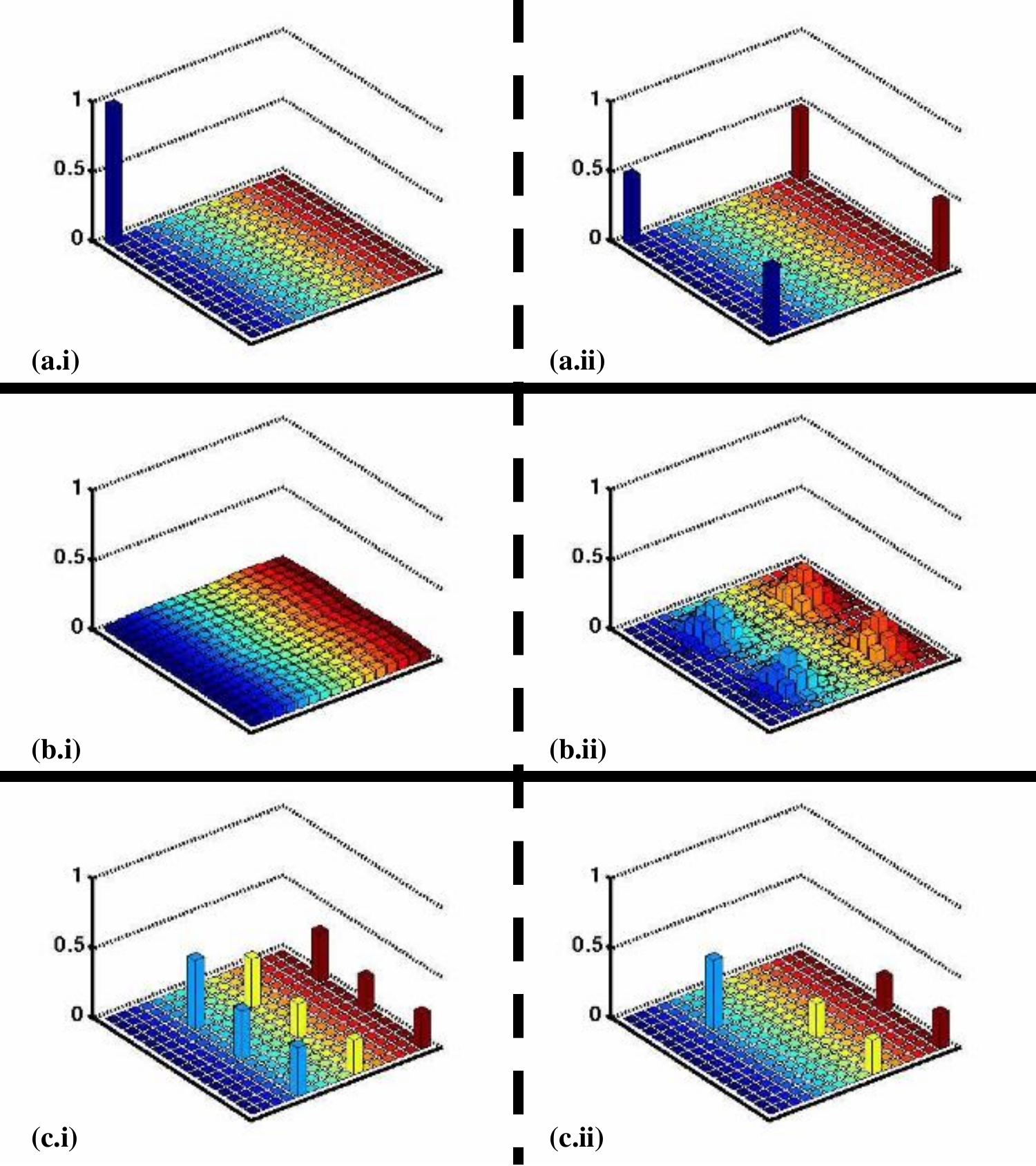}
\caption[Bar chart representations of  tensor product states]{Representations of states with bar charts of the absolute values of the density matrix elements.  (a.i) is a spin coherent state $\ket{\psi}_{ai} =\ket{4,4}$ and (a.ii) is a superposition two oppositely oriented spin coherent states, one for each of the two manifolds,    $\ket{\psi}_{aii} =\frac{1}{\sqrt{2}}(\ket{4,4}+ \ket{3,-3}$.  In (b.i, b.ii) we show the effects of rotations on a superposition of spin squeezed states, each determined as the ground state of $F_z^2-F_y$ in the respective irreducible manifold. Finally, in (c.i) we have a coherent superposition of the state $\ket{4,0}$ and a cat state $\frac{1}{\sqrt{2}}(\ket{3,3} + \ket{3,-3})$ and in (c.ii) we have an incoherent mixture of those two states. }
\label{F:plots1}
\end{center}
\end{figure}

\begin{figure}[t!]
\label{plots2}
\begin{center}
\includegraphics[width=14cm,clip]{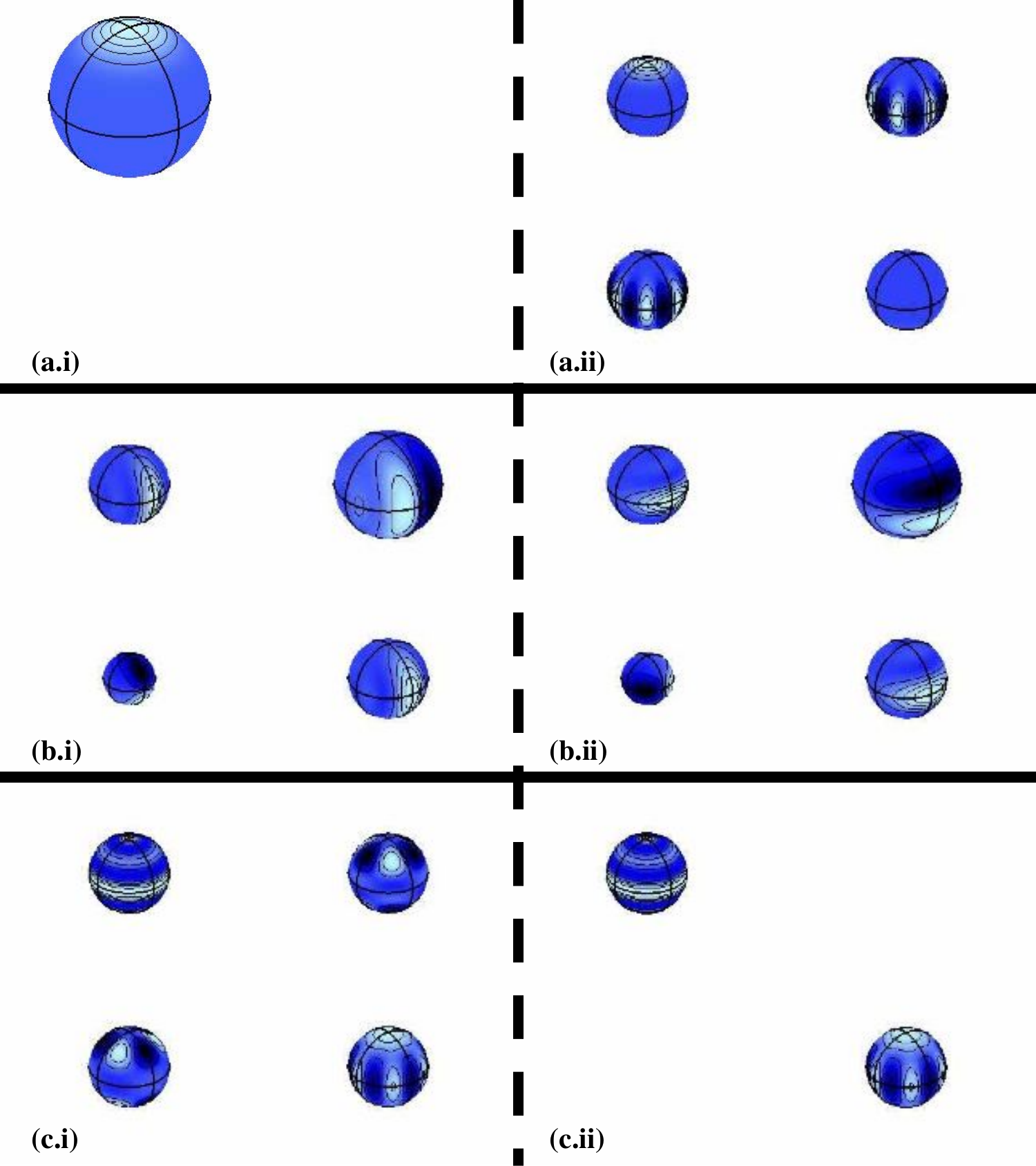}
\caption[Spherical Wigner function representations of tensor product states]{Representations of the six states shown in Fig.~(\ref{F:plots1}) by the generalized spherical Wigner functions. Each state is represented by four spheres.  The spheres on the diagonal are the standard SU(2) Wigner functions in the $F=4$ (upper diagonal) and $F=3$ (lower diagonal) irreducible subspaces. The radius of these spheres, ranging from zero to one, determines the total population in that subspace. The off-diagonal spheres represent the coherences between the two subspaces (see text).}
\label{F:plots2}
\end{center}
\end{figure}

In dealing with high dimensional spin systems, it is useful to be able generate graphical representations of the quantum states which give some geometric intuition.  The spin coherent state Wigner function representation introduced by Agarwal \cite{agarwal81} provides a generalization of the standard Wigner function based on harmonic oscillator coherent states used to describe infinite dimensional systems.  Given a spin $J$, the spin coherent state Wigner function is essentially a multipole representation on the sphere defined as,
  \be
  W_{\hat{\rho}}(\theta, \phi) = \sum_k  \sum_m \textrm{Tr} [\hat{\rho} \hat{T}^{(k)}_q(J)] Y^{(k)}_q(\theta, \phi),
  \ee 
  where $Y^{(k)}_q(\theta, \phi)$ are the spherical harmonics, and   $\hat{T}^{(k)}_q(J)$ are the irreducible spherical tensors given by
\be 
T^{(k)}_q(J) = \sqrt{\frac{2k+1}{2J+1}}\sum_m \braket{J,m+q}{k,q;J,m} \ket{J,m+q}\bra{J,m}.
\ee 
For a given spin, the indices describing non-trivial irreducible tensors run from $0 \leq k \leq 2J$ and $-k \leq q \leq k$.  These plots are useful visualization tools because they capture the effect of geometric rotations on the quantum state.  Two quantum states that differ solely by a SU(2) rotation will generate Wigner functions that also differ from each other by the same physical rotation. 

We seek to generalize this to the case of a tensor product space of two spins (here electron and nuclear), equivalent to the direct sum of two irreducible representations of SU(2) in the hyperfine subspaces, $F$ and $F'$.  We achieve this by considering the expanded set of tensors defined by
  \bea
 &&T^{(k)}_{q}(F,F') = \nonumber\\ 
 &&\sqrt{\frac{2k+1}{2F+1}}\sum_{m} \braket{F,m+q}{k,q;F',m} \ket{F,m+q}\bra{F',m}.\nonumber\\
  \eea 
The range of the indices is now $|F-F'| \leq k \leq F+F'$ and  $-k \leq q \leq k$.  One can easily show that for two spin manifolds, the set of operators $\{T^{(k)}_{q}(F,F), ~$ $T^{(k)}_{q}(F,F'),$$ T^{(k)}_{q}(F',F), $$T^{(k)}_{q}(F',F')$$\}$ comprises a complete orthonormal operator basis for the tensor product space.  We again can map these operators to the spherical harmonics, and for each state get four spherical Wigner functions: one each for the $F$ and $F'$ manifolds, and two for the coherences between manifolds.  We label them $W_{F,F},W_{F,F'},W_{F',F}$ and $W_{F',F'}$.  By the Hemiticity of the density operator, $W_{F,F'}$ and $W_{F',F}$ contain redundant information and are complex, so one need only consider the real and imaginary part of $W_{F,F'}$, yielding four real functions.  

We scale the radii of the spheres over which the Wigner function is plotted.  For the functions that describe a given hyperfine manifold, we let the radius of the sphere equal the population in the subspace, $\textrm{Tr}(P_F \rho P_F)$. In order to set the radii of the spheres corresponding to the coherences between the manifolds, we look at the sum of the singular values of the off-block component of the density matrix, $\sqrt{\sum_m \sum_{m'} |\bra{F,m} \rho \ket{F',m'}|^2}$.  This allows for nonequal dimensions of the two subspace.  Additionally, we scale these ``coherence spheres" by the ratio of the real versus imaginary parts of Wigner function.  The primary purpose of doing this is to be able to distinguish between pure superpositions and incoherent mixtures between the two manifolds.

To gain some intuition, we show examples of different states and different representations.  Figure~(\ref{F:plots1}) shows bar charts of the absolute values of the density matrix elements for the six states: $\ket{\psi}_{ai} = \ket{4,4}$ and $\ket{\psi}_{aii} = (\ket{4,4}+ \ket{3,-3})\sqrt{2}$ are spin coherent states and their superposition;~ $\ket{\psi}_{bi}$, and $\ket{\psi}_{bii}$ are superpositions of spin squeezed states in the two manifolds along different quadratures;~  $\ket{\psi}_{ci}$, and $\ket{\psi}_{cii}$ are coherent superpositions vs. incoherent mixtures of a ``cat state"  $(\ket{3,3}+\ket{3,-3})/\sqrt{2}$ in one manifold, and a Dicke state $\ket{4,0}$ in the other.  The corresponding Wigner functions are shown in Fig.~(\ref{F:plots2}).  From these plots we make the following observations.  When restricted to a subspace corresponding to a given hyperfine manifold, the Wigner functions on the diagonal have the familiar forms of $SU(2)$ Wigner functions, with the radius of the sphere determining the total population in that subspace.  The off-diagonal Wigner functions show the effect of the coherences, had the entire Hilbert space been determined by an irreducible representation.  This is clearly seen in Fig.~(\ref{F:plots2}aii), where the coherences are of the familiar form for a superposition of ``north" and ``south" pole spin coherent states.  The effect of geometric rotation is exhibited in $\ket{\psi}_{bi}$ and $\ket{\psi}_{bii}$.  The bar charts do not indicate any similarity between the states, while the Wigner functions are clearly related by a 90 degree rotation.  Finally, the difference between coherent superpositions and incoherent mixtures of states in the two manifolds is clearly seen in  Fig.~(\ref{F:plots2}c).

\chapter{Controllability code}\label{appen:alg_size_code}

In this appendix I present the Mathmatica Code I used to determine whether a Hamiltonian system is controllable.  This consists of two functions.  The first notebook ``Algsize" computes the iterated commutators of of the initial Hamiltonians and determines whether whey span $\mathfrak{su}(d)$.  The second notebook ``Makebasis" creates a two canonical bases for different representations of $\mathfrak{su}(d)$ and saves them to a file.  The first representation is as irreducible operators on a $d$-dimensional Hilbert space, while the second is the so-called adjoint representation, or the linear operators corresponding to the commutator action of our operators.

\lstset{language=[5.2]Mathematica}

\begin{lstlisting}[breaklines = true]
(* Algsize*)
SetDirectory["/Users/smerkel/research/alg_gen_mathmatica"];
  
  optovec[x_, basis_] := Module[{d, i, jj, temp},
      d = Dimensions[basis][[1]];
      temp = Table[0, {i, (d^2 - 1)}];
      For[jj = 1, jj <= (d^2 - 1), jj++,
        temp[[jj]] = Re[Tr[x.basis[[All, All, jj]]]];
        ];
      1 temp
      ]

AlgSize[hamils_] := Module[{inittime, dims, dh, dalg, dinit, zeroish, 
      fname1, fname2, canbasis, admats, algbasis, baseproject, target, mm, nn,
         test, comaction, numtest, stabil, ttime, lll, ttt, test2, algsize},
      inittime = SessionTime[];
      
      dims = Dimensions[hamils];
      dh = dims[[1]];
      dalg = dh^2 - 1;
      dinit = dims[[3]];
      zeroish = 10.^(-10);
      
      fname1 = "canbasis" <> ToString[dh]; 
      fname2 = "admats" <> ToString[dh]; 
      canbasis = Get[fname1];
      admats = Get[fname2]; 
      
      algbasis = Table[0, {i, dalg}, {j, dalg}];
      baseproject = IdentityMatrix[dalg]; 
      
      target = 1;
      
      For[mm = 1, mm <= dinit, mm++,
        test = optovec[hamils[[All, All, mm]], canbasis];
        If[target > 1, 
          For[nn = 1, nn < target, nn++,
            test = test - (algbasis[[All, nn]].test) algbasis[[All, nn]];
            ];
          ];
        If[test.test > zeroish,
          algbasis[[All, target]] = test / Sqrt[test.test];
          baseproject = baseproject - Outer[Times,
                         algbasis[[All, target]], algbasis[[All, target]]];
          target++;
          ];
        
        
        ];
      
      For[mm = 2, mm <= dalg, mm++,
        
        If[mm == target, Break[]];
        If[target == (dalg + 1), Break[]];
        
        comaction = Table[0, {i, dalg}, {j, dalg}];
        
        For[lll = 1, lll <= dalg, lll++,
          comaction =
               comaction + algbasis[[lll, mm]] admats[[All, All, lll]] ;
          ];
        
        For[nn = 1, nn <= (mm - 1), nn++,
          If[target == (dalg + 1), Break[]];
          test = comaction.algbasis[[All, nn]];
          
          
          test2 = baseproject.test;
          
          If[test2.test2 > zeroish,
            For[ttt = 1, ttt < target, ttt++,
              test = test - (algbasis[[All, ttt]].test) algbasis[[All, ttt]];
              If[test.test < zeroish, Break[]]
              ];
            
            If[test.test > zeroish,
              algbasis[[All, target]] = test / Sqrt[test.test];
              baseproject = baseproject - Outer[
              Times, algbasis[[All, target]], algbasis[[All, target]]];
              target++;
              
              ];
            ];
          ];
        ];
      numtest = Tr[baseproject*Transpose[baseproject]];
      stabil = If[target < dalg, 0,
          If[numtest > 10^(-10), 1, 0]
          ];
      
      algsize = target - 1;
      
      ttime = SessionTime[] - inittime;
      {algsize, stabil, ttime}
      ];

\end{lstlisting}

\begin{lstlisting}[breaklines = true]   

(*saving cannonical basis and comm actions*)
SetDirectory["/Users/smerkel/research/alg_gen_mathmatica"];
    d = 2
    canbasis = Table[0., {ii, d}, {jj, d}, {kk, d^2 - 1}];
    caninc = 1;
    For[ii = 1, ii < d, ii++,
      For[jj = 1, jj < ii + 1, jj++,
        canbasis[[jj, jj, caninc]] = 1/Sqrt[ii^2 + ii];
        ];
      canbasis[[ii + 1, ii + 1, caninc]] = -ii/Sqrt[ii^2 + ii];
      caninc++;
      ];
    For[ii = 1, ii < d, ii++,
      For [jj = ii + 1, jj < (d + 1), jj++,
        canbasis[[ii, jj, caninc]] = 1/Sqrt[2];
        canbasis[[jj, ii, caninc]] = 1/Sqrt[2];
        caninc++;
        ];
      ];
    For[ii = 1, ii < d, ii++,
      For [jj = ii + 1, jj < (d + 1), jj++,
        canbasis[[ii, jj, caninc]] = -I/Sqrt[2];
        canbasis[[jj, ii, caninc]] = I/Sqrt[2];
        caninc++;
        ];
      ];
    
    admats = Table[0., {ii, d^2 - 1}, {jj, d^2 - 1}, {kk, d^2 - 1}];
    
    For[kk = 1, kk < (d^2 - 2), kk++,
      For[jj = kk + 1, jj < (d^2 - 1), jj++,
        For[ii = jj + 1, ii < (d^2), ii++,
          
          temp = Re[Tr[-I(canbasis[[All, 
          All, kk]].canbasis[[
              All, All, jj]] - canbasis[[All, All, jj]].canbasis[[All, All, 
                          kk]]).canbasis[[All, All, ii]] ]];
          
          admats[[ii, jj, kk]] = temp;
          admats[[jj, kk, ii]] = temp;
          admats[[kk, ii, jj]] = temp;
          admats[[ii, kk, jj]] = -temp;
          admats[[kk, jj, ii]] = -temp;
          admats[[jj, ii, kk]] = -temp;
          
          
          ];
        
        
        ];
      
      
      ];
    
    canbasis = SparseArray[canbasis];
    admats = SparseArray[admats];
    canbasis >> "canbasis" <> ToString[d];
    admats >> "admats" <> ToString[d];
\end{lstlisting}

\chapter{State preparation code}\label{appen:sp_cdoe}

In this appendix I present my Matlab code for creating state preparations via gradient search.  This is specifically written for use in  the microwave and rf magnetic field control system.  The code is broken into a number of files which are presented here

The first file is the script, ``opt\_fid" which defines all the system parameters.  The variables are stored in  the data structure opt\_params, which has components:\\
init\_state:  initial state of system (usually $\ket{4,4}$)\\
tot\_time: time of total pulse length\\
samp\_rate:   the sampling rate for out Schrodinger integrator\\
mw\_type:  what type of control fields we use (see make\_hamils\_fields)\\
mw\_amp: rabi frequency\\
mw\_slew:  maximum microwave slew rate\\
rf\_amp: Larmor freq.\\
rf\_slew: maximum rf slew rate\\
hamils: array of hamiltonians (see make\_hamils\_fields)\\
var\_info: this is something else that comes from make\_hamils\_fields.  Basically 	three numbers[number of optimization variables, the number of variables 	that belong to the rf fields, the number of resonant microwave frequencies].\\
fields: control fields found from optimization \\
target: target state, usually just use random states\\
fid: fidelity of optimized field\\

\lstset{language=Matlab}

\begin{lstlisting}[breaklines = true]   
%opt_fid
%this is my script to maximise the fidelity for target states
%full phase and amplitude control for both microwaves and rf both x and
%y coils for the rf.  


%%%%%%%%%%%%%%%%%%%%%
%just some useful folders and loading stuff


load('Units.mat');
%abs_path = '/share/Seth/mwrf_cluster';%only need to worry about this with
%                                        %cluster
data_save_folder = 'fields/test';


%%%%%%%%%%%%%%%%%%%%%
%specifying target states and intitial state, 
%%%%%%%%%%%%%%%%%%%%%

%each target state should be a collumn vector and should have associated 
%with it a name.

%initial state, this is the state |4,4>
init_state = zeros(16,1);
init_state(1,1) = 1;


% Here are two ways to make the target state, either using random targets
% or by using a previously created target library
%%%%%%%%%%%%%%%%%%
% random targets
%%%%%%%%%%%%%%%%%%
n_teststates = 2;% number of targret states
target = zeros(16,n_teststates);
targ_name = cell(1,n_teststates);
for ii=1:n_teststates
    target(:,ii) = random_state(16);
    targ_name{ii} = strcat('rand',int2str(ii));
end;
%%%%%%%%%%%%%%%%%%%%%%%%%%%%
% predefined set of targets 
%%%%%%%%%%%%%%%%%%%%%%%%%%%%
%Can generate random library with make_state_lib.m.

% load targ_lib.mat;
% 
% if size(target,2) ~= length(targ_name)
%     error('need to label targets');
% end;

%%%%%%%%%%%%%%%%%%%%%
%specifying optimization parameters  
%%%%%%%%%%%%%%%%%%%%%
%almost all the optimization parameters can be inputted as vectors.  If you
%put in  vector the code will loop over those parameter values.


%general parameters
tot_time = [50]*us;%time of total pulse length
samp_rate=[0.1]*us; %the sampling rate for out Schrodinger integrator
%not going to loop over samprate
%samp_rate=[5]*us;%fake for test
iters = 20;%For each set of parameters we run the optimizaion iters number usually around 20 
%of times and take the best value(something else we don't loop over)


%mw parameters
mw_type ={'2rfap2struwap'};%look up in make_hamils.m
mw_amp = [40]*kHz;%rabi freq for for stretched state transition
mw_slew = [5]*us;%microwave slew time
%mw_slew = [25]*us;%just to test


%rf parameters
rf_amp = [15]*kHz;%rf Larmor frequency
rf_slew = [10]*us;%rf slew time
%rf_slew = [25]*us;%fake for test





job_size = length(tot_time)*length(mw_type)*length(mw_amp)*length(mw_slew)...
    *length(rf_amp)*length(rf_slew)*size(target,2)*iters;

final_iter = 0;

%%%%%%%%%%%%%%%%%%%%%
%now we loop over everything creating a data structure opt_params that 
%contains all the optimization paramters and saving it to a file
%%%%%%%%%%%%%%%%%%%%%
for ttime = 1:length(tot_time)
for mwt = 1:length(mw_type)
for mwa = 1:length(mw_amp)    
for mws  = 1:length(mw_slew)
for rfa = 1:length(rf_amp) 
for rfs  = 1:length(rf_slew)
    
    
    
    
%%%%%%%%%%%%%%%%%%%%%
%initializing mag field parameters into temp file
%%%%%%%%%%%%%%%%%%%%%
clear opt_params
opt_params.init_state = init_state;
opt_params.tot_time = tot_time(ttime);
opt_params.samp_rate = samp_rate;
opt_params.mw_type = mw_type{mwt};
opt_params.mw_amp = mw_amp(mwa);
opt_params.mw_slew = mw_slew(mws);
opt_params.rf_amp = rf_amp(rfa);
opt_params.rf_slew = rf_slew(rfs);


% making matrices for hamitonians and the vector var_info.  hamils just
% contains the hamiltonians in an array. var_info is three numbers 
%[number of optimization variables, the number of those variablesthat 
% belong to the rf fields, the number of resonant microwave frequencies]. 
[hamils,var_info] = make_hamils_fields(opt_params,0);

opt_params.hamils = hamils;
opt_params.var_info = var_info;

%opt_params.fields is where we will store the optimized waveforms, for now
%zero
opt_params.fields = zeros(4+2*opt_params.var_info(3),...
    1+opt_params.tot_time/opt_params.samp_rate);


for targ = 1:size(target,2)  

    
opt_params.target = target(:,targ);    
opt_params.fid = 0;    
    
final_iter = final_iter+1;%just a counter to see how many fields we'll be finding


%%%%%%%%%%%%%%%%%%%%%
%making file names
%%%%%%%%%%%%%%%%%%%%%

%we give each set of parameters a unique name
waveform_name = strcat(targ_name{targ},'_@',int2str(tot_time(ttime)/us),...
'us_',mw_type(mwt),'_',int2str(mw_amp(mwa)/kHz),'kHz',int2str(10*mw_slew(mws)/us),...
'(usd10)',int2str(rf_amp(rfa)/kHz),'kHz',int2str(rf_slew(rfs)/us),...
'us');

%final_fnames tells us the filenames the fields will be stored in when the
%optimization is finished.  It will look to see if you already have a file
%with that name so as not to erase stuff if you want to try the whole thing
%a second time
final_fnames{final_iter} = strcat(data_save_folder,'/', waveform_name,'.mat');
if exist(char(final_fnames{final_iter}),'file') ~= 2
    save(char(final_fnames{final_iter}),'opt_params');
end



for ii=1:iters
    
    
    
    
    
    
%%%%%%%%%%%%%%%%%%%%%
%temp file names
%%%%%%%%%%%%%%%%%%%%%


% these are temporary filenames tthat we use as an input to our optimizer.
% The optimizer will delete these files after it's don with them.

waveform_name = strcat(targ_name{targ},'_@',int2str(tot_time(ttime)/us),...
    'us_',mw_type(mwt),'_',int2str(mw_amp(mwa)/kHz),'kHz',int2str(10*mw_slew(mws)/us),...
    '(usd10)',int2str(rf_amp(rfa)/kHz),'kHz',int2str(rf_slew(rfs)/us),...
    'us',int2str(ii));


all_fnames{final_iter,ii} = strcat(data_save_folder,'/', waveform_name,'.mat');

save(char(all_fnames{final_iter,ii}),'opt_params');



    
end
end
end
end
end
end
end
end

%%%%%%%%%%%%%%%%%%%%%
%Optimizing
%%%%%%%%%%%%%%%%%%%%%

% This is the part where all the optimization actually happens. There's two 
% chunks of code.  The uncommented bit is for running it on a single machine
% and the commented bit is how I was running things on our cluster.  Since each 
% function call is independent it is a big speed up to parralelize this code.


all_fnames_no_del = all_fnames;% not used for anything except my own diagnostics

%%%%%%%%%%%%%%%%%%%%%
% on a local machine
%%%%%%%%%%%%%%%%%%%%%

for kk = 1:iters
for ii = 1:size(all_fnames,1)
   results = make_optim(char(all_fnames{ii,kk}),char(final_fnames{ii}));
end
end



%%%%%%%%%%%%%%%%%%%%%
% cluster code
%%%%%%%%%%%%%%%%%%%%%

% 
% sched = findResource('scheduler', 'type', 'jobmanager');
% %sched = findResource('scheduler', 'type', 'local');%more diagnostics
%                                           %can run paralel stuff locally 
% 
% 
% fid_max = zeros(1,size(all_fnames,1));
% all_fnames_no_del = all_fnames;
% 
% 
%     
% j = createJob(sched);
% 
% p = {abs_path,strcat(abs_path,'/',data_save_folder)};
% set(j, 'PathDependencies', p);
% for kk = 1:iters
% for ii = 1:size(all_fnames,1)
%    createTask(j, @make_optim2, 2, {char(all_fnames{ii,kk}),char(final_fnames{ii})}) ;
% end
% end
% 
% 
% submit(j);
% 
% waitForState(j);
% results = getAllOutputArguments(j)
% 
% 
% destroy(j);
  
\end{lstlisting}

The function ``make\_hamils\_fields" contains basically all the physics of the problem.  If fflag == 0,  this function creates the Hamiltonians, as well as some variables describing the number of optimization variables.  It does this based on the the physical setup, which I label with "opt\_params.mwtype".  If fflag ==1, this code will take some raw optimization variables and fit them with cubic splines to create physical waveforms with the proper slew rates.  This is also cased out by "mwtype".

\begin{lstlisting}[breaklines = true]
function [hamils_fields,var_info] = make_hamils_fields(opt_params,fflag,x)
%this is the function in which I basically put all of the physics of the
%problem. Basically, when I want to change the physical con(that is by 
%making a new opt_params.mwtype) I only have to change thing here.  I use 
%this function for two different things depending on the fflag.  

%If fflag is 0, this function makes the hamiltonians in an array called 
%hamils and stores some of the relevant variable about the optimization in 
%the vector var_info.

%If fflag is 1, this makes control fields out of the optimization variable
%x using cubic splines.

if isstruct(opt_params)==0
   load(opt_params); 
end

mwtype = opt_params.mw_type;

switch mwtype
    
    case '2rfap2struwap'
    %two spatial rf directions with amplitude and phase control
    %microwaves ressonant on both stretched state transition amplitude and
    %phase control
        if fflag == 0 
        
        all_mw_trans = [3,4;-3,-4];%m_F for microwaves
        rel_amps = [1;1];%scaling factor for rabi frequencies. 
                            %Useful if you want freq other than stretched
                
        fup = 4;
        fdown = 3;
        grel = -1.00321; % just g4/g3

        ntrans = size(all_mw_trans,1);
        hamils_fields = zeros(2*(fup+fdown+1),2*(fup+fdown+1),4+2*ntrans);


        %rf hamiltonians
        upang = make_gen(fup);
        downang = make_gen(fdown);

        hamils_fields(:,:,1) = zeros(2*(fup+fdown+1),2*(fup+fdown+1));
        hamils_fields(:,:,1) = [upang.jx,zeros(2*fup+1,2*fdown+1);
            zeros(2*fdown+1,2*fup+1),grel*(downang.jx)];
        hamils_fields(:,:,2) = zeros(2*(fup+fdown+1),2*(fup+fdown+1));
        hamils_fields(:,:,2) = [upang.jy,zeros(2*fup+1,2*fdown+1);
            zeros(2*fdown+1,2*fup+1),-grel*(downang.jy)];
        hamils_fields(:,:,3) = zeros(2*(fup+fdown+1),2*(fup+fdown+1));
        hamils_fields(:,:,3) = [upang.jy,zeros(2*fup+1,2*fdown+1);
            zeros(2*fdown+1,2*fup+1),grel*(downang.jy)];
        hamils_fields(:,:,4) = zeros(2*(fup+fdown+1),2*(fup+fdown+1));
        hamils_fields(:,:,4) = [upang.jx,zeros(2*fup+1,2*fdown+1);
            zeros(2*fdown+1,2*fup+1),-grel*(downang.jx)];


        %mw hamiltonians

        for ii = 1:ntrans;
            [mw_x,mw_y] = uw_maker_int(all_mw_trans(ii,:));
            hamils_fields(:,:,4+2*ii -1) = rel_amps(ii)*mw_x;
            hamils_fields(:,:,4+2*ii ) = rel_amps(ii)*mw_y;

        end

        %something for the distribution of variable for this optimization
        %let's say var_info is a vector with components (total number of variables
        %needed, number aloocated to rf fields, number of microwave transition)

        nrf_vars = 4*(ceil(opt_params.tot_time/opt_params.rf_slew) - 1);
        nmw_vars = 2*ntrans*(ceil(opt_params.tot_time/opt_params.mw_slew) - 1);
        var_info = [nrf_vars+nmw_vars,nrf_vars,ntrans];
        
        elseif fflag==1
            
            
            ntrans=opt_params.var_info(3);
        nrf_vars = opt_params.var_info(2);


        hamils_fields = zeros(4+2*ntrans,1+opt_params.tot_time/opt_params.samp_rate);

        rf_vars = reshape(x(1:nrf_vars),4,nrf_vars/4);
        mw_vars = reshape(x((nrf_vars+1):end),2*ntrans,(length(x)-nrf_vars)/(2*ntrans)); 

        t = 0:opt_params.samp_rate:opt_params.tot_time;

        rft = 0:opt_params.tot_time/(1+nrf_vars/4):opt_params.tot_time;

        mwt = 0:opt_params.tot_time/(1+(length(x)-nrf_vars)/...
            (2*ntrans)):opt_params.tot_time;

        for ii =1:2
            rf_mags = opt_params.rf_amp*rf_vars(2*ii-1,:);
            rf_thets= 2*pi*cumsum(rf_vars(2*ii,:));
            rfin = rf_mags.*cos(rf_thets);
            rfout = rf_mags.*sin(rf_thets);

            hamils_fields(2*ii-1,:) = spline(rft,[0,rfin,0],t);
            hamils_fields(2*ii,:) = spline(rft,[0,rfout,0],t);

        end

        for ii = 1:ntrans
            mw_mags = opt_params.mw_amp*mw_vars(2*ii-1,:);
            mw_thets= 2*pi*cumsum(mw_vars(2*ii,:));
            mwin = mw_mags.*cos(mw_thets);
            mwout = mw_mags.*sin(mw_thets);

            hamils_fields(4+2*ii-1,:) = spline(mwt,[0,mwin,0],t);
            hamils_fields(4+2*ii,:) = spline(mwt,[0,mwout,0],t);


        end
        
        end
    case '2rfa2struwa'
        %two spatial rf directions with amplitude control and fixed phase
        %microwaves ressonant on both stretched state transition amplitude
        %control and fixed phase
        if fflag == 0
                
        all_mw_trans = [3,4;-3,-4];
        rel_amps = [1;1];%scaling factor for rabi frequencies
                
        fup = 4;
        fdown = 3;
        grel = -1.00321; % just g4/g3

        ntrans = size(all_mw_trans,1);
        hamils_fields = zeros(2*(fup+fdown+1),2*(fup+fdown+1),2+ntrans);


        %rf hamiltonians
        upang = make_gen(fup);
        downang = make_gen(fdown);

        hamils_fields(:,:,1) = zeros(2*(fup+fdown+1),2*(fup+fdown+1));
        hamils_fields(:,:,1) = [upang.jx,zeros(2*fup+1,2*fdown+1);
            zeros(2*fdown+1,2*fup+1),grel*(downang.jx)];       
        hamils_fields(:,:,2) = zeros(2*(fup+fdown+1),2*(fup+fdown+1));
        hamils_fields(:,:,2) = [upang.jy,zeros(2*fup+1,2*fdown+1);
            zeros(2*fdown+1,2*fup+1),grel*(downang.jy)];
        
        %mw hamiltonians

        for ii = 1:ntrans;
            [mw_x,mw_y] = uw_maker_int(all_mw_trans(ii,:));
            hamils_fields(:,:,2+ii -1) = rel_amps(ii)*mw_x;
        end

        %something for the distribution of variable for this optimization
        %let's say var_info is a vector with components (total number of variables
        %needed, number aloocated to rf fields, number of microwave transition)

        nrf_vars = 2*(ceil(opt_params.tot_time/opt_params.rf_slew) - 1);
        nmw_vars = ntrans*(ceil(opt_params.tot_time/opt_params.mw_slew) - 1);
        var_info = [nrf_vars+nmw_vars,nrf_vars,ntrans];


        
        elseif fflag==1
            
            
             ntrans=opt_params.var_info(3);
        nrf_vars = opt_params.var_info(2);


        hamils_fields = zeros(2+ntrans,1+opt_params.tot_time/opt_params.samp_rate);

        rf_vars = reshape(x(1:nrf_vars),2,nrf_vars/2);
        mw_vars = reshape(x((nrf_vars+1):end),ntrans,(length(x)-nrf_vars)/ntrans); 

        t = 0:opt_params.samp_rate:opt_params.tot_time;

        rft = 0:opt_params.tot_time/(1+nrf_vars/2):opt_params.tot_time;

        mwt = 0:opt_params.tot_time/(1+(length(x)-nrf_vars)/...
            (ntrans)):opt_params.tot_time;

        for ii =1:2
            rf_mags = opt_params.rf_amp*rf_vars(ii,:);
            hamils_fields(ii,:) = spline(rft,[0,rf_mags,0],t);
   

        end

        for ii = 1:ntrans
            mw_mags = opt_params.mw_amp*mw_vars(ii,:);
            hamils_fields(2+ii-1,:) = spline(mwt,[0,mw_mags,0],t);
            
        end
            
            
        end
 case '2rfp2struwp'
                
     %two spatial rf directions with fixed amplitude and phase control
    %microwaves ressonant on both stretched state transition fixed amplitude and
    %phase control
    
        if fflag == 0
        all_mw_trans = [3,4;-3,-4];
        rel_amps = [1;1];%scaling factor for rabi frequencies
                
        fup = 4;
        fdown = 3;
        grel = -1.00321; % just g4/g3

        ntrans = size(all_mw_trans,1);
        hamils_fields = zeros(2*(fup+fdown+1),2*(fup+fdown+1),4+2*ntrans);


        %rf hamiltonians
        upang = make_gen(fup);
        downang = make_gen(fdown);

        hamils_fields(:,:,1) = zeros(2*(fup+fdown+1),2*(fup+fdown+1));
        hamils_fields(:,:,1) = [upang.jx,zeros(2*fup+1,2*fdown+1);
            zeros(2*fdown+1,2*fup+1),grel*(downang.jx)];
        hamils_fields(:,:,2) = zeros(2*(fup+fdown+1),2*(fup+fdown+1));
        hamils_fields(:,:,2) = [upang.jy,zeros(2*fup+1,2*fdown+1);
            zeros(2*fdown+1,2*fup+1),-grel*(downang.jy)];
        hamils_fields(:,:,3) = zeros(2*(fup+fdown+1),2*(fup+fdown+1));
        hamils_fields(:,:,3) = [upang.jy,zeros(2*fup+1,2*fdown+1);
            zeros(2*fdown+1,2*fup+1),grel*(downang.jy)];
        hamils_fields(:,:,4) = zeros(2*(fup+fdown+1),2*(fup+fdown+1));
        hamils_fields(:,:,4) = [upang.jx,zeros(2*fup+1,2*fdown+1);
            zeros(2*fdown+1,2*fup+1),-grel*(downang.jx)];


        %mw hamiltonians

        for ii = 1:ntrans;
            [mw_x,mw_y] = uw_maker_int(all_mw_trans(ii,:));
            hamils_fields(:,:,4+2*ii -1) = rel_amps(ii)*mw_x;
            hamils_fields(:,:,4+2*ii ) = rel_amps(ii)*mw_y;

        end

        %something for the distribution of variable for this optimization
        %let's say var_info is a vector with components (total number of variables
        %needed, number aloocated to rf fields, number of microwave transition)

        nrf_vars = 2*(ceil(opt_params.tot_time/opt_params.rf_slew) - 1);
        nmw_vars = ntrans*(ceil(opt_params.tot_time/opt_params.mw_slew) - 1);
        var_info = [nrf_vars+nmw_vars,nrf_vars,ntrans];

        elseif fflag==1
        
            
            
            ntrans=opt_params.var_info(3);
        nrf_vars = opt_params.var_info(2);


        hamils_fields = zeros(4+2*ntrans,1+opt_params.tot_time/opt_params.samp_rate);

        rf_vars = reshape(x(1:nrf_vars),2,nrf_vars/2);
        mw_vars = reshape(x((nrf_vars+1):end),ntrans,(length(x)-nrf_vars)/(ntrans)); 

        t = 0:opt_params.samp_rate:opt_params.tot_time;

        rft = 0:opt_params.tot_time/(1+nrf_vars/2):opt_params.tot_time;

        mwt = 0:opt_params.tot_time/(1+(length(x)-nrf_vars)/...
            (ntrans)):opt_params.tot_time;

        for ii =1:2
            rf_mags = opt_params.rf_amp;
            rf_thets= 2*pi*cumsum(rf_vars(ii,:));
            rfin = rf_mags.*cos(rf_thets);
            rfout = rf_mags.*sin(rf_thets);

            hamils_fields(2*ii-1,:) = spline(rft,[0,rfin,0],t);
            hamils_fields(2*ii,:) = spline(rft,[0,rfout,0],t);

        end

        for ii = 1:ntrans
            mw_mags = opt_params.mw_amp;
            mw_thets= 2*pi*cumsum(mw_vars(ii,:));
            mwin = mw_mags.*cos(mw_thets);
            mwout = mw_mags.*sin(mw_thets);

            hamils_fields(4+2*ii-1,:) = spline(mwt,[0,mwin,0],t);
            hamils_fields(4+2*ii,:) = spline(mwt,[0,mwout,0],t);


        end
        
        end

    
    case '2rfap1struwap'
        %two spatial rf directions with amplitude and phase control
    %microwaves ressonant on one stretched state transition amplitude and
    %phase control
         
        if fflag == 0
        all_mw_trans = [3,4];
        rel_amps = [1];%scaling factor for rabi frequencies
                
        fup = 4;
        fdown = 3;
        grel = -1.00321; % just g4/g3

        ntrans = size(all_mw_trans,1);
        hamils_fields = zeros(2*(fup+fdown+1),2*(fup+fdown+1),4+2*ntrans);


        %rf hamiltonians
        upang = make_gen(fup);
        downang = make_gen(fdown);

        hamils_fields(:,:,1) = zeros(2*(fup+fdown+1),2*(fup+fdown+1));
        hamils_fields(:,:,1) = [upang.jx,zeros(2*fup+1,2*fdown+1);
            zeros(2*fdown+1,2*fup+1),grel*(downang.jx)];
        hamils_fields(:,:,2) = zeros(2*(fup+fdown+1),2*(fup+fdown+1));
        hamils_fields(:,:,2) = [upang.jy,zeros(2*fup+1,2*fdown+1);
            zeros(2*fdown+1,2*fup+1),-grel*(downang.jy)];
        hamils_fields(:,:,3) = zeros(2*(fup+fdown+1),2*(fup+fdown+1));
        hamils_fields(:,:,3) = [upang.jy,zeros(2*fup+1,2*fdown+1);
            zeros(2*fdown+1,2*fup+1),grel*(downang.jy)];
        hamils_fields(:,:,4) = zeros(2*(fup+fdown+1),2*(fup+fdown+1));
        hamils_fields(:,:,4) = [upang.jx,zeros(2*fup+1,2*fdown+1);
            zeros(2*fdown+1,2*fup+1),-grel*(downang.jx)];


        %mw hamiltonians

        for ii = 1:ntrans;
            [mw_x,mw_y] = uw_maker_int(all_mw_trans(ii,:));
            hamils_fields(:,:,4+2*ii -1) = rel_amps(ii)*mw_x;
            hamils_fields(:,:,4+2*ii ) = rel_amps(ii)*mw_y;

        end

        %something for the distribution of variable for this optimization
        %let's say var_info is a vector with components (total number of variables
        %needed, number aloocated to rf fields, number of microwave transition)

        nrf_vars = 4*(ceil(opt_params.tot_time/opt_params.rf_slew) - 1);
        nmw_vars = 2*ntrans*(ceil(opt_params.tot_time/opt_params.mw_slew) - 1);
        var_info = [nrf_vars+nmw_vars,nrf_vars,ntrans];
        
        elseif fflag==1
            
            
            ntrans=opt_params.var_info(3);
        nrf_vars = opt_params.var_info(2);


        hamils_fields = zeros(4+2*ntrans,1+opt_params.tot_time/opt_params.samp_rate);

        rf_vars = reshape(x(1:nrf_vars),4,nrf_vars/4);
        mw_vars = reshape(x((nrf_vars+1):end),2*ntrans,(length(x)-nrf_vars)/(2*ntrans)); 

        t = 0:opt_params.samp_rate:opt_params.tot_time;

        rft = 0:opt_params.tot_time/(1+nrf_vars/4):opt_params.tot_time;

        mwt = 0:opt_params.tot_time/(1+(length(x)-nrf_vars)/...
            (2*ntrans)):opt_params.tot_time;

        for ii =1:2
            rf_mags = opt_params.rf_amp*rf_vars(2*ii-1,:);
            rf_thets= 2*pi*cumsum(rf_vars(2*ii,:));
            rfin = rf_mags.*cos(rf_thets);
            rfout = rf_mags.*sin(rf_thets);

            hamils_fields(2*ii-1,:) = spline(rft,[0,rfin,0],t);
            hamils_fields(2*ii,:) = spline(rft,[0,rfout,0],t);

        end

        for ii = 1:ntrans
            mw_mags = opt_params.mw_amp*mw_vars(2*ii-1,:);
            mw_thets= 2*pi*cumsum(mw_vars(2*ii,:));
            mwin = mw_mags.*cos(mw_thets);
            mwout = mw_mags.*sin(mw_thets);

            hamils_fields(4+2*ii-1,:) = spline(mwt,[0,mwin,0],t);
            hamils_fields(4+2*ii,:) = spline(mwt,[0,mwout,0],t);


        end
        

        end

    case '2rfa1struwa'
        %two spatial rf directions with amplitude control and fixed phase
        %microwaves ressonant on one stretched state transition amplitude
        %control and fixed phase
         
        if fflag == 0
        all_mw_trans = [3,4];
        rel_amps = [1];%scaling factor for rabi frequencies
                
        fup = 4;
        fdown = 3;
        grel = -1.00321; % just g4/g3

        ntrans = size(all_mw_trans,1);
        hamils_fields = zeros(2*(fup+fdown+1),2*(fup+fdown+1),2+ntrans);


        %rf hamiltonians
        upang = make_gen(fup);
        downang = make_gen(fdown);

        hamils_fields(:,:,1) = zeros(2*(fup+fdown+1),2*(fup+fdown+1));
        hamils_fields(:,:,1) = [upang.jx,zeros(2*fup+1,2*fdown+1);
            zeros(2*fdown+1,2*fup+1),grel*(downang.jx)];       
        hamils_fields(:,:,2) = zeros(2*(fup+fdown+1),2*(fup+fdown+1));
        hamils_fields(:,:,2) = [upang.jy,zeros(2*fup+1,2*fdown+1);
            zeros(2*fdown+1,2*fup+1),grel*(downang.jy)];
        
        %mw hamiltonians

        for ii = 1:ntrans;
            [mw_x,mw_y] = uw_maker_int(all_mw_trans(ii,:));
            hamils_fields(:,:,2+ii -1) = rel_amps(ii)*mw_x;
        end

        %something for the distribution of variable for this optimization
        %let's say var_info is a vector with components (total number of variables
        %needed, number aloocated to rf fields, number of microwave transition)

        nrf_vars = 2*(ceil(opt_params.tot_time/opt_params.rf_slew) - 1);
        nmw_vars = ntrans*(ceil(opt_params.tot_time/opt_params.mw_slew) - 1);
        var_info = [nrf_vars+nmw_vars,nrf_vars,ntrans];
        
        
        elseif fflag==1

            ntrans=opt_params.var_info(3);
        nrf_vars = opt_params.var_info(2);


        hamils_fields = zeros(2+ntrans,1+opt_params.tot_time/opt_params.samp_rate);

        rf_vars = reshape(x(1:nrf_vars),2,nrf_vars/2);
        mw_vars = reshape(x((nrf_vars+1):end),ntrans,(length(x)-nrf_vars)/ntrans); 

        t = 0:opt_params.samp_rate:opt_params.tot_time;

        rft = 0:opt_params.tot_time/(1+nrf_vars/2):opt_params.tot_time;

        mwt = 0:opt_params.tot_time/(1+(length(x)-nrf_vars)/...
            (ntrans)):opt_params.tot_time;

        for ii =1:2
            rf_mags = opt_params.rf_amp*rf_vars(ii,:);
            hamils_fields(ii,:) = spline(rft,[0,rf_mags,0],t);
   

        end

        for ii = 1:ntrans
            mw_mags = opt_params.mw_amp*mw_vars(ii,:);
            hamils_fields(2+ii-1,:) = spline(mwt,[0,mw_mags,0],t);
            
        end
            
        end

 case '2rfp1struwp'
     %two spatial rf directions with fixed amplitude and phase control
    %microwaves ressonant on one stretched state transition fixed amplitude and
    %phase control
         
        if fflag == 0
        all_mw_trans = [3,4];
        rel_amps = [1];%scaling factor for rabi frequencies
                
        fup = 4;
        fdown = 3;
        grel = -1.00321; % just g4/g3

        ntrans = size(all_mw_trans,1);
        hamils_fields = zeros(2*(fup+fdown+1),2*(fup+fdown+1),4+2*ntrans);


        %rf hamiltonians
        upang = make_gen(fup);
        downang = make_gen(fdown);

        hamils_fields(:,:,1) = zeros(2*(fup+fdown+1),2*(fup+fdown+1));
        hamils_fields(:,:,1) = [upang.jx,zeros(2*fup+1,2*fdown+1);
            zeros(2*fdown+1,2*fup+1),grel*(downang.jx)];
        hamils_fields(:,:,2) = zeros(2*(fup+fdown+1),2*(fup+fdown+1));
        hamils_fields(:,:,2) = [upang.jy,zeros(2*fup+1,2*fdown+1);
            zeros(2*fdown+1,2*fup+1),-grel*(downang.jy)];
        hamils_fields(:,:,3) = zeros(2*(fup+fdown+1),2*(fup+fdown+1));
        hamils_fields(:,:,3) = [upang.jy,zeros(2*fup+1,2*fdown+1);
            zeros(2*fdown+1,2*fup+1),grel*(downang.jy)];
        hamils_fields(:,:,4) = zeros(2*(fup+fdown+1),2*(fup+fdown+1));
        hamils_fields(:,:,4) = [upang.jx,zeros(2*fup+1,2*fdown+1);
            zeros(2*fdown+1,2*fup+1),-grel*(downang.jx)];


        %mw hamiltonians

        for ii = 1:ntrans;
            [mw_x,mw_y] = uw_maker_int(all_mw_trans(ii,:));
            hamils_fields(:,:,4+2*ii -1) = rel_amps(ii)*mw_x;
            hamils_fields(:,:,4+2*ii ) = rel_amps(ii)*mw_y;

        end

        %something for the distribution of variable for this optimization
        %let's say var_info is a vector with components (total number of variables
        %needed, number aloocated to rf fields, number of microwave transition)

        nrf_vars = 2*(ceil(opt_params.tot_time/opt_params.rf_slew) - 1);
        nmw_vars = ntrans*(ceil(opt_params.tot_time/opt_params.mw_slew) - 1);
        var_info = [nrf_vars+nmw_vars,nrf_vars,ntrans];

        
        elseif fflag==1
            
            ntrans=opt_params.var_info(3);
        nrf_vars = opt_params.var_info(2);


        hamils_fields = zeros(4+2*ntrans,1+opt_params.tot_time/opt_params.samp_rate);

        rf_vars = reshape(x(1:nrf_vars),2,nrf_vars/2);
        mw_vars = reshape(x((nrf_vars+1):end),ntrans,(length(x)-nrf_vars)/(ntrans)); 

        t = 0:opt_params.samp_rate:opt_params.tot_time;

        rft = 0:opt_params.tot_time/(1+nrf_vars/2):opt_params.tot_time;

        mwt = 0:opt_params.tot_time/(1+(length(x)-nrf_vars)/...
            (ntrans)):opt_params.tot_time;

        for ii =1:2
            rf_mags = opt_params.rf_amp;
            rf_thets= 2*pi*cumsum(rf_vars(ii,:));
            rfin = rf_mags.*cos(rf_thets);
            rfout = rf_mags.*sin(rf_thets);

            hamils_fields(2*ii-1,:) = spline(rft,[0,rfin,0],t);
            hamils_fields(2*ii,:) = spline(rft,[0,rfout,0],t);

        end

        for ii = 1:ntrans
            mw_mags = opt_params.mw_amp;
            mw_thets= 2*pi*cumsum(mw_vars(ii,:));
            mwin = mw_mags.*cos(mw_thets);
            mwout = mw_mags.*sin(mw_thets);

            hamils_fields(4+2*ii-1,:) = spline(mwt,[0,mwin,0],t);
            hamils_fields(4+2*ii,:) = spline(mwt,[0,mwout,0],t);


        end    
        
        end
        
        
    case '2rfap1struw0'
        %two spatial rf directions with amplitude and phase control
    %microwaves ressonant on both stretched state transition fixed amplitude and
    %fixed phase "always on"
        
        if fflag == 0
        all_mw_trans = [3,4];
        rel_amps = [1];%scaling factor for rabi frequencies
                
        fup = 4;
        fdown = 3;
        grel = -1.00321; % just g4/g3

        ntrans = size(all_mw_trans,1);
        hamils_fields = zeros(2*(fup+fdown+1),2*(fup+fdown+1),2+ntrans);


        %rf hamiltonians
        upang = make_gen(fup);
        downang = make_gen(fdown);

        hamils_fields(:,:,1) = zeros(2*(fup+fdown+1),2*(fup+fdown+1));
        hamils_fields(:,:,1) = [upang.jx,zeros(2*fup+1,2*fdown+1);
            zeros(2*fdown+1,2*fup+1),grel*(downang.jx)];
        hamils_fields(:,:,2) = zeros(2*(fup+fdown+1),2*(fup+fdown+1));
        hamils_fields(:,:,2) = [upang.jy,zeros(2*fup+1,2*fdown+1);
            zeros(2*fdown+1,2*fup+1),-grel*(downang.jy)];
        hamils_fields(:,:,3) = zeros(2*(fup+fdown+1),2*(fup+fdown+1));
        hamils_fields(:,:,3) = [upang.jy,zeros(2*fup+1,2*fdown+1);
            zeros(2*fdown+1,2*fup+1),grel*(downang.jy)];
        hamils_fields(:,:,4) = zeros(2*(fup+fdown+1),2*(fup+fdown+1));
        hamils_fields(:,:,4) = [upang.jx,zeros(2*fup+1,2*fdown+1);
            zeros(2*fdown+1,2*fup+1),-grel*(downang.jx)];


        %mw hamiltonians

        
        hamils_fields(:,:,5) = uw_maker_int(all_mw_trans(1,:));
     
        %something for the distribution of variable for this optimization
        %let's say var_info is a vector with components (total number of variables
        %needed, number aloocated to rf fields, number of microwave transition)

        nrf_vars = 4*(ceil(opt_params.tot_time/opt_params.rf_slew) - 1);
        nmw_vars = 0;
        var_info = [nrf_vars+nmw_vars,nrf_vars,ntrans];
        
        elseif fflag==1
            
            ntrans=opt_params.var_info(3);
        nrf_vars = opt_params.var_info(2);


        hamils_fields = zeros(5,1+opt_params.tot_time/opt_params.samp_rate);

        rf_vars = reshape(x(1:nrf_vars),4,nrf_vars/4);
        
        t = 0:opt_params.samp_rate:opt_params.tot_time;

        rft = 0:opt_params.tot_time/(1+nrf_vars/4):opt_params.tot_time;

 
        for ii =1:2
            rf_mags = opt_params.rf_amp*rf_vars(2*ii-1,:);
            rf_thets= 2*pi*cumsum(rf_vars(2*ii,:));
            rfin = rf_mags.*cos(rf_thets);
            rfout = rf_mags.*sin(rf_thets);

            hamils_fields(2*ii-1,:) = spline(rft,[0,rfin,0],t);
            hamils_fields(2*ii,:) = spline(rft,[0,rfout,0],t);

        end

        
        
        hamils_fields(5,:) = opt_params.mw_amp*ones(1,1+opt_params.tot_time/opt_params.samp_rate);

        end

    case '2rfa1struw0'
        %two spatial rf directions with amplitude control and fixed phase
    %microwaves ressonant on both stretched state transition fixed amplitude and
    %fixed phase "always on"
          
        if fflag == 0
         all_mw_trans = [3,4];
        rel_amps = [1];%scaling factor for rabi frequencies
                    
        fup = 4;
        fdown = 3;
        grel = -1.00321; % just g4/g3

        ntrans = size(all_mw_trans,1);
        hamils_fields = zeros(2*(fup+fdown+1),2*(fup+fdown+1),2+ntrans);


        %rf hamiltonians
        upang = make_gen(fup);
        downang = make_gen(fdown);

        hamils_fields(:,:,1) = zeros(2*(fup+fdown+1),2*(fup+fdown+1));
        hamils_fields(:,:,1) = [upang.jx,zeros(2*fup+1,2*fdown+1);
            zeros(2*fdown+1,2*fup+1),grel*(downang.jx)];       
        hamils_fields(:,:,2) = zeros(2*(fup+fdown+1),2*(fup+fdown+1));
        hamils_fields(:,:,2) = [upang.jy,zeros(2*fup+1,2*fdown+1);
            zeros(2*fdown+1,2*fup+1),grel*(downang.jy)];
        
        %mw hamiltonians

        hamils_fields(:,:,5) = uw_maker_int(all_mw_trans(1,:));

        %something for the distribution of variable for this optimization
        %let's say var_info is a vector with components (total number of variables
        %needed, number aloocated to rf fields, number of microwave transition)

        nrf_vars = 2*(ceil(opt_params.tot_time/opt_params.rf_slew) - 1);
        nmw_vars = 0;
        var_info = [nrf_vars+nmw_vars,nrf_vars,ntrans];
        
        elseif fflag==1
            
            ntrans=opt_params.var_info(3);
        nrf_vars = opt_params.var_info(2);


        hamils_fields = zeros(5,1+opt_params.tot_time/opt_params.samp_rate);

        rf_vars = reshape(x(1:nrf_vars),2,nrf_vars/2);
        
        t = 0:opt_params.samp_rate:opt_params.tot_time;

        rft = 0:opt_params.tot_time/(1+nrf_vars/2):opt_params.tot_time;

       
        for ii =1:2
            rf_mags = opt_params.rf_amp*rf_vars(ii,:);
            hamils_fields(ii,:) = spline(rft,[0,rf_mags,0],t);
   

        end

        hamils_fields(5,:) = opt_params.mw_amp*ones(1,1+opt_params.tot_time/opt_params.samp_rate);

        end

 case '2rfp1struw0'
         %two spatial rf directions with fixed amplitude and phase control
    %microwaves ressonant on both stretched state transition fixed amplitude and
    %fixed phase "always on"
        
        if fflag == 0
         all_mw_trans = [3,4];
        rel_amps = [1];%scaling factor for rabi frequencies
                                
        fup = 4;
        fdown = 3;
        grel = -1.00321; % just g4/g3

        ntrans = size(all_mw_trans,1);
        hamils_fields = zeros(2*(fup+fdown+1),2*(fup+fdown+1),2+ntrans);


        %rf hamiltonians
        upang = make_gen(fup);
        downang = make_gen(fdown);

        hamils_fields(:,:,1) = zeros(2*(fup+fdown+1),2*(fup+fdown+1));
        hamils_fields(:,:,1) = [upang.jx,zeros(2*fup+1,2*fdown+1);
            zeros(2*fdown+1,2*fup+1),grel*(downang.jx)];
        hamils_fields(:,:,2) = zeros(2*(fup+fdown+1),2*(fup+fdown+1));
        hamils_fields(:,:,2) = [upang.jy,zeros(2*fup+1,2*fdown+1);
            zeros(2*fdown+1,2*fup+1),-grel*(downang.jy)];
        hamils_fields(:,:,3) = zeros(2*(fup+fdown+1),2*(fup+fdown+1));
        hamils_fields(:,:,3) = [upang.jy,zeros(2*fup+1,2*fdown+1);
            zeros(2*fdown+1,2*fup+1),grel*(downang.jy)];
        hamils_fields(:,:,4) = zeros(2*(fup+fdown+1),2*(fup+fdown+1));
        hamils_fields(:,:,4) = [upang.jx,zeros(2*fup+1,2*fdown+1);
            zeros(2*fdown+1,2*fup+1),-grel*(downang.jx)];


        %mw hamiltonians

        hamils_fields(:,:,5) = uw_maker_int(all_mw_trans(1,:));

        %something for the distribution of variable for this optimization
        %let's say var_info is a vector with components (total number of variables
        %needed, number aloocated to rf fields, number of microwave transition)

        nrf_vars = 2*(ceil(opt_params.tot_time/opt_params.rf_slew) - 1);
        nmw_vars = 0;
        var_info = [nrf_vars+nmw_vars,nrf_vars,ntrans];
        
        
        elseif fflag==1

             ntrans=opt_params.var_info(3);
        nrf_vars = opt_params.var_info(2);


        hamils_fields = zeros(5,1+opt_params.tot_time/opt_params.samp_rate);

        rf_vars = reshape(x(1:nrf_vars),2,nrf_vars/2);
        
        t = 0:opt_params.samp_rate:opt_params.tot_time;

        rft = 0:opt_params.tot_time/(1+nrf_vars/2):opt_params.tot_time;

        
        for ii =1:2
            rf_mags = opt_params.rf_amp;
            rf_thets= 2*pi*cumsum(rf_vars(ii,:));
            rfin = rf_mags.*cos(rf_thets);
            rfout = rf_mags.*sin(rf_thets);

            hamils_fields(2*ii-1,:) = spline(rft,[0,rfin,0],t);
            hamils_fields(2*ii,:) = spline(rft,[0,rfout,0],t);

        end

        hamils_fields(5,:) = opt_params.mw_amp*ones(1,1+opt_params.tot_time/opt_params.samp_rate);


        end













end
end




function [mw_x,mw_y] = uw_maker_int(mwtran)
%little function to make the pauli operators between the correct m_F states
fup = 4;
fdown = 3;

mw_x = zeros(2*(fup+fdown+1),2*(fup+fdown+1));
mw_x(fup + 1 + mwtran(2), 2*fup + 1 + fdown + 1 + mwtran(1)) = 1/2;
mw_x(2*fup + 1 + fdown + 1+mwtran(1), fup + 1+mwtran(2)) = 1/2;
mw_y = zeros(2*(fup+fdown+1),2*(fup+fdown+1));
mw_y(fup + 1+mwtran(2), 2*fup + 1 + fdown + 1+mwtran(1)) = -i/2;
mw_y(2*fup + 1 + fdown + 1+mwtran(1), fup + 1+mwtran(2)) = i/2;


end



\end{lstlisting}

``make\_optim" basically takes an input file, optimizes the control waveform with "fmincon", deletes the input file, and conditional on the new waveform being better than previous waveforms save it to the specified save file location. 

\begin{lstlisting}[breaklines = true]
function [t_timing, best_fid] = make_optim(fname,save_name)
%make optim basically does all the optimization.  It takes an input file
%from fname, finds an optimal state preparation and staores it to save_name
%conditional one the new fidelity being higher than whatever was previously
%in save_name.  Function outputs the fidelity as well as the time it took
%the program to run.

init_time = cputime;
load(save_name);
past_fid = opt_params.fid;


if past_fid > 0.99
    
    %%%%%%%%%%%%%%%%%%%%%%%%%%%%%%%%%%%%%%%%%%%%%%%%%%%%%%%%%%%%%%%%%%%%%%%
    %I decided we'd never need a waveform with a fidelity higher than 0.99
    %so if we already have a good waveform from a previous optimization
    %this function doesn't run an optimization  You can change this value to whatever you
    %want, but if you're running a big batch of optimizations some will
    %find good fields before the others so you'd like to not waste
    %resources optimizing something that's already pretty good.
    %%%%%%%%%%%%%%%%%%%%%%%%%%%%%%%%%%%%%%%%%%%%%%%%%%%%%%%%%%%%%%%%%%%%%%%
    best_fid = past_fid;
    delete(fname);%removes temp file
    t_timing = 0;
else
load(fname);


fidforpsi = @(x) -fid_mwrf(opt_params, x);%objective for optimization
lb = -ones(1,opt_params.var_info(1));
ub =  ones(1,opt_params.var_info(1));
rand_vars =  rand(1,opt_params.var_info(1));%random seed
vars_lmax=fmincon(fidforpsi,rand_vars,[],[],[],[],lb,ub,[],...
       optimset('TolX',1e-3,'TolFun',1e-3,'Display','iter'));

%vars_lmax = rand_vars;%just for diagnostics optimization takes a long time    
fields = make_hamils_fields(opt_params,1,vars_lmax);%make fields from optimum 
opt_params.fields = fields;
opt_params.fid = fid_mwrf(opt_params);%calculate fidelity
best_fid = opt_params.fid;


% is this better than the previous optimium?
if best_fid > past_fid
    save(save_name, 'opt_params');
else
    best_fid = past_fid;
end
delete(fname);%remove temp file
t_timing = (cputime-init_time)/60;
end

\end{lstlisting}

``fid\_mwrf" calculates the fidelity of a state preparation.  This can be called either with the data structure opt\_params or a file name as an input. 

\begin{lstlisting}[breaklines = true]
function fid = fid_mwrf(opt_params,x)
%will output fidelity of state preparation.  opt_params is the data
%structure opt_params in the optimization, but can also be the file_name
%where opt_params is stored.  I call this with the filename input after I
%have optimized fields to see that everything checks out.  During the
%optimization, espcially on a cluster architecture, it is fairly exspensive
%to load things over and over again.  x is the optimization variables, and
%can be left out if you're using this function outsid eof the optimization.

if isstruct(opt_params)==0
   load(opt_params); 
end

if nargin > 1
%makes fields out of the optimization variables and puts them in opt_params
fields = make_hamils_fields(opt_params,1,x);
opt_params.fields = fields;
end

%calls the schrodinger evolution
psi_f = unit_evol_mwrf(opt_params);

%calculate fidelity
fid = abs(opt_params.target'*psi_f);
\end{lstlisting}

``unit\_evol\_mwrf" is a Schrodinger integrator.

\begin{lstlisting}[breaklines = true]
function psi_f = unit_evol_mwrf(opt_params)
%just a simple schrodinger integrator

if isstruct(opt_params)==0
   load(opt_params); 
end


psi_f = opt_params.init_state;

for ii = 1:size(opt_params.fields,2)
    ht =0;
    for jj=1:size(opt_params.fields,1)
        ht = ht + opt_params.fields(jj,ii).*opt_params.hamils(:,:,jj);
    end
    psi_f = expm(-i*opt_params.samp_rate*ht)*psi_f;
end;
\end{lstlisting}

``make\_gen" provides generator's of angular momentum on an arbitrary spin.

\begin{lstlisting}[breaklines = true]
function Anggen = make_gen(s)
%generates the irrdeuciable angular momentum operators for a spin-s system

d = 2*s+1;
Anggen.jx=zeros(d);
for m=1:d
   for n=1:d
      if(m+1==n)
          Anggen.jx(m,n)=(1/2)*sqrt((d-m)*m);
           Anggen.jx(n,m)=(1/2)*sqrt((d-m)*m);
      end;
      
   end;
   
end;

Anggen.jy=zeros(d);
for m=1:d
   for n=1:d
      if(m+1==n)
          Anggen.jy(m,n)=-i*(1/2)*sqrt((d-m)*m);
           Anggen.jy(n,m)=i*(1/2)*sqrt((d-m)*m);
      end;
      
   end;
   
end;

Anggen.jz=zeros(d);
for m =0:(d-1)
    Anggen.jz(m+1,m+1) = (d-1)/2 - m;
end;
clear m n d 

\end{lstlisting}

\bibliographystyle{AMS}
\bibliography{MerkelDissertation}

\end{document}